\documentclass{aa}
\usepackage{graphicx}
\usepackage{txfonts}
\usepackage{natbib}
\usepackage{epsfig}
\bibpunct{(}{)}{;}{a}{}{,}
\def\harrow{\mathrel{\hbox{\rlap{\hbox{\raise4pt\hbox{${\rm +H}$}}}\hbox{$\longrightarrow$}}}}
\def\oarrow{\mathrel{\hbox{\rlap{\hbox{\raise4pt\hbox{${\rm +O}$}}}\hbox{$\longrightarrow$}}}}
\def\lesssim{\mathrel{\hbox{\rlap{\hbox{\lower4pt\hbox{$\sim$}}}\hbox{$<$}}}}
\def\gtsim{\mathrel{\hbox{\rlap{\hbox{\lower4pt\hbox{$\sim$}}}\hbox{$>$}}}}

\begin{document}

\title{Physical structure of the envelopes of intermediate-mass protostars}

\author{N.Crimier\inst{1} \and C. Ceccarelli\inst{1} \and T. Alonso-Albi\inst{2} \and A. Fuente\inst{2}
 P. Caselli\inst{3} \and 
D. Johnstone\inst{4,5} \and 
C. Kahane\inst{1} \and 
 B. Lefloch\inst{1} \and 
S. Maret\inst{1} \and 
R. Plume\inst{6} \and 
J.R. Rizzo\inst{7} \and 
M. Tafalla\inst{2} \and 
E. van Dishoeck\inst{8} \and 
F. Wyrowski\inst{9} }
\institute{Laboratoire d'Astrophysique de l'Observatoire de Grenoble, 
BP 53, 38041 Grenoble, Cedex 9, France. \email{ncrimier@obs.ujf-grenoble.fr}
 \and Observatorio Astron\'omico Nacional (OAN,IGN), Apdo 112, E-28803 Alcal\'a de Henares.
 \and School of Physics \& Astronomy, E.C. Stoner Building, The University of Leeds, Leeds LS2 9JT, UK.
 \and National Research Council of Canada, Herzberg Institute of Astrophysics, 5071West Saanich Road, Victoria, BC, V9E 2E7, Canada.
 \and Department of Physics \& Astronomy, University of Victoria, Victoria, BC, V8P 1A1, Canada.
 \and Department of Physics and Astronomy, University of Calgary, Calgary, T2N 1N4, AB, Canada.
 \and Centro de Astrobiolog\'{i}a (CSIC/INTA), Laboratory of Molecular Astrophysics, Ctra. Ajalvir km. 4, E-28850 Torrej\'{o}n de Ardoz, Spain
 \and Leiden Observatory, P.O. Box 9513, 2300 RA Leiden, Netherlands. 
 \and Max-Planck-Institut fur Radioastronomie, Auf dem H\"ugel 69, D-53121 Bonn, Germany.
 }

\offprints{N.Crimier}

\date{}

\abstract{Intermediate mass protostars provide a bridge between low- and
high-mass protostars. Furthermore, they are an important component of the UV
  interstellar radiation field. Despite their relevance, little is known about
  their formation process.}
{We present a systematic study of the physical structure of five
  intermediate mass, candidate Class 0 protostars. Our two goals are
  to shed light on the first phase of intermediate mass star formation and
  to compare these protostars with low- and high-mass sources.}
{We derived the dust and gas temperature and density profiles of the
  sample. We analysed all existing continuum
  data on each source and modelled the resulting SED with 
  the 1D radiative transfer code
  DUSTY. The gas temperature was then predicted by means of a modified
  version of the code CHT96.}
{We found that the density profiles of five out of six studied intermediate mass envelopes are consistent with the predictions of the ``inside-out'' collapse theory. We compared several physical parameters, like the power law index of the density profile, the size, the mass, the average density, the density at 1000 AU and the density at 10 K of the envelopes of low-, intermediate, and high-mass protostars. When considering these various physical parameters, the
  transition between the three groups appears smooth, suggesting that the
  formation processes and triggers do not substantially differ.} 
{} 

\titlerunning{The dust physical structure of an Intermediate mass protostars sample}
\authorrunning{N.Crimier et al.}

\maketitle

\section{Introduction}\label{sec:introduction}

Stars are usually catalogued in three groups according to their mass:
low-mass ($\lesssim$2 M$_\odot$), intermediate mass (IM) and high-mass
($\gtsim$8 M$_\odot$). The high-mass stars are usually found to form
in tight clusters \citep[e.g.][]{Hillenbrand1998}, including many low-mass
stars as well. Low-mass stars on their own can be
formed in isolation or in loose groups of a few objects per cubic parsec
\citep{Gomez1993}.
While there is reasonable consensus on the overall process leading
to the formation of isolated, low-mass stars \citep[see e.g.][]{Shu1987},
the situation for  high-mass star formation is much more
confusing. The major problem is that a central object with M $\gtsim$ 8
M$_\odot$ produces a radiative pressure stronger than the
gravitational pressure, which should stop the accretion. Therefore
stars with a mass greater than $\sim$8 M$_\odot$ cannot be formed in a
simple-minded fully gravitationally driven process and other
mechanisms need to be considered. Studying the differences in observed
environments, several authors have proposed that high-mass stars are
formed by the coalescence of lower-mass stars, while other authors
favor ``monolithic'' formation \citep[see for example the recent
review by][]{Beuther2007}, where turbulence or winds/outflows can play a major role \citep{Mckee2003,Krumholz2005}.

The differences between low- and high-mass protostars are not only restricted to the dynamics 
of the formation mechanism. Notable differences also exist when it comes 
to the chemistry. \citet{Bottinelli2006} compared the abundances
of O-bearing complex organic molecules (COMs) present in Hot Cores surrounding high-mass
protostars and in Hot Corinos around low-mass protostars. They
concluded that the abundances of the O-bearing COMs, compared to
methanol or formaldehyde, are about one order of magnitude higher in
Hot Corinos than Hot Cores. In addition, significant differences in
the abundances of molecules constituting the grain mantles (H$_2$O,
CO, CO$_2$,CH$_3$OH, OCN$^-$) were observed between low- and high-mass
protostars \citep{Boogert2004}. Finally, in contrast to high-mass
star forming regions, low-mass protostars present an extremely
enhanced molecular deuteration, the so-called Super-Deuteration
\citep[e.g.][]{Ceccarelli2007}, not observed in high-mass star forming
regions.

The IM stars, namely stars whose mass lies in the 2 to 8 M$_\odot$
range, are crucial in studies of star formation because they 
provide the link between low- and high-mass stars
\citep{DiFrancesco1997,Mannings1997,Mannings2000} and therefore
can help to determine whether different formation processes are required.
IM stars are also found in clusters
\citep[e.g.][]{Testi1998,Neri2007,Fuente2007}, with a smooth
transition towards the low-mass star, loose cluster regime for star
masses around 3.5 M$_\odot$ \citep{Testi1999}. These authors also
conclude that IM stars mark the transition from the low-density
aggregates of $\lesssim$ 10 stars per cubic parsec of T Tauri stars to
the dense clusters of $\gtrsim$ 10$^3$ stars per cubic parsec associated
with massive stars.
Finally, IM stars are among the most dominant sources for the
interstellar FUV field \citep[e.g.][]{Habing1968,Gondhalekar1975},
which regulates the phases of the ISM in the Galaxy and, in turn, the
overall Galaxy star formation process and history.
In this context, the study of the formation of the IM stars is not only
 interesting in itself, but it can also greatly help the debate on
the formation of high-mass stars. Indeed, due to their intermediate
position, the study of IM protostars provides crucial information
about the transition between the two formation regimes, as well as on the
limits of the low-mass and high-mass formation scenarios.

Despite their importance, very little is known about the formation and
first evolutionary stages of IM stars. Although some individual IM protostars have been studied in detail \citep{Fuente2007,Neri2007,Hogerheijde1999,Schreyer2002}, the situation is so bad that to
date we neither have a satisfying sample of Class 0 IM objects, namely
objects representing the first phases of stellar formation where the
protostar is embedded in its envelope and its luminosity is dominated
by the accretion luminosity, nor a systematic study of their physical
structure, as in the case for low-mass and high-mass protostars.
On the contrary, several previous studies have considered samples of low-
and high-mass protostars. Relevant similar studies to the present
IM one are those reported by \citet{Jorgensen2002} and \citet{Shirley2002},
 who targeted 7 and 11 low-mass Class 0 protostars respectively,
and the systematic studies by \citet{Mueller2002}, \citet{Hatchell2003},
\citet{Williams2005} and \citet{Vandertak2000}, who
targeted 31, 10, 32, and 9 high-mass protostars respectively. These
authors derived the physical structure of the envelope for each protostar
using continuum observations coupled with a radiative transfer code
assuming a 1D spherical model.  Combined, these studies cover a luminosity
range of 1 $<$ L$_{\odot} <$ 22, and 1$\times$10$^{3}$ $<$ L$_{\odot}
<$ 2$\times$10$^{6}$ for low- and high-mass protostars respectively,
leaving a significant gap in luminosity and  knowledge.

This paper aims to fill this gap, providing a systematic study of the
physical structure of the envelopes of a sample of five IM
protostars. In addition, this study will provide a valuable input for future work on
the chemical structure of IM protostars \citep[e.g.][ in prep]{Alonso-albi2009in_prep}. The
present work follows the method presented in \citet{Crimier2009a} to
study the IM protostars OMC2-FIR4. Namely, the combination of
sub-millimeter to infrared observations with radiative transfer codes are used
to derive the physical structure of each envelope. We selected a
sample of five representative candidate Class 0 IM protostars:
CB3-mm, Cep E-mm, IC1396 N BIMA 2, NGC7129 FIRS 2, and Serpens FIRS
1. These sources were selected using the following criteria:
\begin{itemize}
\item the luminosity is in the 30 to 1000 L$_{\odot}$ range (corresponding to a $\sim$ 3 to 9 M$_{\odot}$ of ZAMS stars);
\item the continuum maps at submillimetre and/or millimetre wavelengths are available;
\item the source is closer than 2.5 kpc to the Sun and isolated, namely no source of similar brightness in the submillimetre within an angular distance of $\sim$100$''$.
\end{itemize}
The coordinates, the distances and the bolometric luminosities of the selected sources are listed in Table \ref{basic_param}.

\begin{table*}[h]
\begin{tabular}{|l|lllll|} \hline
Source               &   CB3       & Cep E       & IC1396 N    & NGC7129     & Serpens     \\ 
                     & mm          & mm          & BIMA 2      & FIRS 2      & FIRS 1      \\  \hline 
RA (2000)      & 00:28:42.1  & 23:03:12.7  & 21:40:41.8  & 21:43:01.5  & 18:29:49.8  \\  
Dec (2000)          & 56:41:59.4  & 61:42:27.4  & 58:16:13.5  & 66:03:25.0  & 01:15:18.4  \\  
Distance (pc)        & 2500   & 730 & 750 & 1250 & 230   \\
Luminosity (L$_{\odot}$) & 1000   & 100 & 150 & 500 & 33     \\ \hline 
\end{tabular}
\caption{The coordinates, the distances and the bolometric luminosities of the sources of the sample.\label{basic_param}}
\end{table*}

This article is organised as follows. Section
\ref{sec:General_method} describes the observations used and the
method and tools employed to derive the dust and gas density and
temperature profiles.
Section \ref{sec:Results} first summarises the general results obtained
toward the IM protostar sample. Next, for each source, background information, 
data available, and the results obtained are presented.
In Sect. \ref{sec:Discussion}, the physical structure of
the envelopes of IM protostars obtained in this study are compared
with those obtained in previous studies of low- and high-mass protostar
envelopes. In this section we also discuss the limitations of our modeling method. Finally, Sect. \ref{sec:Conclusions} concludes the article.

\section{General method}\label{sec:General_method}

In this section we present the general method used to derive the dust
density and temperature profiles by modelling the continuum maps of
each region, plus the Spectral Energy Distribution (SED) from the
millimetre to the Mid-Infrared (MIR). We first
describe the observations used in the analysis and how the brightness
profiles and the SED are derived (Sect. \ref{subsec:observational_data}), and
then we present the modelling (Sect. \ref{subsec:Modelling}).

\subsection{Continuum emission : observational data}\label{subsec:observational_data}

In our analysis we used the maps of the continuum emission at 850 and
450 $\mu$m obtained at the 15 m James Clerk Maxwell Telescope (JCMT). In
addition, we took into account the SED 
of each source from 24 $\mu$m to 1300 $\mu$m data obtained with the 
Spitzer, IRAS, JCMT and IRAM telescopes.
Note that specific details (datasets,
pipeline, uncertainties, contamination ... etc) are given separately
for each source in Sect. \ref{sec:Results}.

\noindent
{\it a) Continuum brightness profiles}\\
For each source we retrieved from  the JCMT Science Archive\footnote{{\it
  http://www.jach.hawaii.edu/JCMT/archive/}} 
  the 450 $\mu$m and 850 $\mu$m maps obtained 
  by the focal-plane instrument SCUBA (Submillimetre Common-User
Bolometer Array). The data were reduced
with the JCMT pipelines SURF V1.6-13, and KAPPA 1.9-7. When multiple
observations were available in the archive, we used the most sensitive
ones. The angular resolution of the maps is approximately 7.5\arcsec and 14.8\arcsec at 450
and 850 $\mu$m respectively 
(see Sect. \ref{subsec:Modelling} for more details). 
Based on the many previous JCMT published
observations, the calibration uncertainty and noise levels are
$\lesssim$ 10\% and 0.04 Jy beam$^{-1}$ at 850 $\mu$m and $\lesssim$
30\% and 0.3 Jy beam$^{-1}$ at 450 $\mu$m, respectively. To evaluate
the continuum brightness profile (the brightness as a function of the
distance from the envelope centre), we averaged the observed flux over
annuli with the same distance from the centre of the source. The final
uncertainties were evaluated taking into account the calibration
uncertainty, the noise levels of the used maps, and the non-sphericity of the source.
The non-sphericity of the source is taken into account via the computation of the standard deviation relative to the azimuthal average for each averaged flux of the brightness profile.

\noindent
{\it b) SED}\\
The SED points at 850 $\mu$m and 450 $\mu$m were obtained integrating the
continuum flux profiles previously derived over the full envelope. We
also considered the integrated flux at 24 $\mu$m and 70 $\mu$m extracted from
the Spitzer Space Telescope Multiband Imaging Photometer (MIPS)
maps. To this end, we retrieved the observations from the Spitzer
Science archive\footnote{{\it http://ssc.spitzer.caltech.edu/archanaly/}}. The
data reduction was performed using the pipeline S16.1.0.  The
evaluation of the fluxes was done with the same method employed for
the maps at 850 $\mu$m and 450 $\mu$m, namely deriving the brightness
profiles from the maps and integrating over each annuli. We evaluated
the uncertainties according to the instrument uncertainties and
varying the outer radius of the envelope to account for the
uncertainty in the envelope size. To do this we first set the outer radius equal to where the flux profile becomes flat (because dominated by the noise or the flux of the parent cloud) and varied this value up to 100 $\%$. In general, the uncertainties remain relatively constant for variations larger than 30-50$\%$. Note that we checked a posteriori that our adopted radius and variation range is constitent with the outer radius obtained after the modelling. Finally, we also considered the IRAS
fluxes at 60 $\mu$m and 100 $\mu$m, respectively, taken from the literature
or the IRAS Point Source Catalog v2.0\footnote{\textit{http://irsa.ipac.caltech.edu/cgi-bin/Gator/nph-dd?catalog=iraspsc}}.
The SED fluxes of each source and their uncertainties are reported in
Table \ref{Integrated_fluxes}.

\begin{table*}[h]
\begin{tabular}{|l|lllll|} \hline \hline 
Source               &   CB3       & Cep E       & IC1396 N    & NGC7129     & Serpens     \\ 
                     & mm          & mm          & BIMA 2      & FIRS 2      & FIRS 1      \\  \hline 
F$_{24\mu m}$ (Jy) - Spitzer   & $1.4\pm0.8$ & $6.0\pm3.0$ & $2.3\pm2.0$ & $6.0\pm3.0$ & $3.0\pm1.0$ \\ 
F$_{60\mu m}$ (Jy) - IRAS   & $31\pm15$   & $61\pm30$   & -           & -           & -           \\ 
F$_{70\mu m}$ (Jy) - Spitzer   & $58\pm20$   & $100\pm40$  & -           & $110\pm60$  & $135\pm50$  \\ 
F$_{100\mu m}$ (Jy) - IRAS   & $110\pm55$  & $110\pm55$  & -           & -           & -           \\ 
F$_{450\mu m}$ (Jy) - JCMT & $30\pm10$   & $44\pm10$   & $100\pm30$  & $80\pm40$   & $130\pm40$  \\ 
F$_{850\mu m}$ (Jy) - JCMT  & $2.0\pm1.0$ & $4.1\pm1.3$ & $17\pm7.0$  & $4.8\pm2.0$ & $14\pm4.0$  \\ 
F$_{1300\mu m}$ (Jy) - IRAM30m  & -           & $1.0\pm0.2$ & -           & $2.2\pm0.5$ & -           \\  \hline
\end{tabular}
\caption{Continuum fluxes integrated over the envelope of each source and used in the SED analysis.\label{Integrated_fluxes}}
\end{table*}

\subsection{Continuum emission : modelling}\label{subsec:Modelling}
To derive the dust temperature and
density profiles of the envelope, we used the 1D radiative transfer code DUSTY
\citep{Ivezic1997}.  Briefly, giving as input the temperature and size of the
central object and a dust density profile, DUSTY self-consistently computes
the dust temperature profile and the dust
emission. The comparison between the computed 450 $\mu$m, 850 $\mu$m and 1300 $\mu$m
brightness profiles (namely the brightness versus the distance from
the centre of the envelope) and integrated SED against the observed brightness
profiles and integrated SED (see previous paragraph) allows one to 
constrain the density profile and, consequently, the temperature
profile of the envelope.

To be compared against the observations, the theoretical emission must be
convolved with the beam pattern of each telescope. Following the
recommendations for the JCMT, the beam is assumed to be a combination of
three Gaussian curves: at 850 $\mu$m we used HPBWs of 14.5\arcsec, 60\arcsec,
and 120\arcsec, with amplitudes of 0.976, 0.022, and 0.002 respectively;
at 450 $\mu$m the HPBWs are 8\arcsec, 30\arcsec, and 120\arcsec. with amplitude
ratios of 0.934, 0.06, and 0.006, respectively \citep{Sandell2001}.

In all sources we assumed that the envelope density follows a single
index power law
\begin{equation} \label{density_power_law}
  n(r)=n_0 \times \left( \frac{r_0}{r} \right)^\alpha,
\end{equation}
where the power law index, $\alpha$, and the density, $n_0$, at $r_0$ 
are free parameters of the model. The envelope is assumed to start
at a radius r$_{in}$ and extends to r$_{out}$. Both r$_{in}$ and
r$_{out}$ are additional free parameters of the model. The last input into DUSTY is
the temperature of the central source, T$_*$, here assumed to be 5000
K. We verified that the choice of this last parameter does not
influence the results. Indeed, given the high optical thickness of the envelopes at the wavelengths 
where the central source emits, the model outputs are not sensitive to the T$_*$ value. 
Finally, the opacity of the dust as function
of the wavelength is another parameter of DUSTY. Following numerous
previous studies \citep{vandertak1999,Evans2001,Shirley2002,Young2003}, 
we adopted the dust
opacity calculated by \citet{Ossenkopf1994}, specifically their OH5 dust
model, which refers to grains coated by ice. We also obtained a
lower limit to T$_{in}$ (the temperature at R$_{in}$) of 300 K: 
any higher value for T$_{in}$ would give similar results. 
Again, given the high optical thickness of the envelope
 at short wavelengths and the low contribution from the dust at T $>$ 300 K
 to the sub-millimetre emission ( as the NIR emission is mainly from
 the very inner part, i.e. a small volume), increasing T$_{in}$ does not make a
 difference to the resulting best-fit.

In summary, the output of DUSTY depends on four free parameters: $\alpha$,
$n_0$, r$_{in}$ and r$_{out}$. In practice, the DUSTY input parameters
are the power law index, $\alpha$, the optical thickness at 100
$\mu$m, $\tau_{100}$, the ratio between the inner and outer radius, Y
(=r$_{out}$/r$_{in}$) and the temperature at the inner radius
T$_{in}$. The optical thickness is, in turn, proportional to the dust
column density which depends on $n_0$ and the physical thickness of
the envelope.  Note that because the beam sizes of the available maps are
relatively large ($\geq 7.5$\arcsec, which corresponds to a radius of $\geq
3250$ AU for a source at a distance of 1000 pc), the inner regions of
the envelopes are relatively unconstrained by the available
observational data.  Finally, as explained in \citet{Ivezic1997}, DUSTY
gives scaleless results (which make it very powerful because the same
grid of models can be applied to many different sources). 
To compare the DUSTY output with actual observations it is necessary
to scale the output by the source bolometric luminosity, L$_{bol}$, and
the distance to the source. Note that the bolometric luminosity 
is estimated by integrating the emission across the full spectrum. 
By definition, this can only be done when the entire SED is known. This is 
exactly one of the outputs of the modelling. Therefore, we 
re-evaluate the luminosity of each source iteratively from the best-fit model, 
by minimising the
$\chi^2$.

We ran a grid of models to cover the parameter space as reported in
Table \ref{DUSTY_input}. 
\begin{table}[h] \centering
\begin{tabular}{|lll|} \hline \hline 
Parameter   & Range   & step \\ \hline
$\alpha$    & 0.2-2.5  &  0.1    \\ 
Y           & 50-2000   & 10  \\ 
$\tau_{100}$ & 0.1-10.  & 0.1 \\
T$_{in}$     & 300 K    & \\
T$_*$       & 5000 K    & \\ \hline 
\end{tabular}
\caption{Range of the input parameters to DUSTY covered in the present
  study. The range of the $\alpha$, Y and $\tau_{100}$ parameters are
  covered by increasing by a constant value at each step of the grid
  reported in the third column. Note that T$_{in}$ and T$_{_*}$ remain
   fixed as they do not influence the results (see
  text). \label{DUSTY_input}}
\end{table}
The best-fit model is found by minimising the $\chi^2$ with an
iterated two-steps procedure.  First, we used the observed brightness
profiles at 450 $\mu m$ and 850 $\mu m$ to constrain $\alpha$ and Y, assuming
a value for $\tau_{100}$. The $\chi^2_{red}$ computed during this first step are reported as $\chi^2_{red}$-maps in Table \ref{best_fit_phy_param}. Second, we constrain the optical thickness
$\tau_{100}$ by comparing the computed and observed SED, assuming the
$\alpha$ and Y of the previous step. The new $\tau_{100}$ is used for
the next iteration and the process is repeared. The $\chi^2_{red}$ computed during this second step are reported as $\chi^2_{red}$-SED in Table \ref{best_fit_phy_param}.
In practice, the iteration converges in two
steps. This occurs because the normalised brightness profiles are only 
very weakly dependent on $\tau_{100}$, while they are very dependent 
on the sizes of the envelope and on the slope of the density profile 
\citep[see also][]{Jorgensen2002,Schoier2002,Crimier2009a}. On
the contrary, the optical thickness depends mostly on the absolute
column density of the envelope, which is constrained by the SED.

\subsection{Gas temperature profile : model description}\label{subsec:Gas_temperature_profile}

\citet{Ceccarelli1996}, \citet{Crimier2009a} and \citet{Doty1997}
showed that the gas can be thermally decoupled from the dust in the
inner regions of low-, intermediate and high-mass protostellar
envelopes. The decoupling occurs mainly in the inner part of the envelopes: for example whereas T$_{dust} \sim$ 150 K at 200 AU,  T$_{gas} \sim$ 90 K. Note that in the outer envelope, where T$_{dust} \lesssim$ 100K, gas and dust temperature only differ by a few percents \citep[see discussion in][]{Crimier2009a}. The reason for this decoupling is the high water
abundance in the gas phase caused by the sublimation of grain
mantles. We therefore explicitly computed the gas temperature
profile of the envelope surrounding each source by finding the
equilibrium temperature obtained by equating the gas cooling and heating
terms at each radius. Following the method described by
\citet{Ceccarelli1996}, we considered heating from the gas compression
(due to the collapse), dust-gas collisions, photo-pumping of H$_2$O
and CO molecules by the IR photons emitted by the warm dust close to
the centre, and cosmic rays ionisation which is a minor heating term in
 protostellar envelopes. The cooling is mainly due to the
rotational lines from H$_2$O and CO, plus the fine structure lines
from O. The gas temperature therefore depends on the abundance of
these three species. In practice, only the water abundance is
an important parameter of the model, because the CO and O lines are
optically thick and LTE populated in the range of CO and O abundances typical of
protostellar envelopes, while the water levels are sub-thermally populated (non-LTE). In the non-LTE regime the water levels are excited by collisional processes and de-excited by radiation while in LTE regime the levels are mainly excited and de-excited by collisions. Therefore, a photon emitted in LTE regime in an optically thick region will be absorbed and then the absorber will be de-excited by collisions while in non-LTE regime the absorbed photon will possibly be re-emitted. In this case we say that the water lines are effectively optically thin. For this reason we computed two cases for the
water abundance, as it is generally poorly constrained in protostellar
envelopes and totally unconstrained in intermediate mass
protostars. We adopted a step function for the water abundance profile
to simulate the jump caused by ice sublimation. The jump is
assumed to occur at 100 K. We considered the H$_2$O abundance (with
respect to H$_2$) X(H$_2$O)$_{in}$ in the inner envelope, where
T$\geq$ 100 K, equal to 10$^{-5}$ and 10$^{-6}$, fixing the water
abundance in the outer region, X(H$_2$O)$_{out}$, at 10$^{-7}$. The CO
and O abundances were fixed at the standard values found in molecular clouds, i.e.  $1\times10^{-4}$ \citep{Frerking1982} and $5\times10^{-4}$ \citep{Caux1999,Vastel2000},
respectively. Note that because the O and CO lines are mostly optically
thick the exact value of their abundance is not important. To
compute the cooling from the lines we used the code described in
\citet{Ceccarelli2003,Ceccarelli1996} and \citet{Parise2005}.  
Briefly, the line cooling is computed with an escape
probability method, which takes into account the dust level pumping and
the line optical depths at each point of the envelope by integrating
over the solid angle. A recent description of the code is reported in 
\citet{Crimier2009a,Crimier2009b}. The same code has been used in several past
studies, whose results have been substantially confirmed by other
groups \citep[e.g. the analysis on IRAS16293-2422 by][]{Schoier2002}.

For the collisional coefficients of water with hydrogen molecules, we
used the data by \citet{Faure2007} and \citet{Faure2008} available for the temperature
range 20-5000K. Because the ortho-to-para conversion process of H$_2$ is
chemical rather than radiative, the Ortho-to-Para Ratio (OPR) H$_2$,
which the water population depends on, is highly uncertain. The recent
analysis of H$_2$CO observations towards a cold molecular cloud by
\citet{Troscompt2009} confirms theoretical estimates \citep[e.g.][]{Flower2006}
 that in molecular clouds the H$_2$ OPR is lower than
1. Lacking specific observations towards protostars, here we assume
that H$_2$ OPR is in Local Thermal Equilibrium and therefore follows
the Boltzmann distribution
\begin{equation}
OPR =  \frac{(2I_o+1) \Sigma (2J+1) \exp(-\frac{E_o(J)}{kT})}
            {(2I_p+1) \Sigma (2J+1) \exp(-\frac{E_p(J)}{kT})},
\end{equation}
where $I_o$ and $I_p$ are the total nuclear spin, corresponding to
whether the hydrogen nuclear spins are parallel ($I_o = 1$,
$\uparrow\uparrow$) or anti-parallel ($I_p = 0$,
$\uparrow\downarrow$). The sum in the numerator and denominator
extends over all ortho and para levels J, respectively. Similarly to
H$_2$, water comes in the ortho and para forms. In these cases, because
the water is the dominant gas coolant only in the regions where the
dust temperature exceeds 100 K, we assumed an OPR equal to 3, strictly
valid for gas temperatures higher than 60 K. Because the water lines
are optically thick, the cooling depends on the velocity field,
assumed to be that of an envelope collapsing in free-fall towards a
central object with a mass M$_\star$. Here, we assumed that the entire envelope is in infall.
 The masses M$_\star$ used for each
source are reported in Table \ref{best_fit_phy_param}. They were
derived analytically, following the equation in
\citet{Stahler1986} which links the mass of the central object
to the bolometric luminosity, the accretion rate, the mass and the radius of the
envelope. Basically, the equation assumes that the luminosity is
entirely due to the gravitational energy released in the collapse and
uses the computation of the hydrostatic core radius of the protostar
by \citet{Stahler1986}. We checked the influence of our
results against this assumption, running cases with M$_\star$ fixed at
2 M$_{\sun}$. The difference in the gas temperature between the two
cases never exceeds 1$\%$.

\subsection{Water line observations}\label{Water_line_observations}

In order to constrain the water abundance, which is very important for computing the
gas temperature, we considered observations obtained by the Long
Wavelength Spectrometer (LWS) aboard the Infrared Space Observatory
(ISO) in the 45 $\mu$m to 200 $\mu$m range, where several water lines emit.
All sources excepted CB3 were observed with the LWS. In two sources,
Cep E-mm and Serpens FIRS 1, several water lines were detected and their
analysis has been reported by \citet{Moro-Mart'in2001} and \citet{Larsson2002},
 respectively. As discussed by these authors, given the
relatively large beam of the LWS, the water line fluxes are due to the
combination of many components along the line of sight:
outflows, multiple sources and Photo-Dissociation Regions. 
The measured H$_2$O line fluxes are thus upper limits
to the fluxes from the envelopes and we checked that our predictions do
not exceed the observed fluxes.  For two other sources, IC1396 N BIMA 2
and NGC7129 FIRS 2, we retrieved the LWS grating spectra (spectral
resolution $\Delta \nu / \nu \sim$200) from the ISO Data Archive\footnote{{\it
 http://iso.esac.esa.int/ida/}} and extracted the upper limits to the
flux of the brightest H$_2$O lines. Table \ref{ISO_upper} summarises
the water line ISO observations for each source.

\section{Results : Dust and gas density and temperature profiles}\label{sec:Results}

The general results of our analysis are:
\begin{itemize}
\item The continuum brightness profiles and the SEDs at wavelengths
  larger than 60 $\mu$m for the five IM-protostars of our sample can be
  reproduced by spherical, single index, power law density models. On the
  contrary, in all sample sources, this class of models fails to
  reproduce the 24 $\mu$m flux, underestimating it by 1 to 3 orders of
  magnitude. The possible causes and implications of this failure are 
  discussed in detail in Sect. \ref{sec:Discussion}. The $\chi^2_{red}$-maps and $\chi^2_{red}$-SED obtained for each source are reported in Table \ref{best_fit_phy_param}. The $\chi^2_{red}$-maps ranges from 0.11 to 0.55 for the sources modelled in this paper. The $\chi^2_{red}$-SED ranges from 0.8 to 4.4 for the sources modeled in this paper. Note that because the flux at 24 $\mu$m is underestimated by several orders of magnitude, the $\chi^2_{red}$-SED value is mainly driven by this point.
\item The power law index for the five sample sources varies between
  1.2 and 2.2, with an average value equal to 1.6.
\item The envelope radius varies between 6000 AU for the lowest
  luminosity and closest source, Serpens FIRS 1, and $1\times10^{5}$ AU
  for the brightest and farthest source, CB3-mm.
\item The radius where T$_{dust}$=100 K lies between 100 (Serpens FIRS 1)
  and 700 (CB3-mm) AU.
\item The density at T$_{dust}$=100 K varies, from 0.4
  to $2\times10^{8}$ cm$^{-3}$.
\item The envelope mass ranges from 5 M$_\odot$ (Serpens FIRS 1) to 120 (CB3-mm)
  M$_\odot$.
\item The mass of the central object is estimated to be between 0.1 M$_\odot$
  (Serpens FIRS 1) and 6 (CB3-mm) M$_\odot$.
\item The gas and the dust are thermally coupled across the envelope
  in three of our sources, with temperature differences lower
  than 5 K.  For IC1396 N BIMA 2 (Sect. \ref{subsec:IC1396N}) and
  NGC7129 FIRS 2 (Sect. \ref{subsec:NGC 7129 FIRS 2}) the difference can
  reache 40 K.
\item The predicted H$_2$O lines are consistent with the ISO upper limits of Sect. \ref{Water_line_observations}.
\end{itemize}
Table \ref{best_fit_phy_param} summarises the best-fit parameters and
some relevant physical quantities derived from the dust radiative
transfer analysis of each source, Table \ref{ISO_upper} lists the water line predictions. Note that given the relatively large beam, the ISO observations are contaminated by the emission from outflows, multiples sources, and  Photo-Dissociation Regions and are therefore considered only as upper limits. 

The appendix describes in detail the source background,
the data included in the analysis, and the derived physical structure (gas and dust
density and temperature profiles) for each source.
\begin{table*}
 $$
 \begin{array}{p{0.5\linewidth}rrrrrr|r}
   \hline
   \multicolumn{2}{c}{\mathrm{Fixed\ input\ parameters}} \\
   \hline
   \hline
   \noalign{\smallskip}
   Stellar temperature, $T_{\star}$    & 5000\,\mathrm{K}\\
   Dust temperature at $r_{\mathrm{in}}$, $T_{\mathrm{in}}$   & 300\,\mathrm{K}\\
   Dust opacity (OH5) at 100\,$\mu$m, $\kappa_{100}$ & 86.5\,\mathrm{cm}^2\mathrm{g}^{-1} \\
 \hline
 \noalign{\smallskip}
 \multicolumn{2}{c}{\mathrm{Best\textendash fit\ parameters\ \&\ physical\ quantities}} \\
 \hline
 \hline
 \noalign{\smallskip}
 Source                                                   & CB3\ & Cep\ E\ & IC1396\ N\ & NGC7129\ & Serpens\ & OMC2\ \\
                                                          & mm\  & mm\     & BIMA2\  &  FIRS\ 2\  & FIRS\ 1    &  FIR4\       \\  \hline
 Distance, ($\mathrm{pc}$)	                          & 2500   & 730 & 750 & 1250 & 230  & 450 \\
 Luminosity, ($\mathrm{L}_{\sun}$)                        & 1000   & 100 & 150 & 500 & 33    & 1000 \\
 Dust optical depth at 100\,$\mu$m, $\tau_{100}$          & 5.8	   & 5.0 & 1.4 & 2.3 & 3.0   & 0.6 \\
 Density power law index, $\alpha$                        & 2.2	   & 1.9 & 1.2 & 1.4 & 1.5   & 0.6 \\
 Envelope thickness, Y (= $r_{\mathrm{out}}$/$r_{\mathrm{in}}$)  & 400	   & 500 & 630 & 180 & 200   & 120 \\
$\chi^2_{red}$-maps         & 0.43 & 0.11  & 0.55  & 0.21 & 0.20 & 1.2 \\
$\chi^2_{red}$-SED                & 0.85 & 0.80 & 1.2  & 0.87  & 4.4 & 0.55 \\
 \hline
 Inner envelope radius, $r_{\mathrm{in}}$, ($\mathrm{AU}$)            & 260    &  70   & 50    & 100   & 30   & 100 \\
 Outer envelope radius, $r_{\mathrm{out}}$, ($\mathrm{AU}$)           & 103000 & 35800 & 29600 & 18600 & 5900 & 12000 \\
 Radius at T$_{dust}$ = 100 K,   $r_{\mathrm{100K}}$, ($\mathrm{AU}$) & 700             & 223   & 180             & 373             & 102  & 440 \\
 H$_2$ density at $r_{\mathrm{100K}}$, $n_0$, ($\mathrm{cm}^{-3}$)    & 7.5\times10^{7} & 2.0\times10^{8} & 4.3\times10^{7} & 4.4\times10^{7} & 2.2\times10^{8}  & 4.3\times 10^{6} \\
 Envelope mass, $M_{\mathrm{env}}$, ($\mathrm{M}_{\sun}$) 	      & 120    &  35   & 90    & 50    & 5.0  & 30 \\
(T$_{dust}$-T$_{gas}$)$_{max}$, (K) 	      & 6  &   5  & 40   &   40   &  10 & 150  \\
 \hline
 \noalign{\smallskip}
 \end{array}
 $$
 \caption[]{Summary of the dust radiative transfer analysis for
   the five IM protostars. The top section lists the fixed input parameters, 
   the lower section reports the best-fit parameters, and some relevant 
   physical quantities corresponding to the best-fit model. 
   The last column reports the same parameters for the IM protostar 
   OMC2-FIR4, found with a similar analysis \citep{Crimier2009a},
   for comparison.}
 \label{best_fit_phy_param}
\end{table*}

\begin{table*}[h] \centering
\begin{tabular}{|cc|cc|cc|cc|cc|cc|} \hline 

Source               &                                  & CB3 & & Cep E & & IC1396 N && NGC7129 && Serpens & \\  \hline 
                     &                                  & mm  & & mm    & & BIMA 2   && FIRS 2  && FIRS 1  & \\  \hline 
$\lambda$    &  Transition    &    Model &  LWS  & Model &  LWS  & Model &  LWS  & Model &  LWS  & Model &  LWS  \\
  ($\mu$m)     & J$_{K_- K_+} -$ J'$_{K'_- K'_+}$ &  & & & & & & & & & \\  \hline  
ortho & & & & & & & & & & &  \\
75.38  & 3$_{2 1} -$  2$_{1 2}$ & 0.3  & -- & 0.2  & 1.4$\pm$0.3 & 0.08 & $<$3.1 & 0.2  & $<$1.7 & 0.5 & 2$\pm$0.4 \\ 
108.07 & 2$_{2 1} -$  1$_{1 0}$ & 0.1  & -- & 0.1  & 1.8$\pm$0.5 & 0.07 & $<$2.3 & 0.1  & $<$1.5 & 0.4 & 1.5$\pm$0.4 \\ 
113.54 & 4$_{1 4} -$  3$_{0 3}$ & 0.1  & -- & 0.1  & 2.2$\pm$0.5 & 0.04 & $<$1.2 & 0.08 & $<$1.0 & 0.2 & 2.9$\pm$0.9 \\  
174.62 & 3$_{0 3} -$  2$_{1 2}$ & 0.06 & -- & 0.07 & 1.6$\pm$1   & 0.06 & $<$2.6 & 0.08 & $<$0.8 & 0.2 & 2$\pm$0.2  \\ 
179.53 & 2$_{1 2} -$  1$_{0 1}$ & 0.09 & -- & 0.1  & 2.9$\pm$0.3 & 0.2  & $<$2.2 & 0.2  & $<$1.1 & 0.5 & 1.4$\pm$0.2 \\  
para  & & & & & & & & & & & \\
89.99  & 3$_{2 2} -$  2$_{1 1}$ & 0.2  & -- & 0.1  & 1$\pm$0.2   & 0.04 & $<$2.4 & 0.1  & $<$1.5 & 0.3 & 2.4$\pm$0.6 \\
100.98 & 2$_{2 0} -$  1$_{1 1}$ & 0.1  & -- & 0.1  & $<$0.5      & 0.04 & $<$1.0 & 0.09 & $<$0.6 & 0.3 & 2.6$\pm$0.5 \\  \hline  
\end{tabular}
\caption{The brightest water lines predicted by the our model
  considering X(H$_2$O)$_{in} =$ 1$\times$10$^{-5}$ and
  X(H$_2$O)$_{out} =$ 1$\times$10$^{-7}$ compared with the ISO LWS
  observations. The upper limits refer to 3$\sigma$. All fluxes are in
  10$^{-12}$erg s$^{-1}$ cm$^{-2}$ units. \label{ISO_upper}}
\end{table*}

\section{Discussion}\label{sec:Discussion}

\subsection{The link between low- and high-mass protostars}

One of the major goals of this work is to verify whether intermediate
mass protostars provide a link between low- and high-mass star formation. In this section, we analyse whether the
parameters describing the protostellar envelope structure (power
law index, dust temperature at a given distance, envelope mass...)
depend on the luminosity, and hence the mass, of the future
central star.

\begin{figure*} \centering
\includegraphics[width=15cm,angle=0]{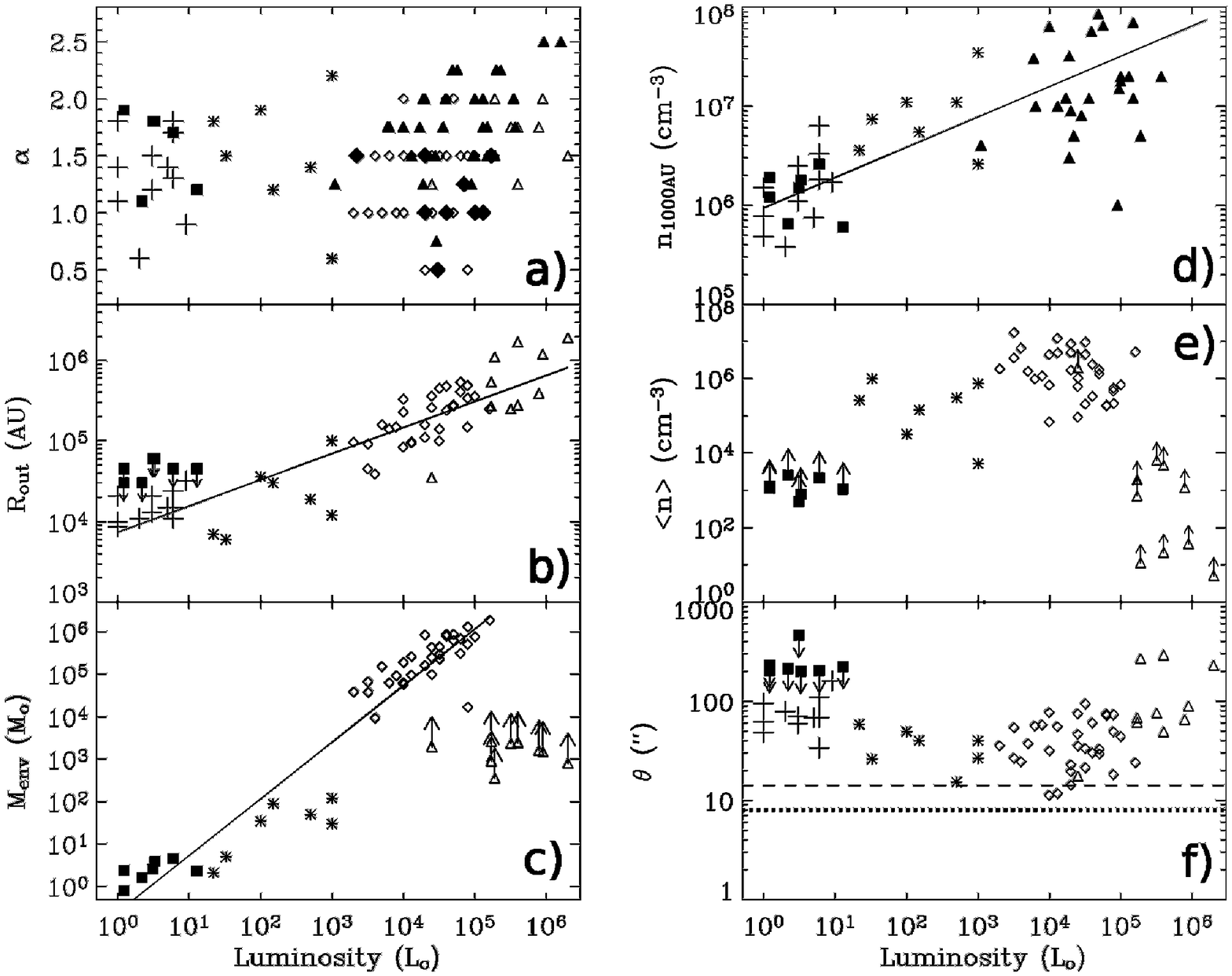}
\caption{Plot of physical quantities within the envelopes surrounding
  low-, intermediate, and high-mass protostars as a function of their
  bolometric luminosity. Each symbol represents a different group of
  sources as follows: crosses: \citet{Jorgensen2002}; filled
  squares: \citet{Shirley2002}; stars: 
  this study and \citet{Crimier2009a,Crimier2009b}; 
  diamonds: \citet{Williams2005}; filled diamonds: \citet{Vandertak2000}
; triangles: \citet{Hatchell2003}; filled
  triangles: \citet{Mueller2002}. Symbols with arrows represent
  upper/lower limits. The solid line represents the linear fit of the points obtained when the correlation is significant (Table \ref{coef_correl}). The dashed line and the dotted line on panel f) represent the angular resolution of the SCUBA maps at 850 and 450 $\mu$m, respectively, which were used in all the in the studies reported in this panel.\label{comparison_all}}
\end{figure*}

Figure \ref{comparison_all} plots key parameters of the envelope
structure (the power law index of the density profile, $\alpha$, the
total mass, M$_{env}$, the outer radius, $r_{out}$, and the average
density, $<$n$>$ of the envelope) for low-, intermediate, and high-mass
protostars as a function on the bolometric luminosity of each
source. The ensemble of the plotted sources covers six orders of
magnitude in luminosity, from about 1 L$_\odot$ to $\sim10^{6}$
L$_\odot$. The plotted data are from the present study
(Table \ref{best_fit_phy_param}) and \citet{Crimier2009a} for the
intermediate mass protostars, \citet{Jorgensen2002}, \citet{Shirley2002}
and \citet{Crimier2009b} for the low-mass protostars, and \citet{Vandertak2000},
\citet{Hatchell2003}, \citet{Williams2005} and
\citet{Mueller2002} for the high-mass protostars. 
Note that \citet{Jorgensen2002}, \citet{Williams2005}, \citet{Hatchell2003}, 
and \citet{Crimier2009a,Crimier2009b},
as well as the present study, use the DUSTY code in the
analysis. The correlation coefficients and probability for a chance correlation between the pair of parameters considered in Fig. \ref{comparison_all} are reported in Table \ref{coef_correl}. The plots and quantities shown in Fig. \ref{comparison_all} and Table \ref{coef_correl}, respectively, lead to the
following remarks.\\

\noindent
{\it Density power law index $\alpha$:}\\
The density power law index $\alpha$ is similar for low, intermediate
and high-mass protostars. In all three cases the average $\alpha$ is
1.5. We find that about 60 \% of the protostars are well modelled by
envelopes with 1.5 $\lesssim \alpha \lesssim$ 2.0. This indicates that
60 \% of the protostar envelopes are consistent with the standard
model of free-fall collapse from an initially singular isothermal
sphere, the so-called inside-out model \citep{Shu1977}. However, it appears
that 35\% of the sample are reproduced by envelopes with much smaller
indexes, namely 0.5 $\lesssim \alpha \lesssim$ 1.5. The theoretical
interpretation of theses low $\alpha$ values is not straightforward. The
phenomenon was already noted in previous studies of low-mass
protostars \citep[e.g.][]{Andre1993,Chandler1998,Motte2001}. 
Various hypotheses to explain theses low $\alpha$ values have been
evoked in the literature. One possibility is that the
  envelopes with low $\alpha$ are described by the collapse of an
  initially logotropic sphere, rather than a singular isothermal
  sphere \citep{Lizano1989,McLaughlin1996,McLaughlin1997,Andre2000}.
 Basically, the logotropic model assumes that
  the gas pressure across the initial condensation depends
  logarithmically on the density, giving rise to a flatter density
  profile in the static part and at the infall/static interface of the
  envelope. Because the structure of prestellar cores is well described 
  by a flat density profile in the inner region ($\alpha \sim$ 0) and a 
  power law index of $\sim$ 2 in the outer part \citep{Visser2002,Andre2000}, 
  it has been mentioned in the past that a lower value of $\alpha$ could probe a 
  younger protostar. However, later systematic studies have not supported this 
  interpretation \citep[e.g.][]{Jorgensen2002}. 
Another possible explanation is that the envelope is flattened for example because of the presence of a magnetic field \citep[e.g.][]{Li1996,Hennebelle2008}. However, testing this hypothesis is not trivial as it requires to solve the radiative transfer in a 2D geometry. A simple toy model which assumes constant temperature and optically thin emission suggests that $\alpha$ cannot be lower than 1 even in the extreme case of a flattened structure with axis ratio of 1:10 seen face-on.
The few cases with $\alpha \geq 2$ 
  are easier to explain and may be due to the presence of one or more 
  high-density structures, like discs \citep{Jorgensen2007}, embedded within 
  the envelope.\\

\noindent
{\it Envelope mass M$_{env}$:}\\
It is very clear from Fig. \ref{comparison_all} that more luminous sources
have larger envelope mass. \citet{Hatchell2003} report the mass of the envelopes
  within 1 pc (empty triangles in Fig.
  \ref{comparison_all}). Therefore, the envelope masses 
  from \citet{Hatchell2003} are considered as lower limits 
  for M$_{env}$, and are not taken into account in the correlation 
  coefficient computation. The correlation coefficient is 0.96 with a probability for a chance correlation of $\sim$ 10$^{-20}$, pointing out a strong relation between the two variables. Assuming that the luminosity is entirely
due to the gravitational energy released, the luminosity-envelope mass
relation suggests that a similar relation exists between the mass
accretion rate and the mass of the envelope. However, while the
result that more luminous sources
have larger envelope mass agrees with theoretical expectations, it should be
considered with caution. Indeed, the observed correlation between two
quantities does not necessarily imply a physical correlation. 
The envelope-mass-derivation differs between
studies. Furthermore, uncertainties in the distance to sources,
particularly in the high-mass cases (uncertainties of several kpc),
and resolution limits of distant objects can introduce errors and 
significant observational bias.\\

\noindent
{\it Envelope radius $r_{out}$:}\\
Similar to M$_{env}$, the outer radius of the envelope $r_{out}$
increases with increasing luminosity, varying from $\sim 10^{4}$ AU for
low-mass protostars to $\sim 10^{6}$ AU for high-mass protostars. The
correlation coefficient between $r_{out}$ and the luminosity is very
high, 0.90, with a probability for a chance correlation of 10$^{-22}$. Note that we checked that the relation is real and not simply due
to the distance to the source combined to the angular resolution limits. This point is illustrated by the plot of the measured angular size, $\theta$, of the sources as function of the luminosity in Fig. \ref{comparison_all} (panel f). The figure shows that $\theta$ is not decreasing with luminosity, which excludes the possibility of an observational bias due to the limited angular resolution of the observations.\\

\noindent
{\it Average density $<n>$:}\\
The average density $<n>$ in the envelope is derived from $r_{out}$
and M$_{env}$ for each source. Unfortunately, $<n>$ could be derived
only for intermediate and some high-mass sources, giving an average of
about 4$\times$10$^{5}$ and 3$\times$10$^{6}$ cm$^{-3}$, respectively,
over a spread in luminosity of about four orders of magnitude. Although
the average $<n>$ is one order of magnitude higher in high-mass protostars
versus intermediate mass protostars, the correlation coefficient between
$<n>$ and the luminosity is only 0.55. The lower limits for low-mass
and the highest mass protostars do not allow any firm conclusion,
but it seems that there is little difference in the average
density of these envelopes.\\

\noindent
{\it Density at 1000 AU:}\\
There is an apparent increase of the density at 1000 AU with
increasing luminosity, going from about $\sim10^{6}$ to $\sim10^{8}$
cm$^{-3}$ for source luminosity varying from 1 to $\sim10^{5}$
L$_\odot$. There are, however, some exceptions on the high-mass
side. Indeed, the increase of the density at a given distance is
consistent with the finding of an  increasing envelope radius and
approximately constant average density for the envelope (see the two
items above).\\

\begin{figure*} \centering

\includegraphics[width=6cm,angle=90]{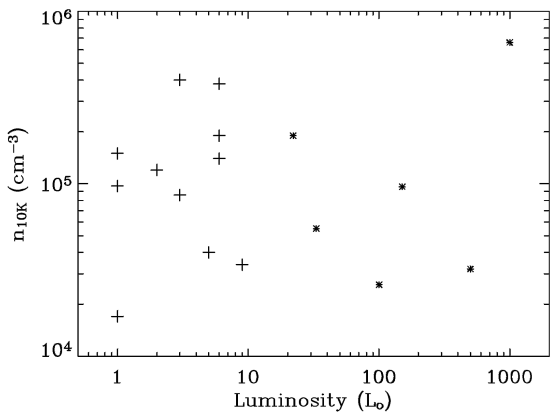}

\caption{Plot of the density at 10 K as function of the bolometric
  luminosity of a sample of low- and intermediate mass protostars.
  Data are from \citet{Jorgensen2002}, (crosses), and the present study,
  (stars).\label{comparison_low}}
\end{figure*}

\noindent
{\it Density $n_{10K}$ at 10 K: }\\
For a smaller sample, formed by low- and intermediate mass protostars
only, it is possible to compare the density $n_{10K}$ at 10 K, which
is an indication of the density of the parental cloud (the gas shielded by UV photons). 
Figure \ref{comparison_low} shows that $n_{10K}$ is between $2\times10^{4}$
and $6\times10^{5}$ cm$^{-3}$. This quantity, however, is not correlated with the
source luminosity (varying by three orders of magnitude). Therefore there is no
evidence that the outer density plays a large role in the
determination of the final star mass, a rather important and
surprising result, which needs further confirmation.\\

\noindent
{\it Summary:}\\
To summarise, the major result of this section is that the protostar
luminosity (namely the mass of the final star) seems preferentially 
linked to the size (or mass) of the envelope, rather than to
the parental cloud density, and that most of the envelope ends up
having a centrally condensed, free-fall density distribution.
Furthermore, and maybe even more important, there is a continuity in
the parameters of the envelopes, going from low- to high-mass
protostars. It appears that there is no important difference to the
trigger or process of star formation for these two mass regimes. The
intermediate mass protostars have allowed a bridge between the
low- and high-mass sources, with no apparent observational discontinuity. 
However, one has to keep in mind that these results are based on single 
dish observations and are therefore driven mainly by the outer region 
of the envelopes ($\gtrsim$ 10''). A more accurate analysis of the smaller 
scale structure ( cavities, density power law index changes...) will require 
interferometric observations.

\begin{table*}[h]
\begin{center}
\begin{tabular}{|c|ccc|c|} \hline 
Physical parameter     & Linear correlation & Number of  &  Probability for       & Linear fit    \\ 
vs. Log(L$_\odot$)     & coefficient        & objects     &   a chance correlation & slope        \\  \hline 
Log($\alpha$)       & 0.12 & 107 & 1.  & $--$ \\
Log(R$_{out}$)      & 0.90 & 60 & 10$^{-20}$  &  0.3 \\
Log(M$_{env}$)      & 0.96 & 46 & 10$^{-22}$   & 1.3 \\
Log(n$_{1000AU}$)    & 0.72 & 56 & 10$^{-8}$   & 0.3 \\
Log($<$n$>$)     & 0.25 & 39 & 0.4   & $--$ \\
Log(n$_{10K}$)    & -0.26 & 18 & 0.5   &  $--$ \\
Log($\theta$)      &  0.88     & 60    &  10$^{-19}$    &   -6$\times$10$^{-3}$   \\
\hline
\end{tabular}
\caption{Linear correlation coefficients and probability for a chance correlation between the pair of parameters plotted on Fig. \ref{comparison_all} and \ref{comparison_low}. \label{coef_correl}}
\end{center}
\end{table*}
\subsection{The problem of the underestimated 24 $\mu$m flux}

As mentioned in the previous sections, our modelling fails to reproduce
the observed flux at 24$\mu$m by several orders of magnitude.
This is certainly not due to a numerical problem in the computation but
is rather a real problem: our model misses some key element. A possibility 
evoked in the literature is the presence of a large spherical cavity within 
the envelope \citep{Jorgensen2005} which could significantly 
reduce the optical thickness at 24$\mu$m.

\begin{figure*} \centering
\includegraphics[width=5cm,angle=90]{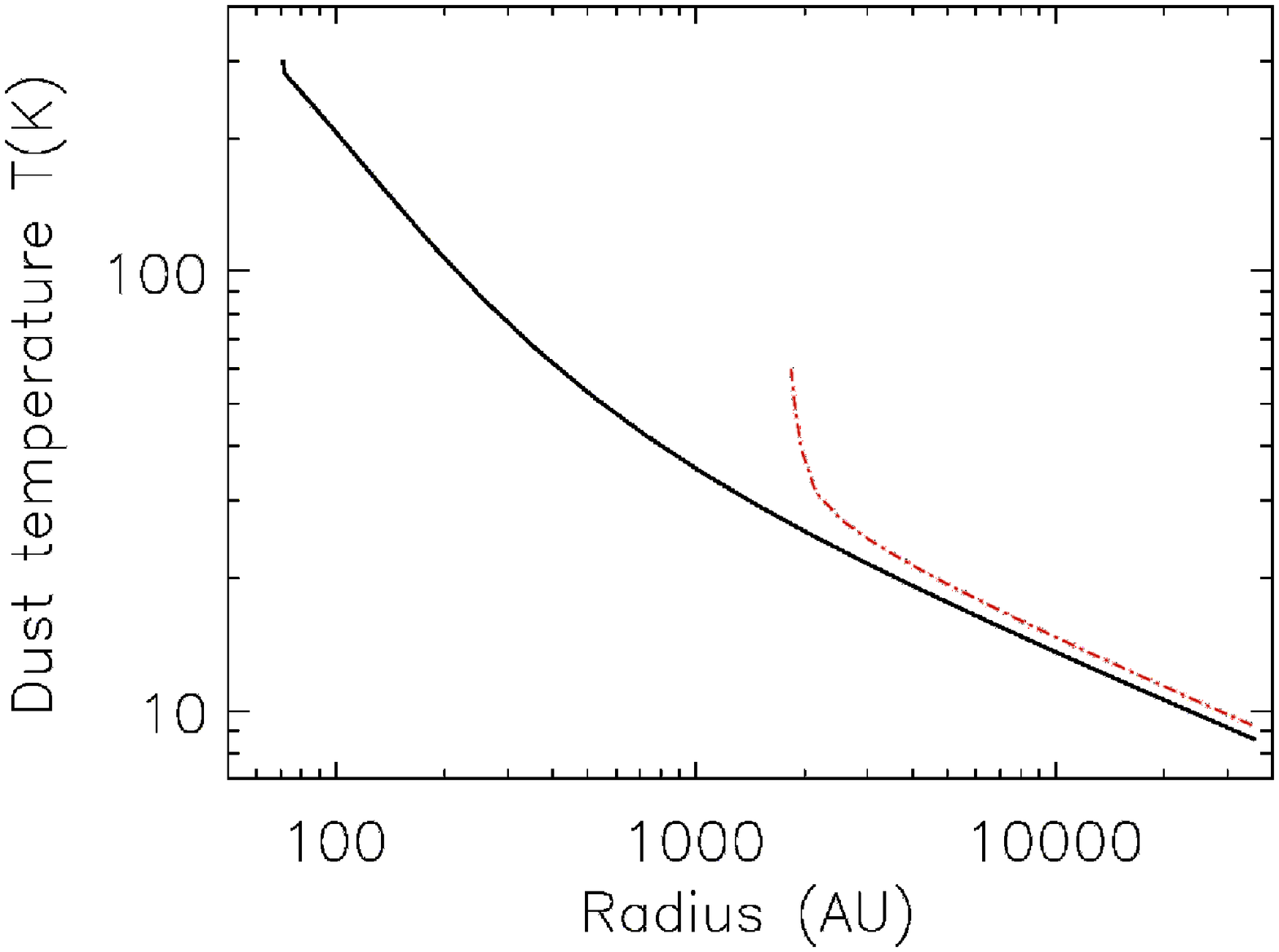}
\includegraphics[width=5cm,angle=90]{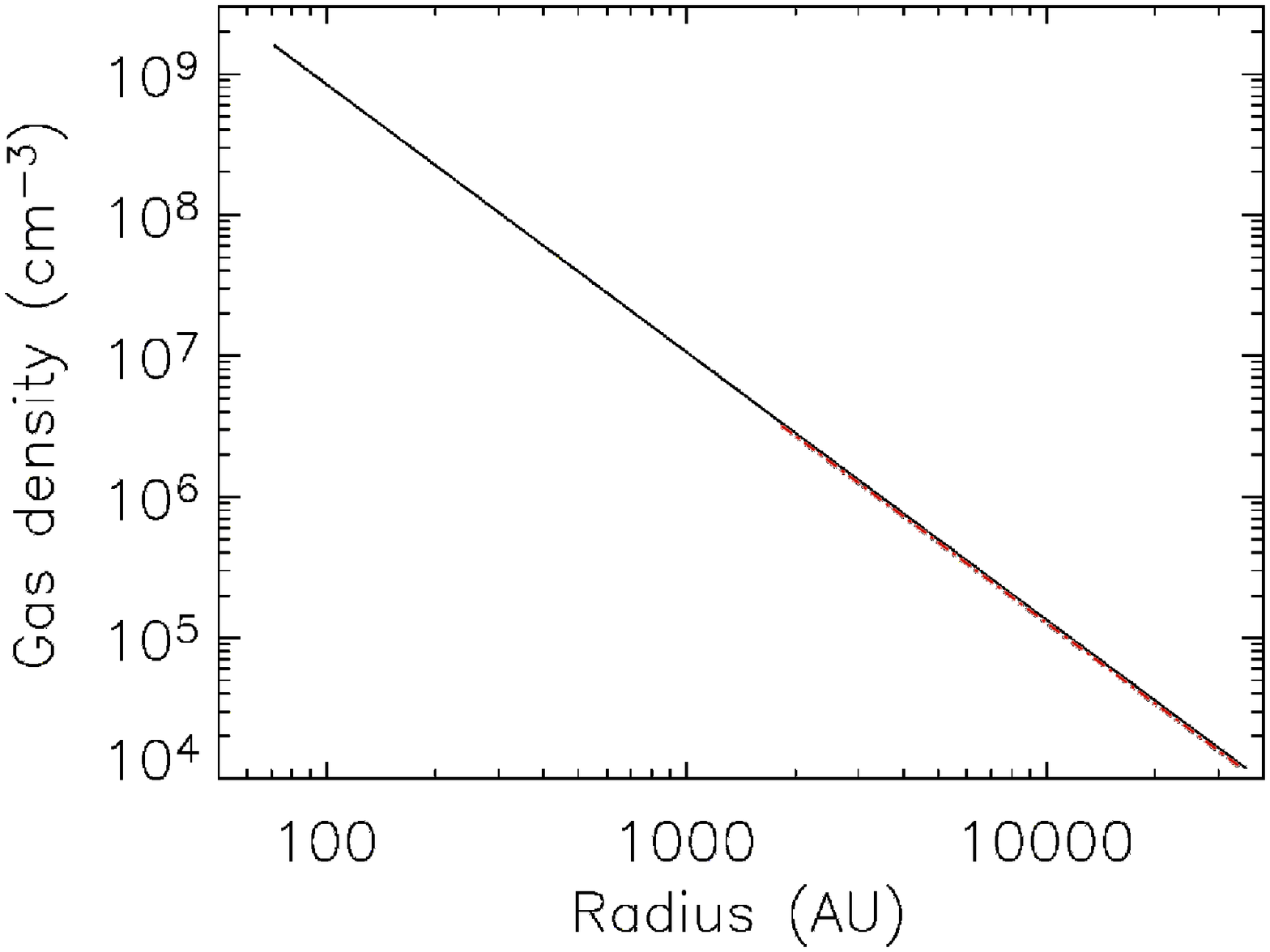}
\includegraphics[width=5cm,angle=90]{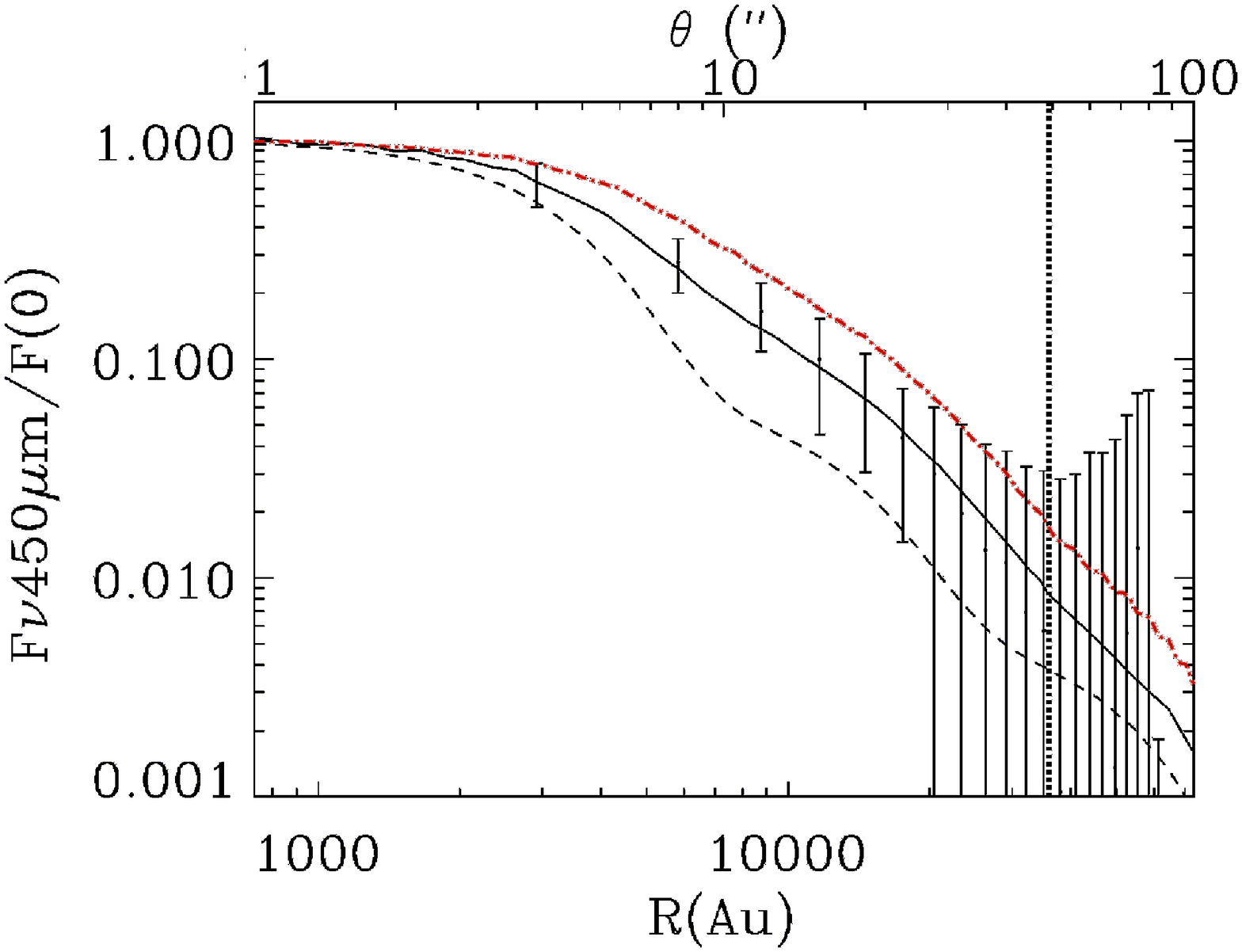}
\includegraphics[width=5cm,angle=90]{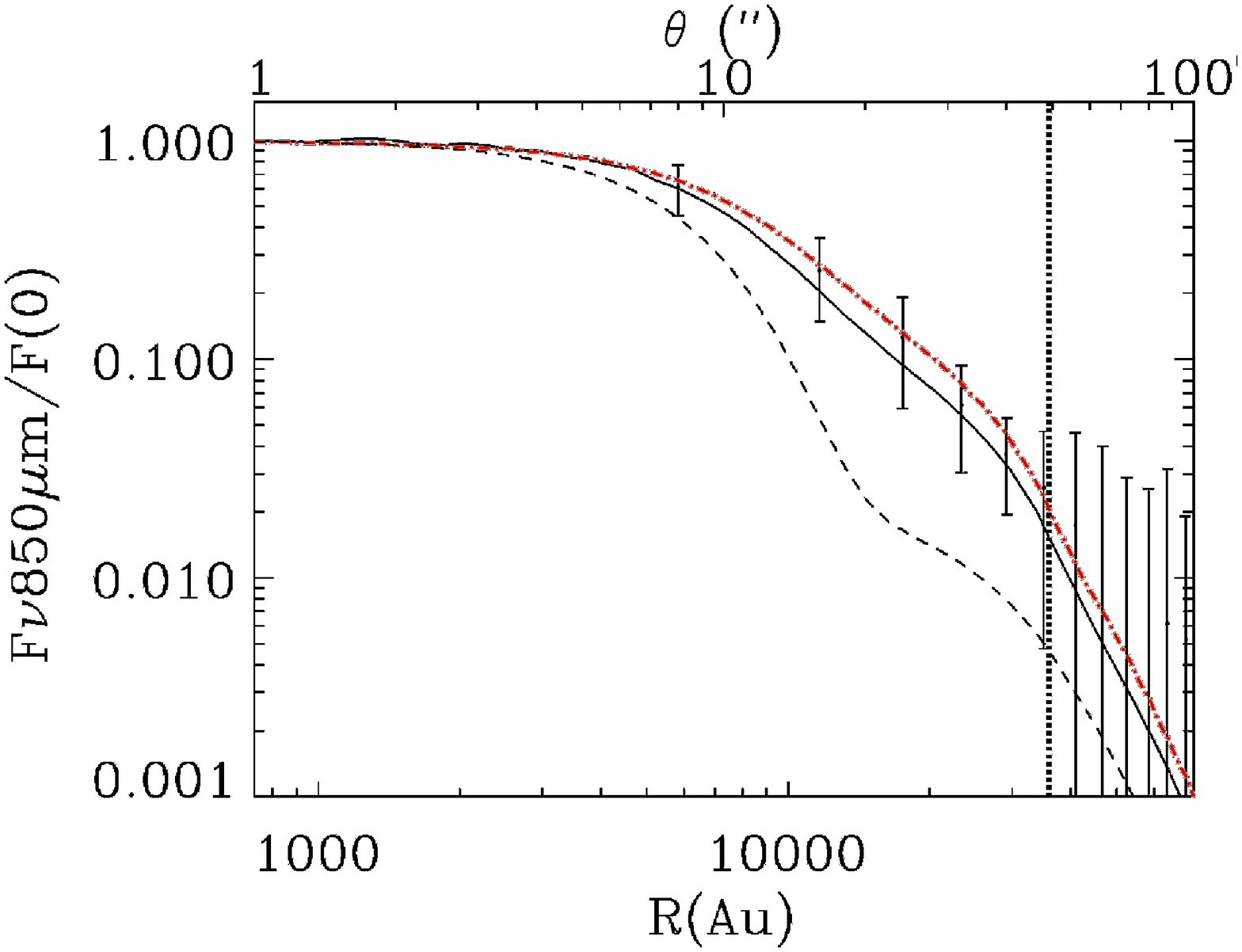}
\includegraphics[width=5cm,angle=90]{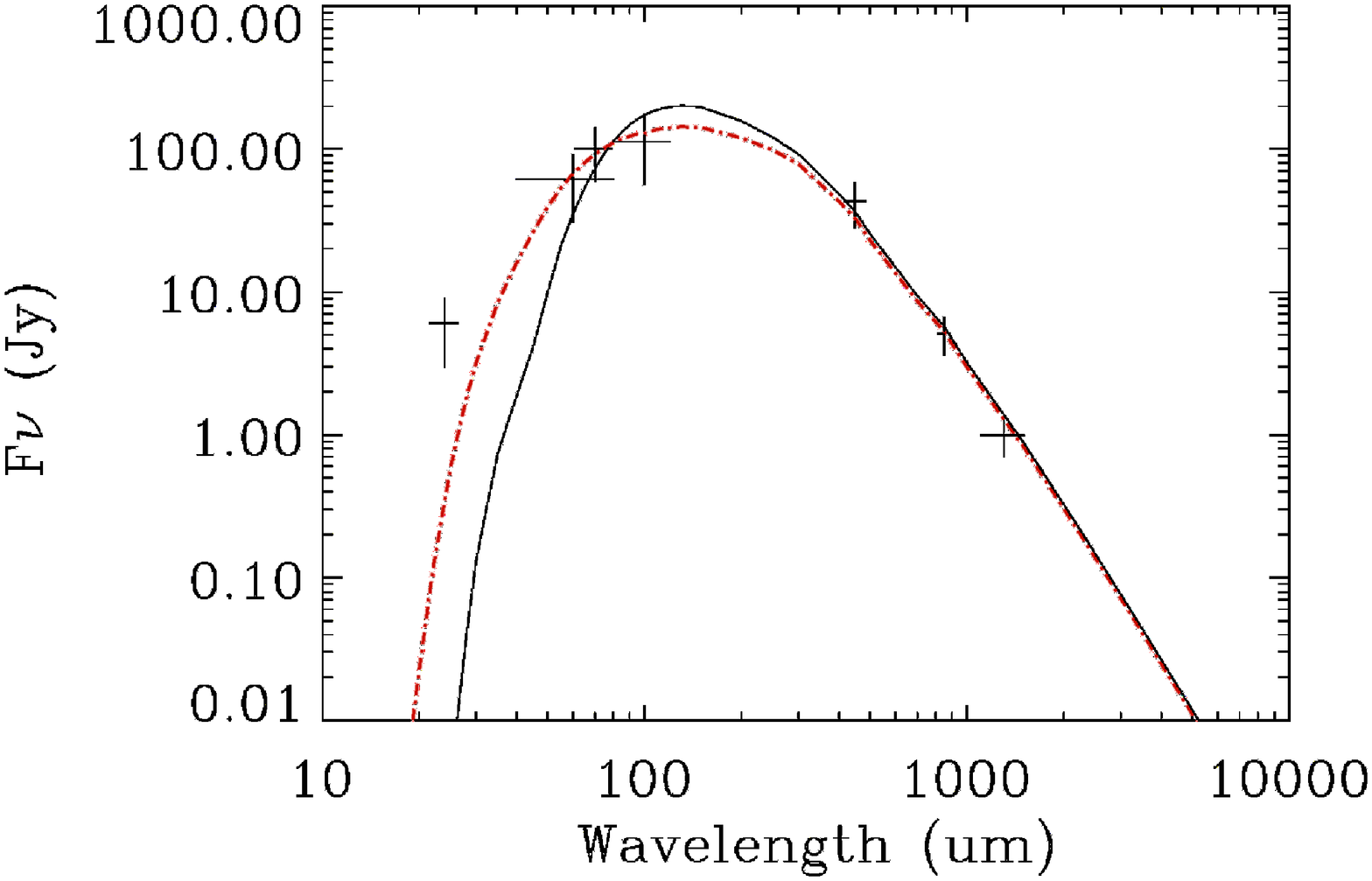}
\caption{ Cep E-mm physical structure, brightness profiles, and SED from the best-fit model (black solid lines) and 
after adding an 1800 AU radius cavity to the best-fit model (red dashed-dotted lines). The dust temperature and H$_{2}$ 
density profiles are reported in the upper left panel and upper right panel, respectively. The observed 
brightness profiles at 450 $\mu$m, 850 $\mu$m and the SED, observed and modelled, are shown in the 
middle left panel, the middle right panel, and the lower panel, respectively. 
The dashed lines represent the beam pattern of the telescope at 450 $\mu$m and 850
  $\mu$m. The vertical black dotted lines show the outer radius of the best-fit model.\label{CepE_24um}}
\end{figure*}

To check this hypothesis, we carried out a few tests
using Cep E-mm as a representative case. The results
of the tests are shown in Fig. \ref{CepE_24um}.  First, we added a
1800 AU radius cavity to the best-fit model of
Table \ref{best_fit_phy_param}.  As expected, the emission at 24 $\mu$m
is increased by several orders of magnitude. However, this model badly
fails to reproduce the brightness profiles (as shown in the
figure). When a correct procedure is carried out that takes into
account the variation in the brightness profiles, namely a minimisation of the
$\chi2$ varying all the envelope parameters, the situation returns 
to the original (best-fit) underestimation of the 24 $\mu$m flux. In
fact, the introduction of a cavity leads to an increase of the
density power law index, $\alpha$, to compensate for the resulting
flattening of the model brightness profiles. As a result, the mass 
found in the outer envelope is lower so that the the overall
dust optical depth must be increased to reproduce the integrated
sub-millimetre fluxes. The dust opacity at 24 $\mu$m again becomes
large and very little emission is able to escape at these wavelengths.
These tests suggest that to solve the 24 $\mu$m flux problem it is
necessary to have a low-opacity escape route for the 24 $\mu$m photons
{\it and} a thick enough envelope to fit the brightness profiles : a simple cavity does not suffice. 

\citet{Larsson2000,Larsson2002} similarly found that the 24 $\mu$m flux is underestimated toward Serpens-FIRS1,
 using different tools and observations. In their first study, Larsson et al. used a 1D radiative
transfer code assuming a spherical envelope model with a single power
law density to reproduce the SED (similar to our approach). Their model
underestimates the observed flux near 24 $\mu$m by several orders of
magnitude. In their second study, they modelled the envelope with a 2D
radiative transfer code, including a biconical cavity. This model
reproduces fairly well the SED from the mid-IR to millimetre
wavelengths, but again underestimates the observed flux at 24 $\mu$m
by about a factor 3-10. The biconical cavity treatment alleviates but does not resolve the 
missing flux problem.
Systematic studies of high-mass protostars by \citet{Vandertak2000}
 and \citet{Williams2005} have led to similar results.
\citet{Williams2005} modelled 36 high-mass protostellar
objects at 850 $\mu$m, using DUSTY, as in the present study.
 The majority of their best-fits fail to reproduce the flux around 24 $\mu$m.  
 They discuss the possibility of the contribution of high accretion
rates, which would significantly increase the near-IR flux density
\citep{Osorio1999}. Another contribution could come from the
presence of circumstellar disks or the stochastic heating of small grains,
which would alter the emission of the envelope and produce more short
wavelength photons \citep[e.g.][]{Sellgren1983,Draine2001}. 
However, the envelope is optically thick at these wavelengths, making
the emission very sensitive to deviations from the assumed spherical
shape. These additional processes still require a low-opacity
escape route to exists \citep[also mentioned by][]{Vandertak2000} to explain
the missing 24 $\mu$m flux, for example a biconical cavity 
excavated by the outflow. Note
that all the intermediate mass protostars studied here are associated
with outflows, except OMC2-FIR4 for which the model indeed fits the
observed flux at 24$\mu$m \citep[see][]{Crimier2009a}. Finally, Van Der
Tak et al. also suggest the possible evaporation of  grain ice
mantles close to the star, which would decrease the 20 $\mu$m optical
depth by 30$\%$ \citep{Ossenkopf1994} in the T$_{dust}$ $>$ 100 K
region.

\section{Conclusions}\label{sec:Conclusions}

We have derived the physical structure of the envelopes of five IM
protostars, with luminosities between 30 to 1000 L$_\odot$.  The
envelope dust density and temperature profiles were determined
by means of the 1D radiative transfer code DUSTY,
using all continuum observations from the literature. The analysis assumed
that the density profiles follow a single index power law and obtained
self-consistently the temperature profile.  The  best-fit
envelope models well reproduce the observations, namely the
sub-millimetre radial brightness profiles and the SED 
between $\sim$60 $\mu$m and 1.3 mm, for each source. 
However, the model underestimates the
24$\mu$m emission by several orders of magnitude. 
We ran test models to better
understand what ingredient is missing and conclude that a
``simple'' cavity is not enough to reproduce the 24$\mu$m
observations. Apparently, the missing ingredient is a low-opacity
escape route plus a warm dust contribution inside the envelope
(circumstellar disc, warm outflow-excavated cavity...).

The gas density and temperature profiles were derived by assuming
a constant dust-to-gas ratio and by computing the gas thermal balance
at each point within the envelope. Because the gas equilibrium temperature
strongly depends on the water abundance in the interiors of the
envelopes, we also computed the expected water emission for each
source.  We found that the gas and dust are thermally coupled across
the envelope with differences less than 5 K in three out of five
sources.  In IC1396 N BIMA 2 and NGC7129 FIRS 2 the gas is colder
than the dust by at most 40 K, in a small region just where the icy mantles
are predicted to sublimate. The predicted water line fluxes are consistent
with the upper limits derived by the ISO observations.

One of the major goals of the present study was to ``use'' the IM
protostars as a bridge between the low- and high-mass protostars with
the hope that this will aid our understanding of the star formation
process at either end. When comparing the characteristics derived by
the modelling of the envelopes of low-, intermediate, and high-mass
protostars, it appears that there is a smooth transition between the
various groups. This suggests that {\it there are basically no different
  triggers or processes between these mass regimes}. The
power law index $\alpha$ is similar in all three groups of objects. The
majority of the sources have $\alpha$ between 1.5 and 2. This is 
consistent with the theory of isothermal collapse from an initially
singular isothermal sphere, the so-called inside-out expansion-wave
collapse (Shu 1977).  Regardless of the mass
group, a few sources have $\alpha$ lower than 1.5, pointing perhaps to the  
collapse of an initially logotropic, virialised sphere (Lizano \& Shu 1989, McLaughlin
96, and McLaughlin 97). Finally, the luminosity (mass) of the star
depends on the size of the envelope, but does not
depend on the density at a given temperature (for example at 10 K or
100 K).

\bigskip

{\it Acknowledgements.}
We warmly thank Patrick Hennebelle for helpful discussions.
One of us (N.Crimier) is supported by a fellowship of the Minist\`ere de l'Enseignement Sup\'erieur et de la Recherche.
We acknowledge the financial support by PPF and the Agence Nationale pour la Recherche (ANR), France (contract ANR-08-BLAN-0225).
The James Clerk Maxwell Telescope is operated by the Joint Astronomy Centre on behalf of the Science and Technology Facilities Council of the United Kingdom, the Netherlands Organisation for Scientific Research, and the National Research Council of Canada.
This paper has been partially supported by  MICINN, within the programme CONSOLIDER INGENIO 2010, under grant  "Molecular Astrophysics: The Herschel and Alma Era -- ASTROMOL" ( ref.:  CSD2009-00038)
Doug Johnstone is supported by a Natural Sciences and Engineering Research Council of Canada (NSERC) Discovery Grant. 
\bigskip

\bibliographystyle{aa}
\bibliography{all}
\begin{appendix}
\section{Results for individual sources}

\subsection{CB3-mm}\label{subsec:CB3-mm}
\subsubsection{Source background}\label{subsubsec:CB3-mm_Background}

The CB3 Bok globule is located at $\sim$ 2.5 kpc \citep{Launhardt1997,Wang1995}
 on the near side of the Perseus
arm of the Galaxy. Using ISO data, \citet{Launhardt1998} carried out a
multi-wavelength study of the CB3 globule and derived a total mass
and bolometric luminosity of the order of 400 M$_{\sun}$ and $\sim$
1000 L$_{\sun}$, respectively, for the entire globule. The globule hosts
$\sim$ 40 NIR sources, 22 of which are likely low- and intermediate mass
protostars in different stages of evolution \citep{Launhardt1998,Yun1995,Yun1994}.
 CB3-mm, the brightest millimetre source
in the globule, was first detected by \citet{Launhardt1997} and
subsequently observed in the sub-millimetre by \citet{Huard2000}.
The high luminosity of the source evaluated by \citet{Launhardt1997}, 
L $\sim$ 900 L$_{\sun}$, suggests that CB3-mm is an intermediate mass
object. A recent interferometric study by \citet{Fuente2007} showed
that CB3-mm contains two compact cores at 3 mm separated by 0.3 pc
($\sim$ 0.43\arcsec). \citet{Yun1994} also detected a molecular
bipolar outflow in CO, elongated in the northeast-southwest direction,
associated with H$_2$O masers \citep{deGregorio-Monsalvo2006}. This
outflow has been mapped in various molecular lines by \citet{Codella1999},
 who concluded that it originates from CB3-mm. The
same authors concluded that CB3-mm is probably a Class 0 source. In
the present study we re-evaluated the luminosity of this object, L
$\sim$ 1000 L$_{\sun}$, making CB3-mm an intermediate mass
protostar.

\subsubsection{Analysis}\label{subsubsec:CB3-mm_Data_used}

\begin{figure*} \centering
\includegraphics[width=6cm,angle=270]{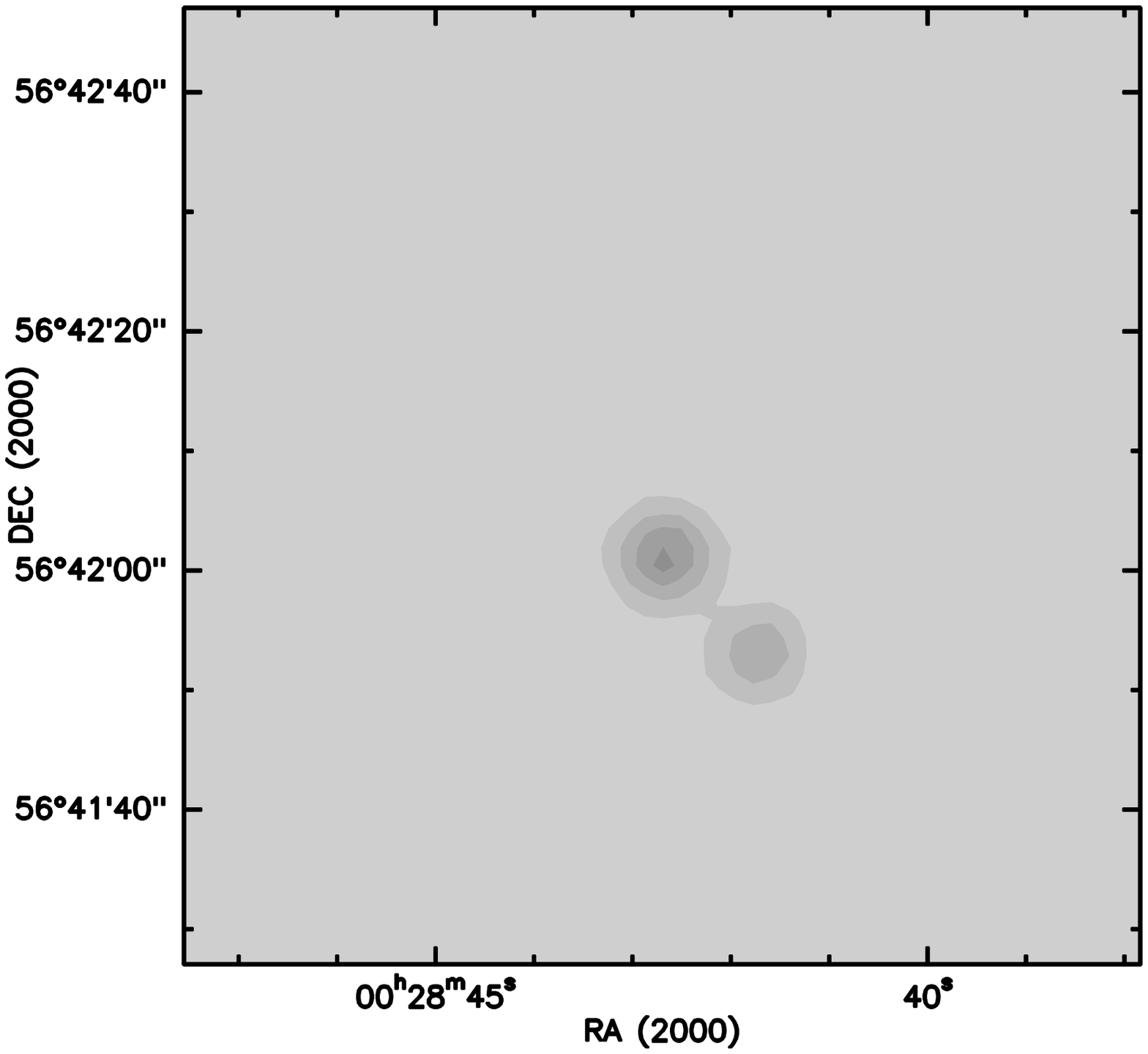}
\includegraphics[width=6cm,angle=270]{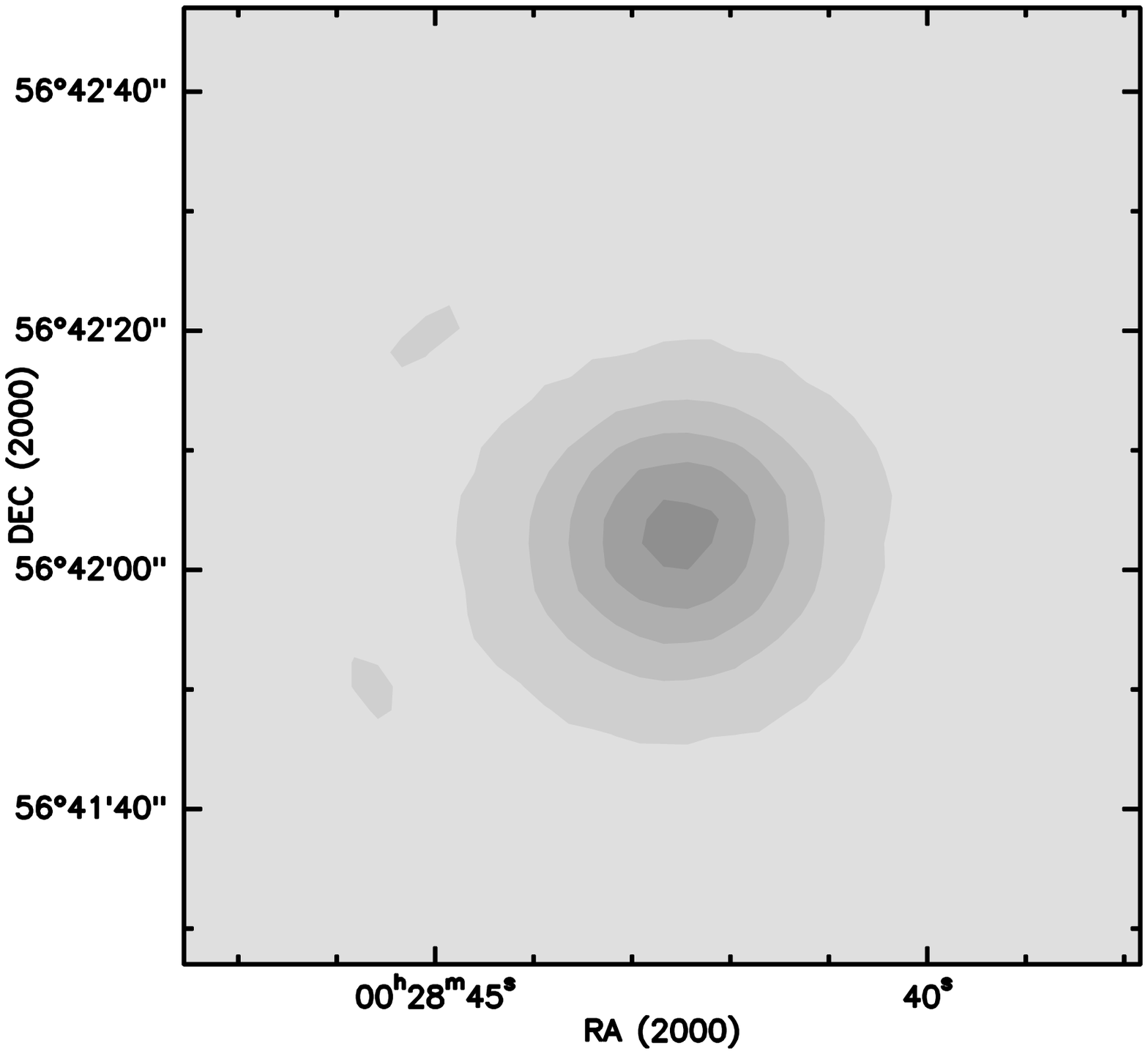}
\includegraphics[width=6cm,angle=270]{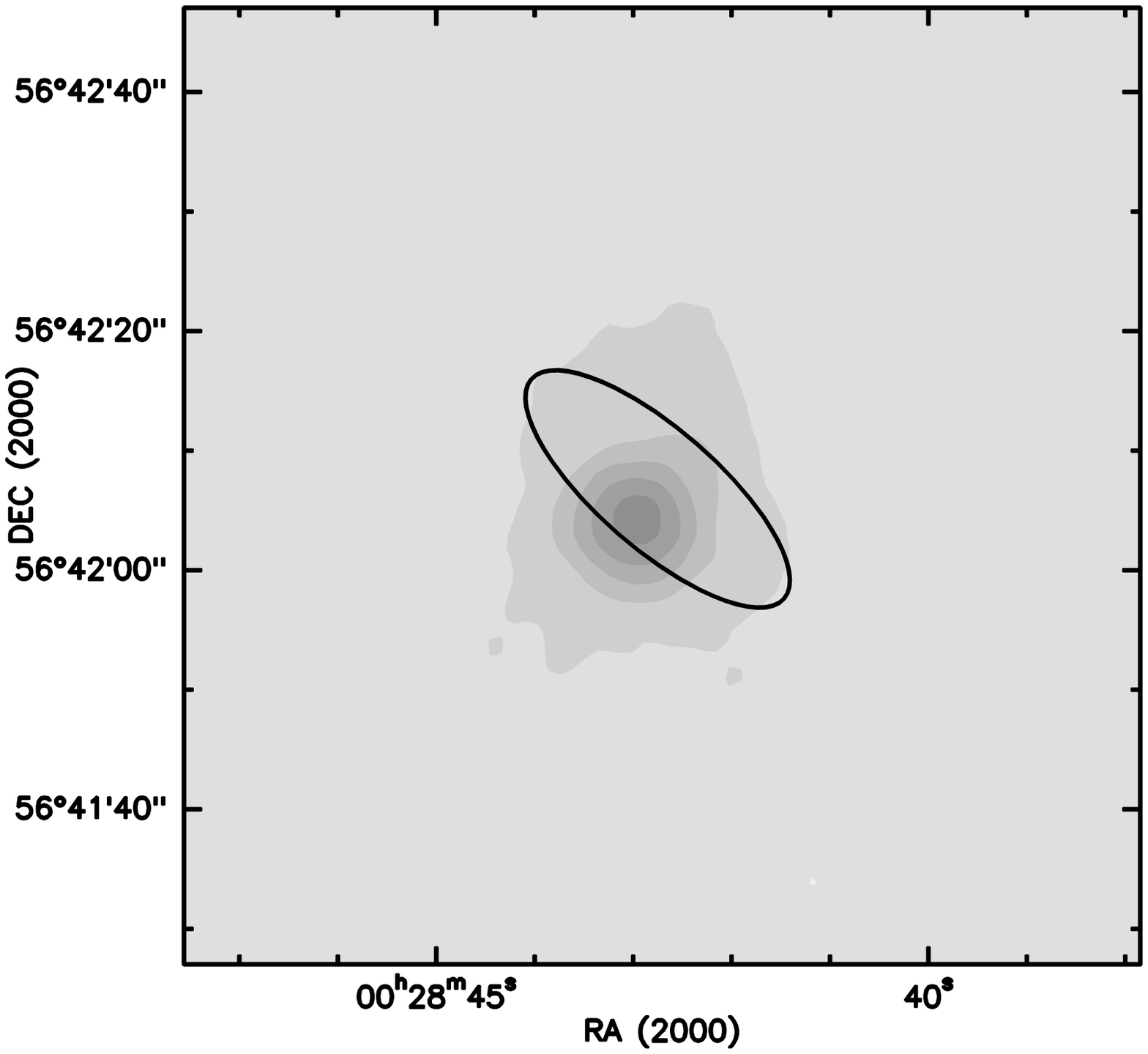}
\includegraphics[width=6cm,angle=270]{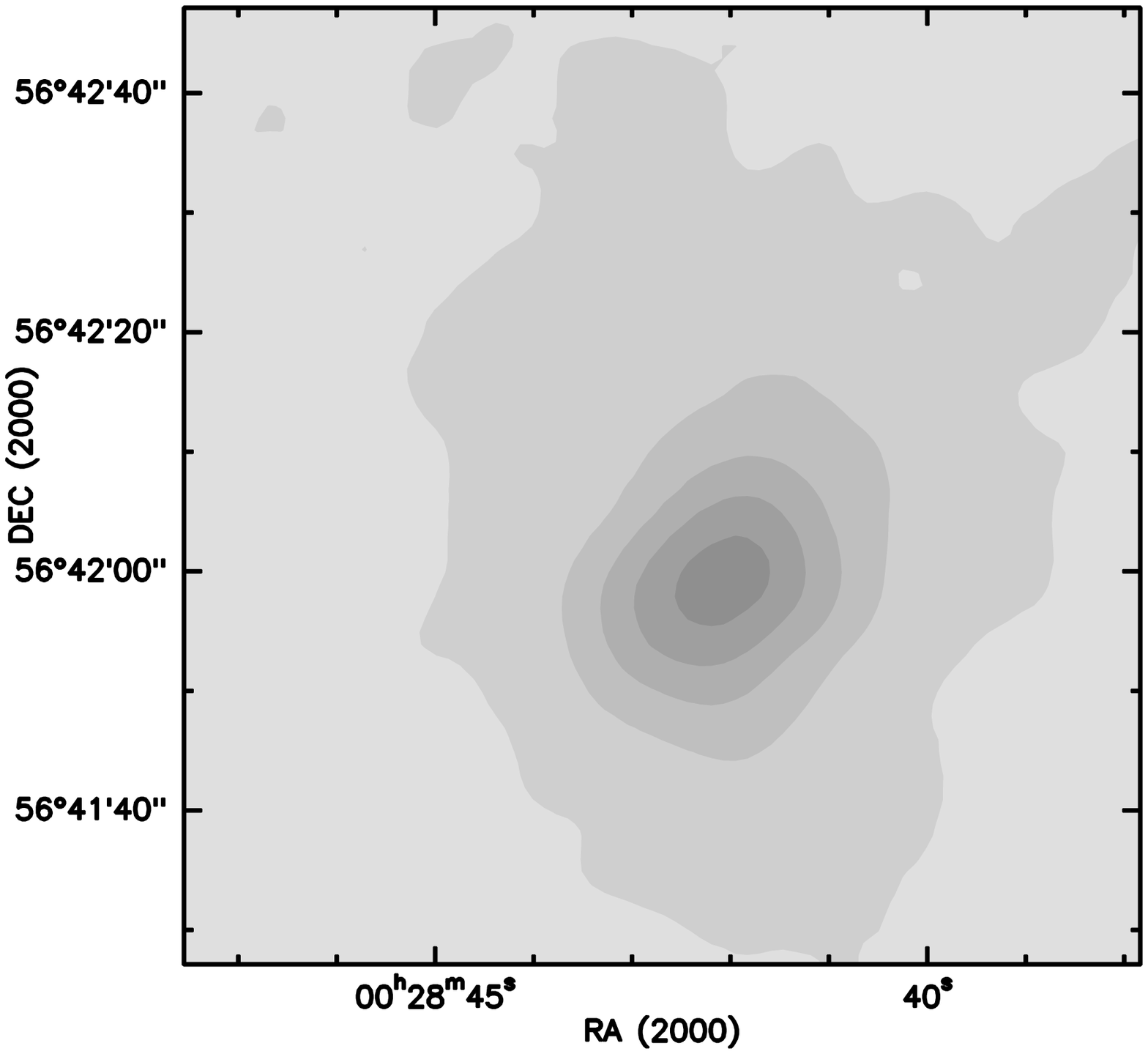}
\caption{Continuum emission maps around CB3-mm at 24 $\mu$m (upper
  left panel), 70 $\mu$m (upper right panel), 450 $\mu$m (lower left
  panel), and 850 $\mu$m (lower right panel). The grey scale contours mark the
  continuum flux from 10 \% to 90 \% of the peak emission by steps of
  20 \%. The uncertainty ellipse position of the IRAS observations
  toward CB3-mm are represented by the ellipse in the 450 $\mu$m
  map. \label{CB3_maps}}
\end{figure*}

The continuum maps used for the CB3-mm analysis are presented in
Fig. \ref{CB3_maps}. The maps at 450 $\mu$m and 850 $\mu$m were
obtained on 1997 December 18 as a part of  project m96bi28 and
on 1998 August 10 as a part of project m98bc21,
respectively. The flux profiles obtained at each wavelength are shown
in Fig. \ref{CB3_fits}.  The Spitzer observations were obtained on the
20th September 2004 as part of the programme ``Comparative Study of
Galactic and Extragalactic HII Regions'' (AOR: 63, PI: James
R. Houck). The integrated fluxes used for the analysis are reported in
Table \ref{Integrated_fluxes} and in Fig. \ref{CB3_fits}. The fluxes at
70 $\mu$m, 450 $\mu$m and 850 $\mu$m were obtained by integration over a $\sim$
50\arcsec radius. The uncertainty ellipse position of the IRAS
observations are reported on the 450 $\mu$m maps in
Fig. \ref{CB3_maps}. Note that the map at 24 $\mu$m presents two
sources separated by $\sim$12\arcsec, centred on the sub-millimetre
source emission (see Fig. \ref{CB3_maps}). Therefore, the flux
$1.4\pm0.8$ Jy at 24 $\mu$m was obtained by adding the integrated
flux over the two sources separately. We also report on the SED the
lower limits from Plateau de Bure (PdB) interferometric fluxes at 1.3 mm
and 3 mm \citep{Fuente2007}.

\subsubsection{Best-fit}\label{subsubsec:CB3-mm_Best_Fit}

\begin{figure*} \centering
\includegraphics[width=5cm,angle=90]{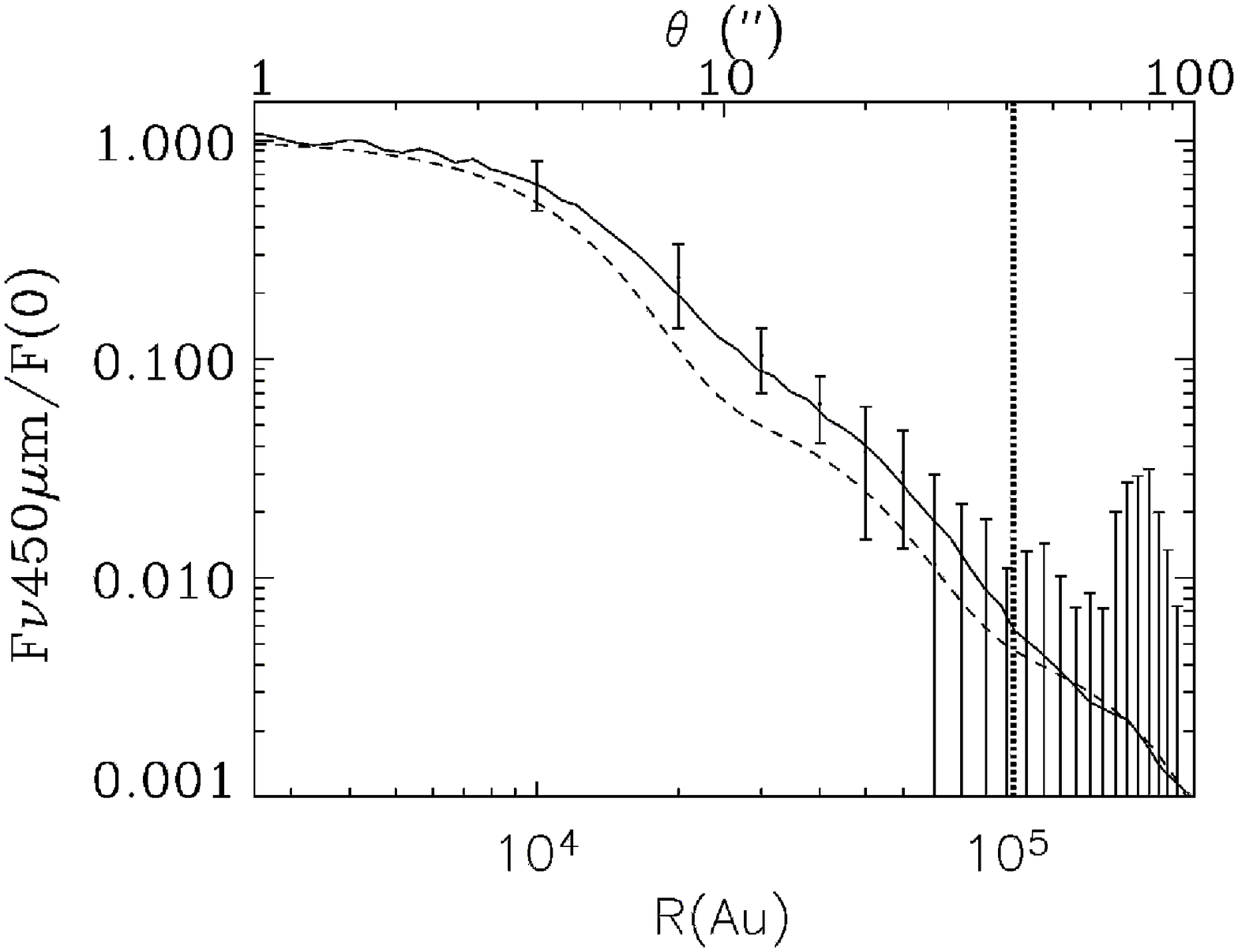}
\includegraphics[width=5cm,angle=90]{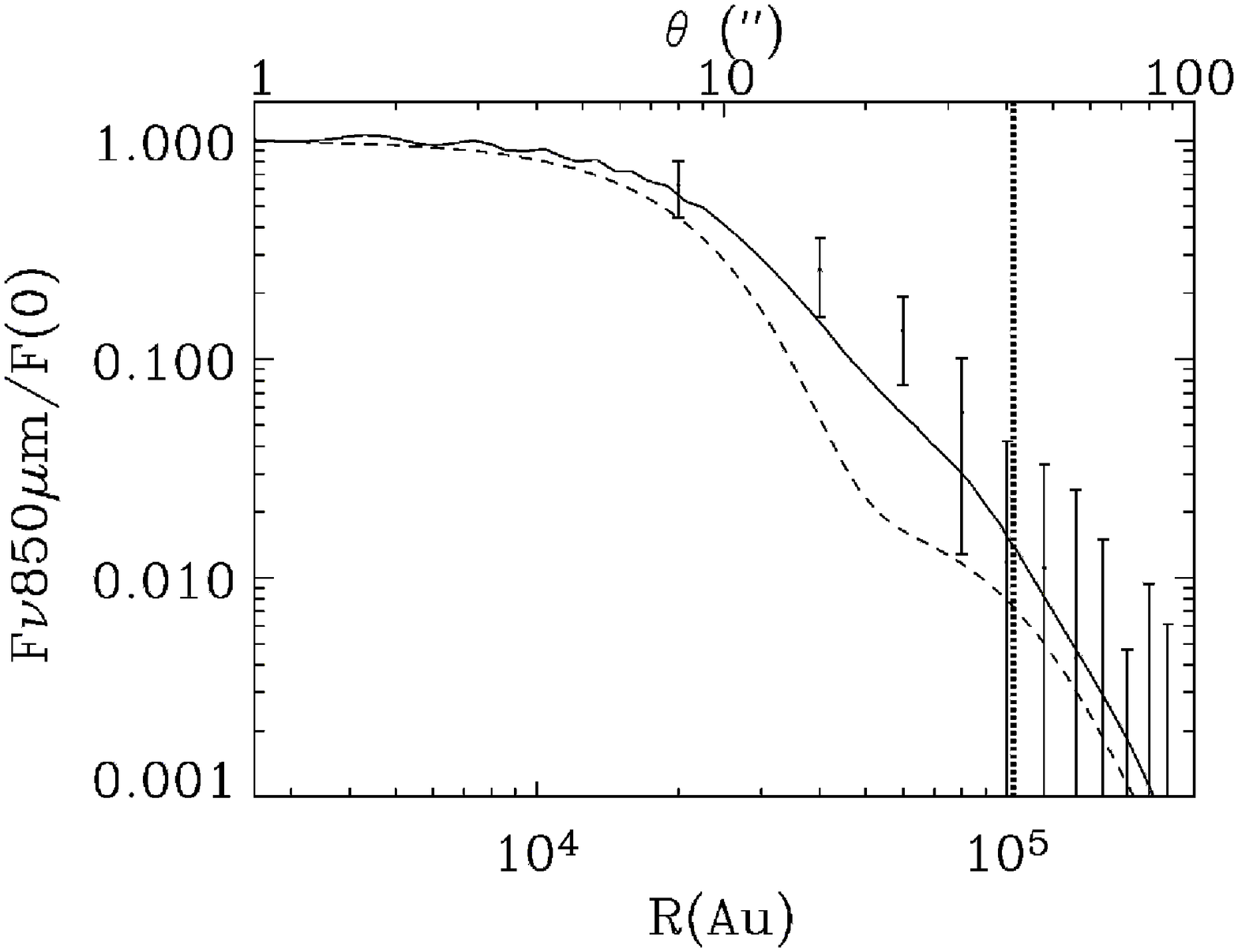}
\includegraphics[width=5cm,angle=90]{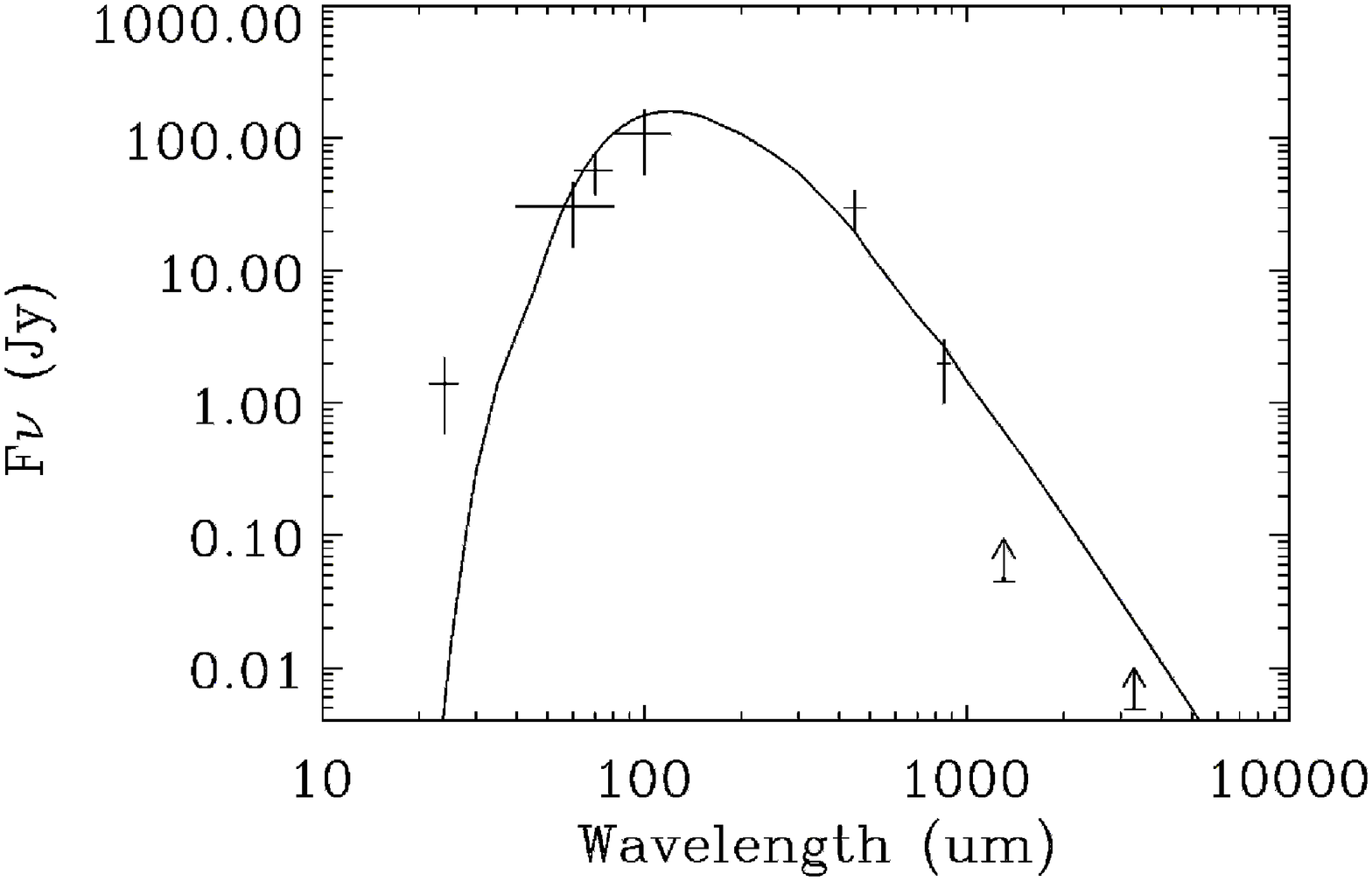}
\caption{Observed brightness profiles at 450 $\mu$m (upper left panel), 850
  $\mu$m (upper right panel) and SED (lower centred panel) toward
  CB3-mm. The solid lines report the best-fit model. The dashed lines
  represent the beam pattern of the telescope at 450 $\mu$m and 850
  $\mu$m. The vertical dotted lines show the outer radius of the best-fit model (Table \ref{best_fit_phy_param}). The lower limits from Plateau de Bure (PdB)
  interferometric fluxes at 1.3 mm and 3 mm \citep{Fuente2007} are
  also reported on the SED. \label{CB3_fits}}
\end{figure*}

\begin{figure*} \centering
\includegraphics[width=5cm,angle=90]{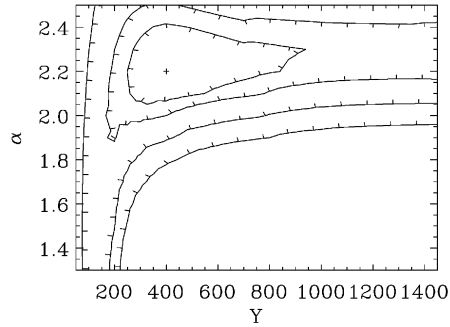}
\caption{$\chi ^{2}_{red}$ contour plots (Y,$\alpha$) for the CB3-mm
  model. In these computations, $\tau_{100}$ is equal to 5.8. The
  contours show the loci of the $\chi ^{2}_{red}$ values equal to 1.1,
  1.2, 1.5 and 2 times the minimum $\chi ^{2}_{red}$ (10, 20, 50 and
  100\%). The cross shows the position of the best-fit. \label{CB3_X2_tot}}
\end{figure*}

\begin{figure*} \centering
\includegraphics[width=5cm,angle=90]{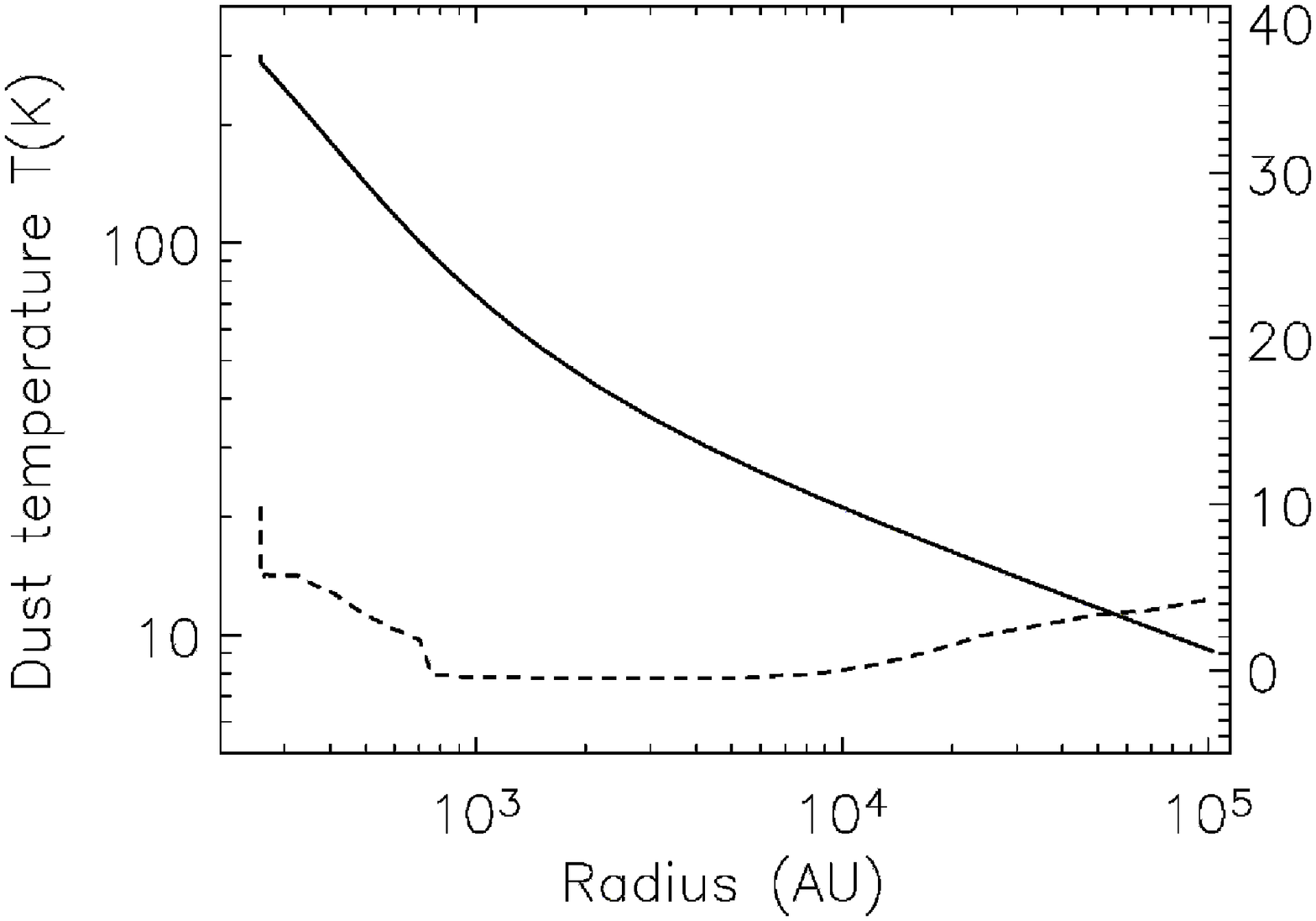}
\includegraphics[width=5cm,angle=90]{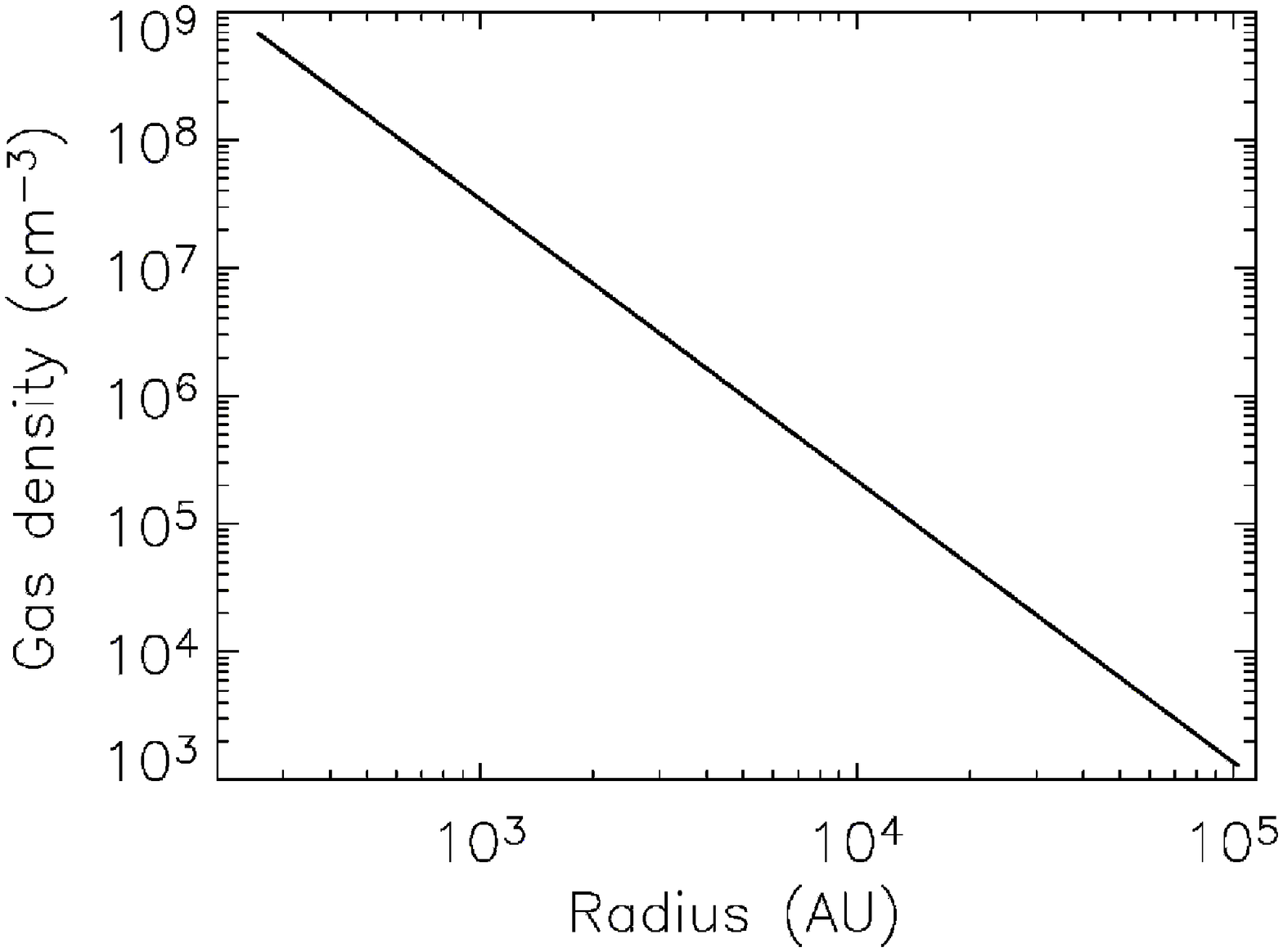}
\caption{CB3-mm dust temperature (left panel) and H$_{2}$ density
  (right panel) profiles from the best-fit model. The dashed line
  represents the difference between the dust and gas temperature
  derived using X(H$_2$O)$_{in} =$ 1$\times$10$^{-5}$ and
  X(H$_2$O)$_{out} =$ 1$\times$10$^{-7}$ (scale on right axis).\label{CB3_T_N}}
\end{figure*}

Table \ref{best_fit_phy_param} presents the set of parameters $\alpha$,
Y, and $\tau_{100}$, which best reproduce the observations, and
summarises some relevant physical quantities of the model. Figure
\ref{CB3_fits} shows the best derived brightness profiles and SED
against the observed ones, while Fig. \ref{CB3_X2_tot} shows the
$\chi^2$ contour plots. The dust density and temperature profiles of
the best-fit model are reported in Fig. \ref{CB3_T_N}. Although the
observed flux profiles and SED fluxes from 60 $\mu$m to 850 $\mu$m are well
reproduced by the model, the observed flux at 24 $\mu$m is
underestimated by about two orders of magnitude. The model
agrees well with the lower limits from Plateau de Bure
interferometric fluxes at 1.3 mm and 3 mm \citep{Fuente2007}, however.

The best-fit model, obtained with an $\alpha$ of 2.2, 
is the steepest among the
five sources modelled in the paper and 
leads to a strong gradient in the density
profile, from $\sim$ 10$^9$ to $\sim$ 10$^3$ cm$^{-3}$ with a density
and radius at 100 K of 7.5$\times$10$^7$ cm$^{-3}$ and 700 AU,
respectively.

Although CB3-mm is the largest and brightest of the five sources
investigated in this paper, its distance of 2500 pc makes it the 
least resolved source. Indeed, the brightness
profiles at each wavelength 
presented in Fig. \ref{CB3_fits} are close to the beam
patterns of the telescope, showing that the source
is barely resolved. Consequently, the value of the envelope radius
$r_{out}$ suffers a relatively large uncertainty. Considering the
$\chi ^{2}_{red}$ contour at 10\% of the minimum $\chi ^{2}_{red}$
(see Fig.  \ref{CB3_X2_tot}) in order to estimate the uncertainty, we obtain
$r_{out}$ = (1.0 $\pm^{1.3}_{3.9}$)$\times10^{5}$ AU.  On the
contrary, the envelope column density is quite insensitive to
$r_{out}$, due to the high power law index of the density
profile. Finally, in minimising the $\chi _{SED}^{2}$, we also varied
the source luminosity from 800 L$_\odot$ to 1200 L$_\odot$ and found that the
best-fit is obtained with source luminosity equal to 1000 L$_\odot$.

\subsection{Cep E-mm}\label{subsec:Cep_E-mm}

\subsubsection{Source background}\label{subsubsec:CepE-mm_Background}

Located in the Cepheus OB3 association at a distance of 730 pc
(Sargent 1977; Crawford $\&$ Barnes 1970), Cepheus E is the second
most massive and dense clump of this region (Few $\&$ Cohen
1983). Cepheus E hosts the source IRAS 23011+6126, which is 
associated with Cep E-mm, catalogued as a Class 0 protostar by \citet{Lefloch1996}.
Targeted by several continuum and lines studies, Cep E-mm
has been observed with IRAS \citep{Palla1993}, IRAM 30m 
\citep{Lefloch1996,Chini2001}, SCUBA \citep{Chini2001}, 
ISO \citep{Froebrich2003}, and Spitzer \citep{Noriega-Crespo2005}. All these
studies confirm the Class 0 status of Cep E-mm and constrain
the  total mass and bolometric luminosity of the source to 7--25
M$_{\sun}$ and $\sim$ 80--120 L$_{\sun}$, respectively. Finally, a bipolar
molecular outflow, first reported by \citet{Fukui1989}, is associated with
Cep E-mm. The properties of the outflow have been thoroughly analysed
by \citet{Eisloffel1996}, \citet{Ayala2000}, \citet{Moro-Mart'in2001}, and 
\citet{Smith2003}. The H$_{2}$ and [FeII] study by \citet{Eisloffel1996} 
shows a quadrapolar outflow morphology,
suggesting that the driving source is a binary. This outflow
morphology has been confirmed by sub-mm and near-IR observations 
\citep{Ladd1997}, and by Spitzer observations \citep{Noriega-Crespo2005}.

\subsubsection{Analysis}\label{subsubsec:CepE-mm_Data_used}

\begin{figure*} \centering
\includegraphics[width=6cm,angle=270]{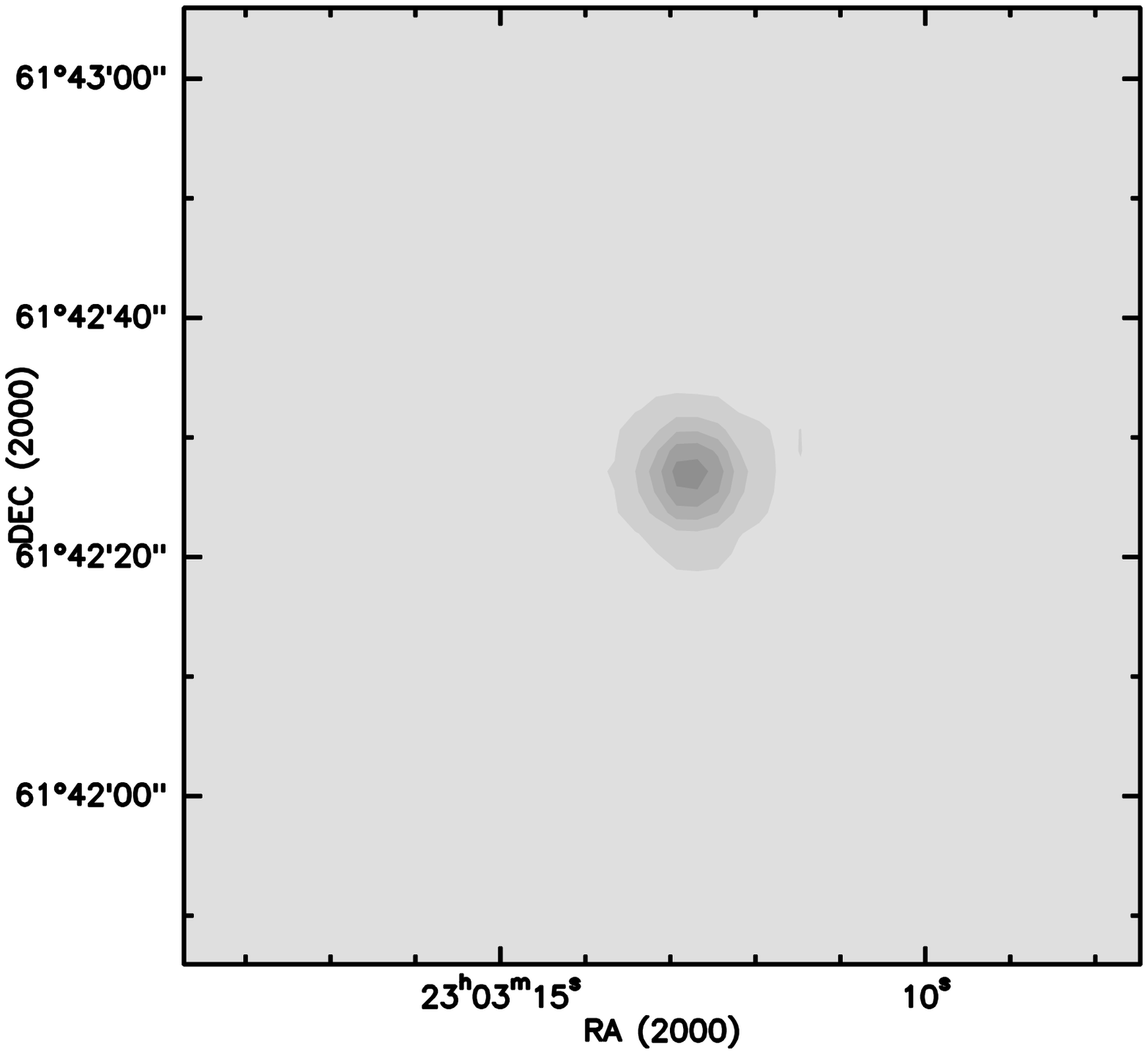}
\includegraphics[width=6cm,angle=270]{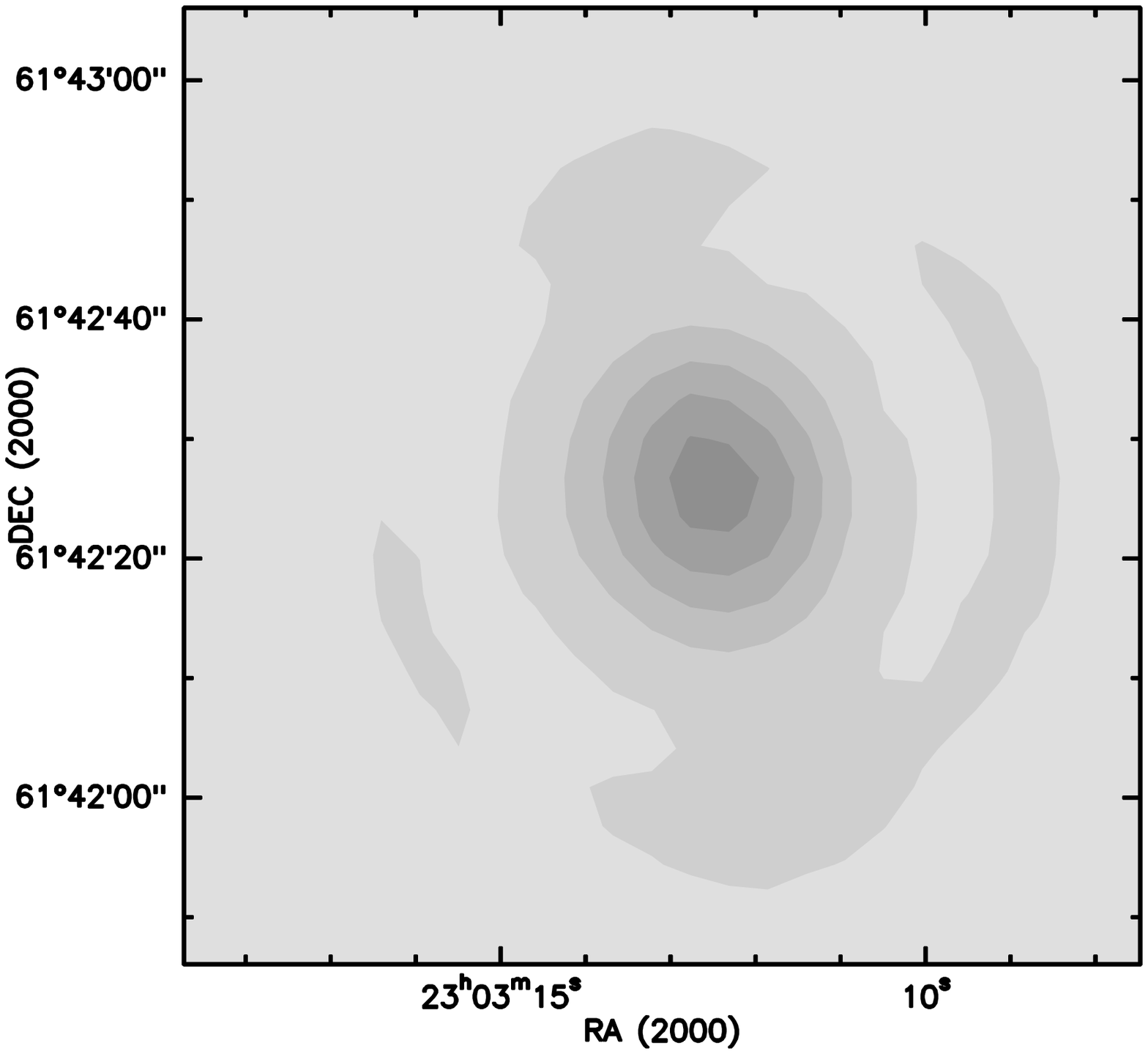}
\includegraphics[width=6cm,angle=270]{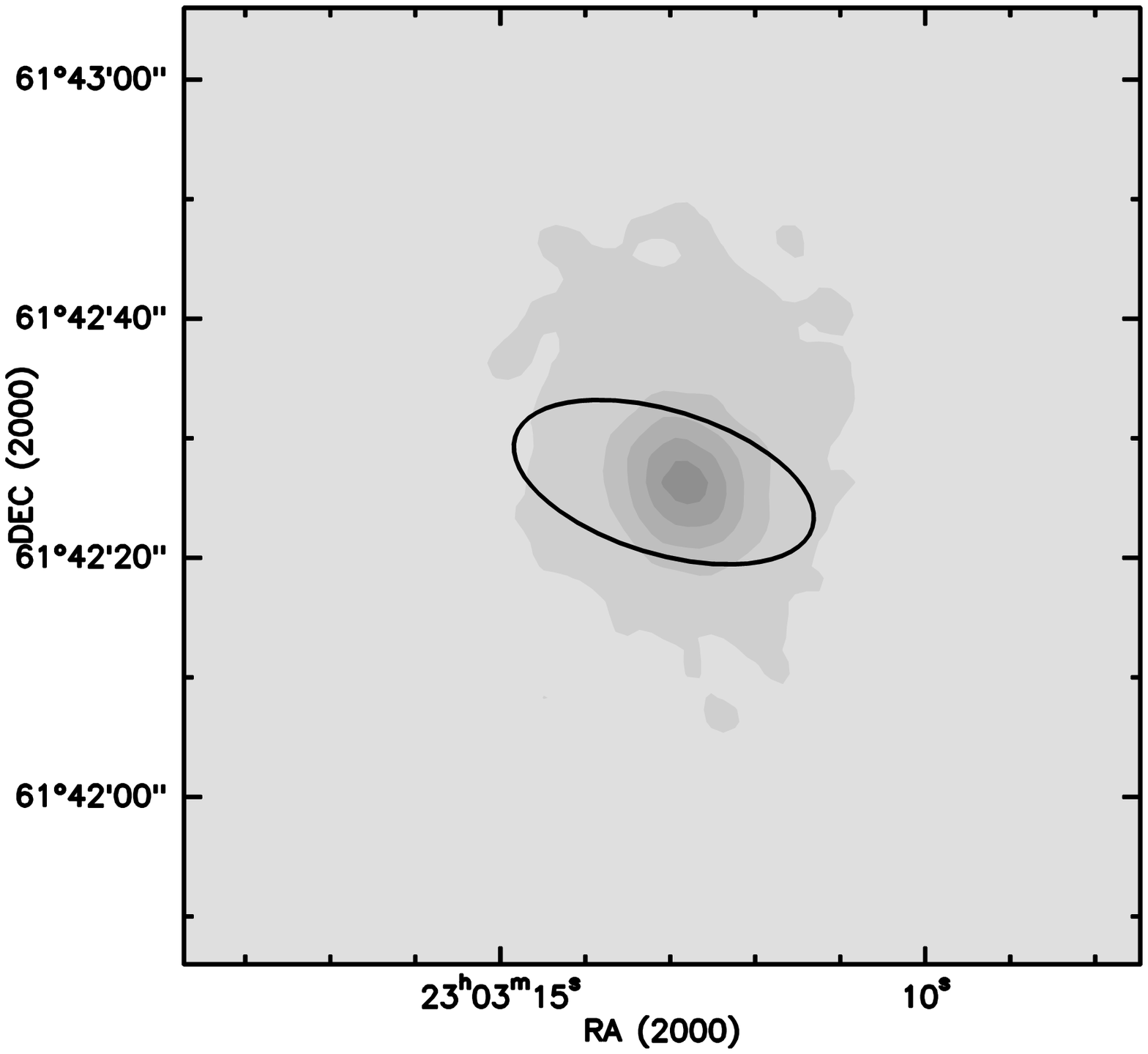}
\includegraphics[width=6cm,angle=270]{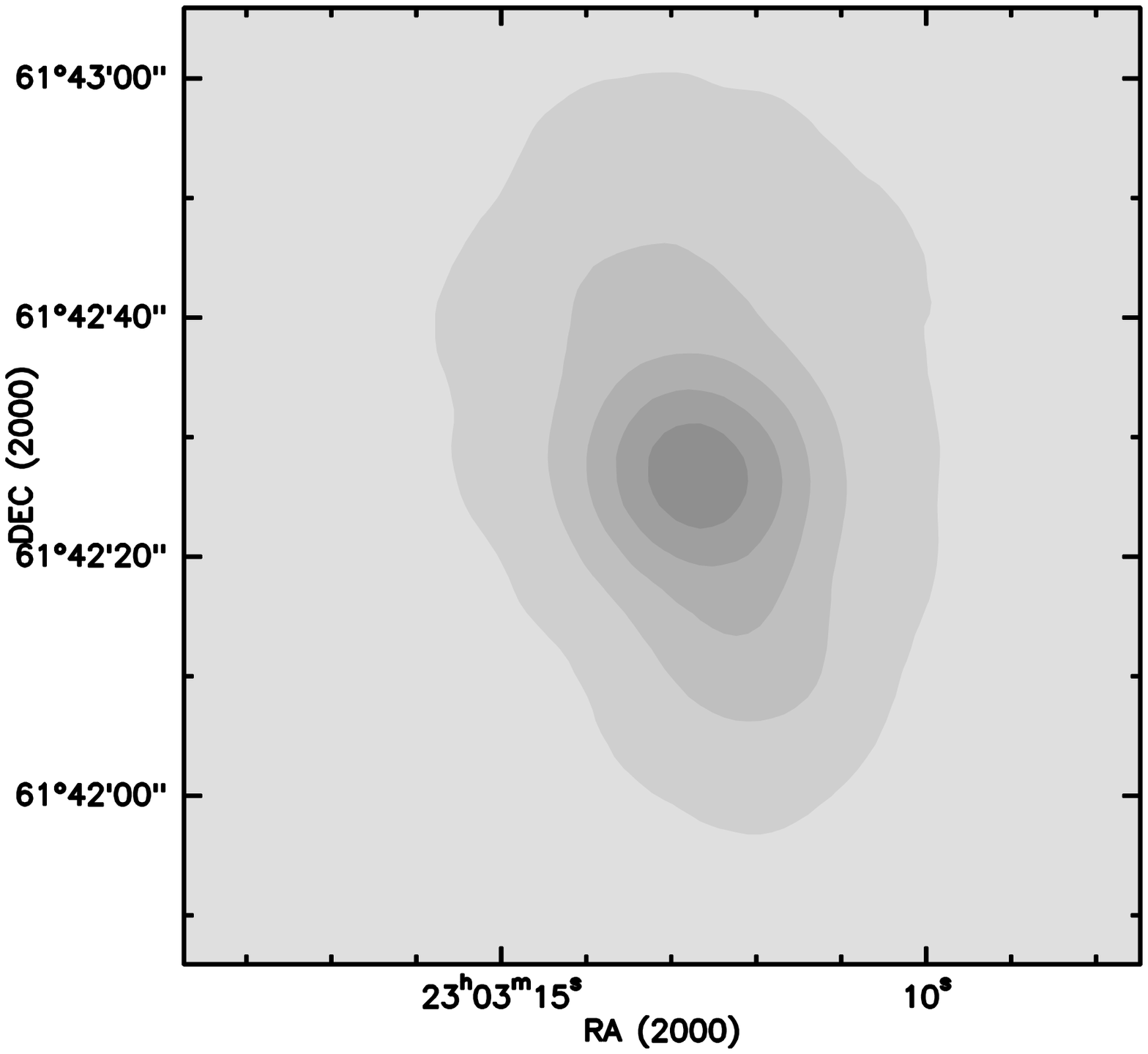}
\caption{Continuum emission maps around Cep E-mm at 24 $\mu$m (upper
  left panel), 70 $\mu$m (upper right panel), 450 $\mu$m (lower left
  panel), and 850 $\mu$m (lower right panel). The grey scale contours mark the
  continuum flux from 10 \% to 90 \% of the peak emission in steps of
  20 \%. The uncertainty ellipse position of the IRAS observation
  toward Cep E-mm is reported on the 450 $\mu$m
  map. \label{CepE_maps}}
\end{figure*}

\begin{figure*} \centering
\includegraphics[width=5cm,angle=90]{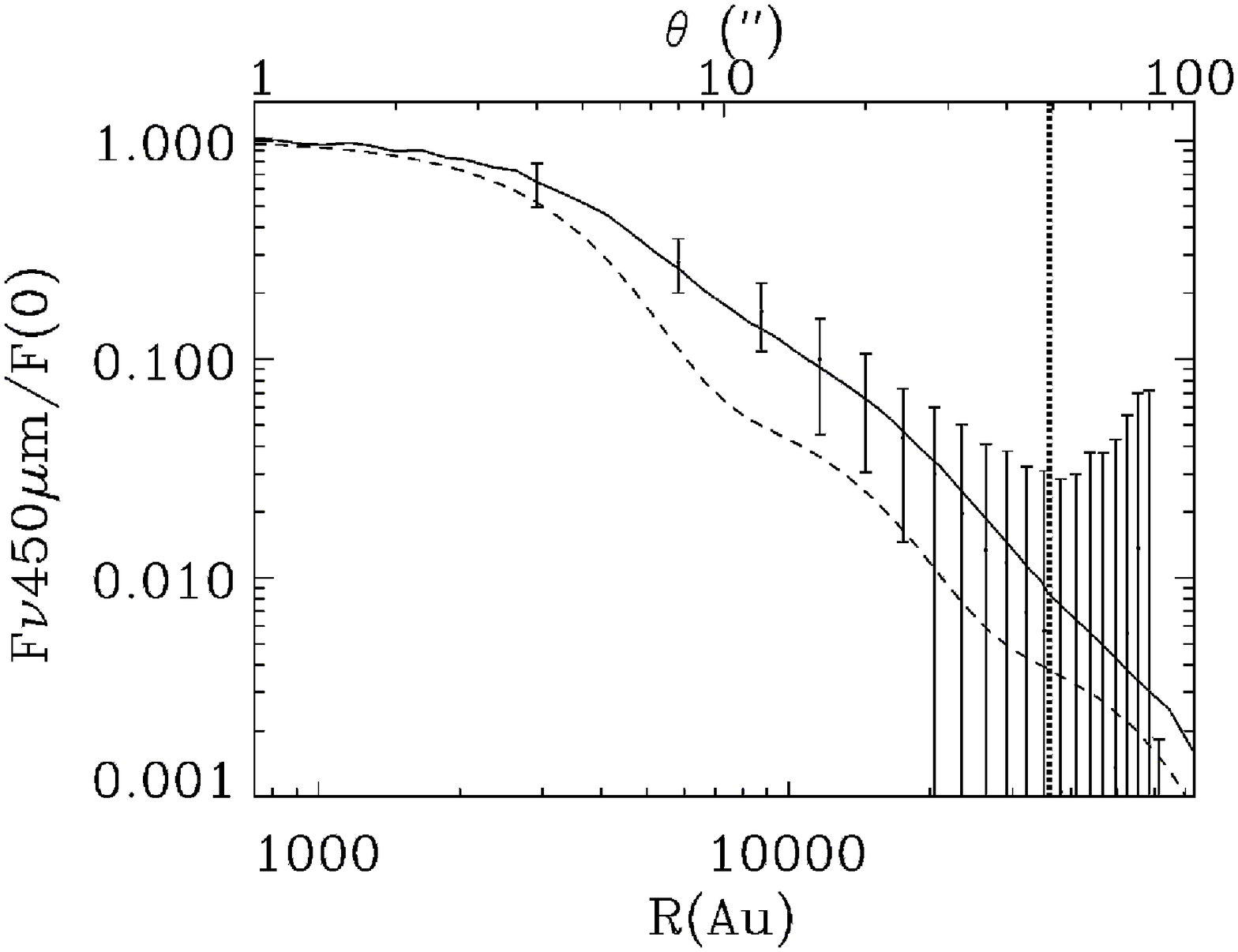}
\includegraphics[width=5cm,angle=90]{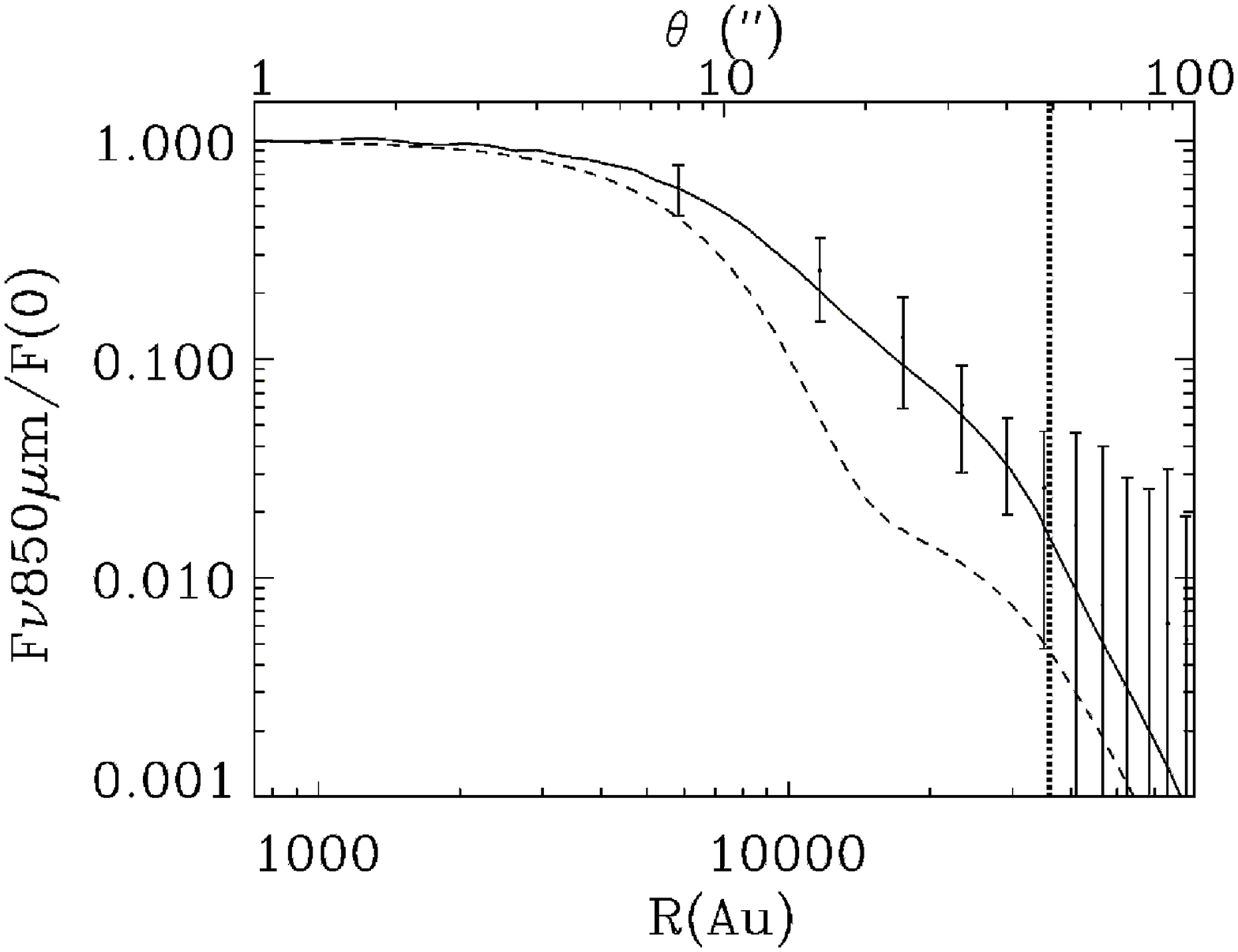}
\includegraphics[width=5cm,angle=90]{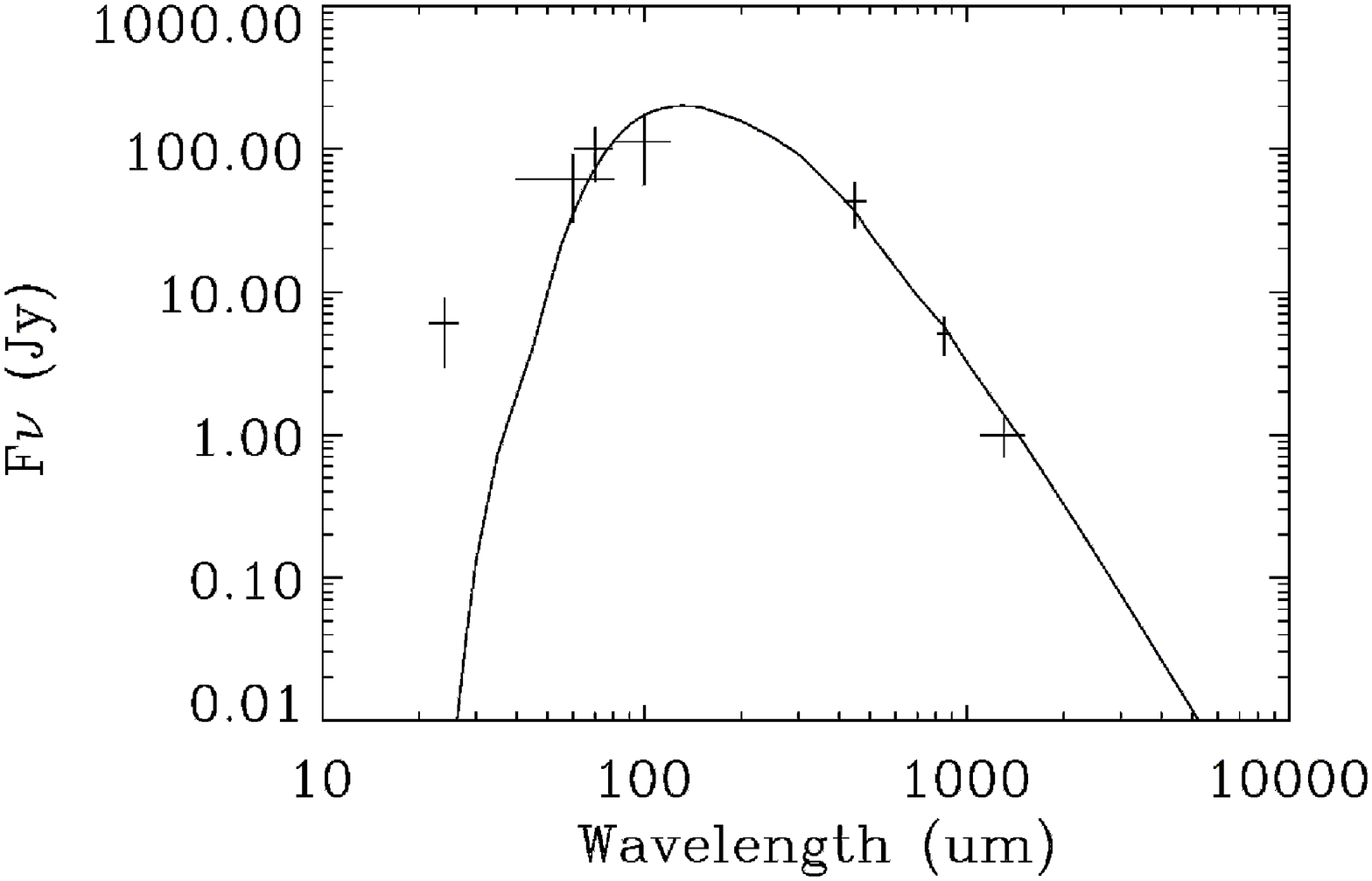}
\caption{Observed brightness profiles at 450$\mu$m (upper left panel), 850
  $\mu$m (upper right panel), and the SED (lower centred panel) toward
  Cep E-mm. The solid lines report the best-fit model (Table \ref{best_fit_phy_param}). 
  The dashed
  lines represent the beam pattern of the telescope at 450 $\mu$m and 850
  $\mu$m. The vertical dotted lines show the outer radius of the best-fit model. \label{CepE_fits}}
\end{figure*}

The continuum maps used for the Cep E-mm analysis are presented in
Fig. \ref{CepE_maps}. The maps at 450 $\mu$m and 850 $\mu$m were
obtained in August 1997 as a part of the project m97bu87 and the observations
are described in detail by \citet{Chini2001}. The flux profiles
obtained at each wavelength are shown in Fig. \ref{CepE_fits}.  The
Spitzer observations were obtained on the 2003 September 29 as part of
the programme ``MIPS/IRAC imaging of protostellar jet HH 212'' (AOR:
1063, PI: Alberto Noriega-Crespo) and are described in detail by
\citet{Noriega-Crespo2001}.  The integrated fluxes at 450 $\mu$m, 850 $\mu$m, and
1300 $\mu$m were retrieved from the literature \citep{Chini2001}. 
The uncertainty ellipse position of the IRAS observation is reported on the 450
$\mu$m map in Fig. \ref{CepE_maps}. The fluxes at 24 $\mu$m and 70 $\mu$m
were obtained by integration over a $\sim$ 30\arcsec and $\sim$ 60\arcsec
radius, respectively. Note that to compute the flux uncertainties, we
varied the integration radius of about 50\% and found variations of
$\lesssim$ 30\%. The integrated fluxes used for the analysis
are reported in Table \ref{Integrated_fluxes} and in
Fig. \ref{CepE_fits}.

\subsubsection{Best-fit}\label{subsubsec:CepE-mm_Best_Fit}

\begin{figure*} \centering
\includegraphics[width=5cm,angle=90]{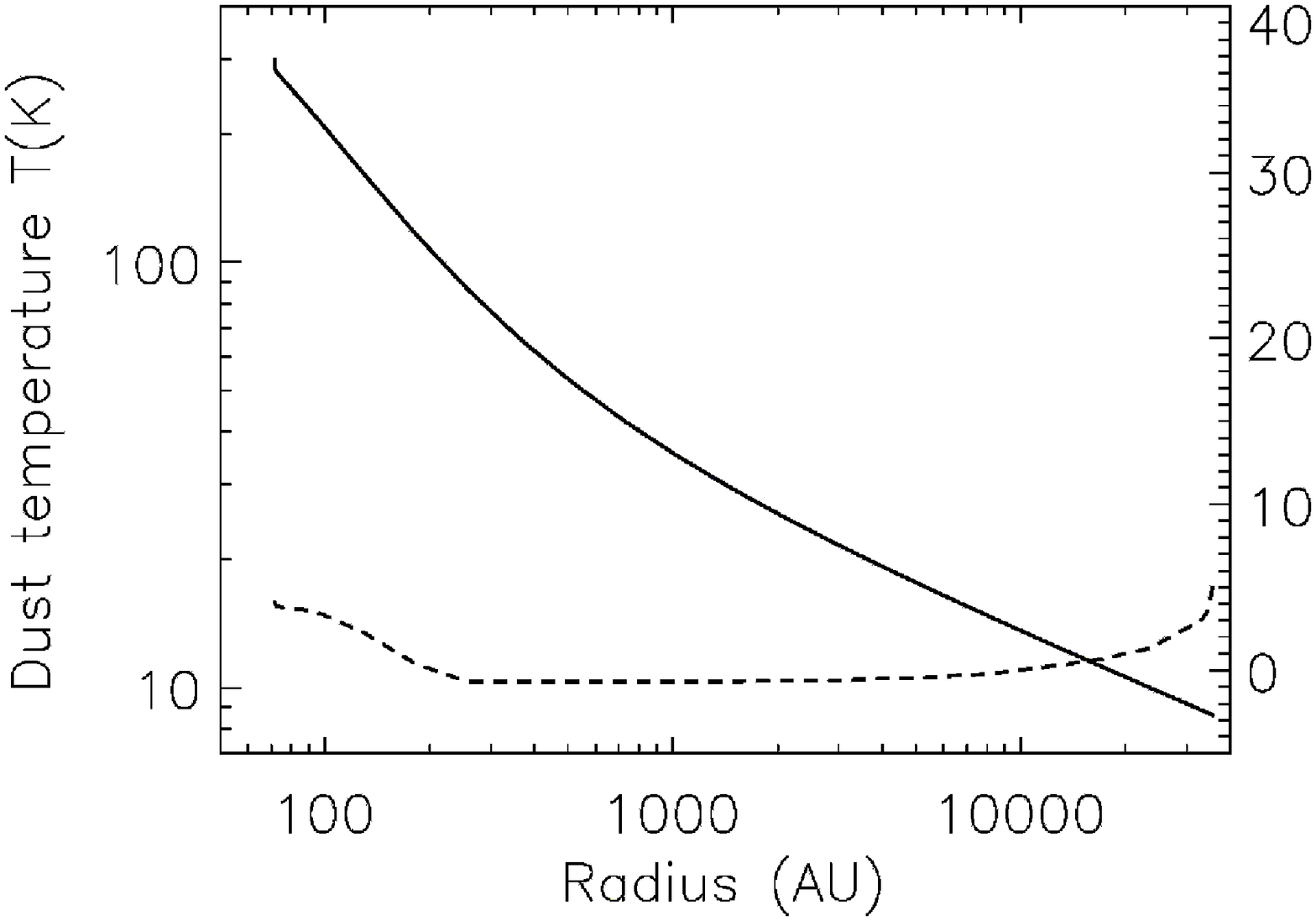}
\includegraphics[width=5cm,angle=90]{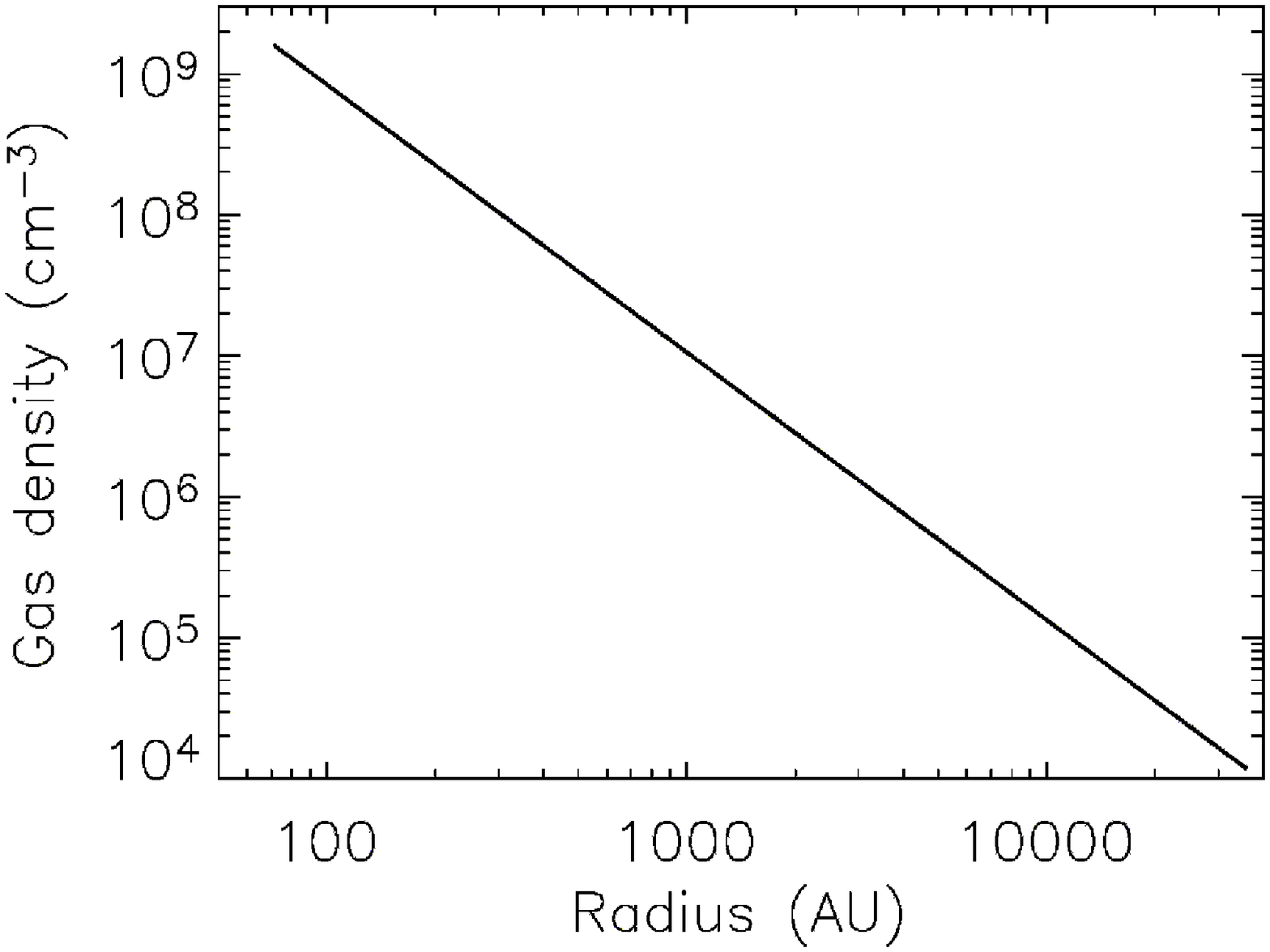}
\caption{Cep E-mm dust temperature (left panel) and H$_{2}$ density
  (right panel) profiles from the best-fit model. The dashed line
  represents the difference between the dust and gas temperature
  derived using X(H$_2$O)$_{in} =$ 1$\times$10$^{-5}$ and
  X(H$_2$O)$_{out} =$ 1$\times$10$^{-7}$ (scale on right axis). \label{CepE_T_N}}
\end{figure*}
Table \ref{best_fit_phy_param} presents the set of parameters $\alpha$,
Y, and $\tau_{100}$, which best reproduce the observations, and
summarises some relevant physical quantities of the model. Figure
\ref{CepE_fits} shows the derived brightness profiles and SED
against the observations. The dust density and temperature profiles
of the best-fit model are reported in Fig. \ref{CepE_T_N}. Although
the observed flux profiles and SED fluxes from 60 $\mu$m  to 1300 $\mu$m are
well reproduced by the model, the observed flux at 24 $\mu$m is
underestimated by about $\sim$3 orders of magnitude. 
Similarly to CB3-mm, the envelope model derived
for Cep E-mm with an $\alpha$ of 1.9 is very peaked, leading to
a strong gradient in the density profile, from $\sim$ 10$^9$ to $\sim$
10$^4$ cm$^{-3}$ with a density and radius at 100 K of
2.0$\times$10$^8$ cm$^{-3}$ and 223 AU, respectively. Considering the
$\chi ^{2}_{red}$ contour at 10\% of the minimum $\chi ^{2}_{red}$ in order to
estimate the uncertainty, we obtained $r_{out}$ = (3.6$\pm$0.7)
$\times10^{4}$ AU. Finally, in minimising the $\chi _{SED}^{2}$, we
also varied the source luminosity from 70 L$_\odot$ to 130 L$_\odot$ and
found the best-fit for a source luminosity equal to 100 L$_\odot$.

\subsection{IC1396 N BIMA 2}\label{subsec:IC1396N}
\subsubsection{Source background}\label{subsubsec:IC1396N_Background}

IC1396 N is a bright globule located 750 pc \citep{Matthews1979} from the Sun, near
the border of the IC1396 extended HII region \citep{Osterbrock1957,Weikard1996} and 
 at a projected distance of 11 pc north of the O6.5 star HD
206267 which ionises the region. IC1396 N is associated with the source
IRAS21391+5802. Its strong submillimetre and millimetre
continuum emission \citep{Wilking1993,Sugitani2000,Codella2001},
high-density gas \citep{Serabyn1993,Cesaroni1999,Codella2001,Beltr'an2004},
and water maser emission \citep{Felli1992,Tofani1995,Patel2000} reveal that IC1396
 N is an active site of star formation. The bolometric luminosity
is estimated to range from 235 L$_{\sun}$ \citep{Saraceno1996a} to 440
L$_{\sun}$ \citep{Sugitani2000}.

Using BIMA interferometric millimetre observations, \citet{Beltr'an2002}
 detected three sources (BIMA 1, BIMA 2 and BIMA 3) deeply
embedded in the globule. BIMA 2 has the strongest millimetre emission
and is also the most massive object. The authors concluded that BIMA 2 is
most likely an IM protostar, while BIMA 1, and BIMA 3, located at
$\sim$15\arcsec west and south-east respectively, are less massive and/or
more evolved. The recent studies by \citet{Neri2007} and \citet{Fuente2007},
 using Plateau de Bure interferometric observations,
provide a detailed study of BIMA 2 and showed that BIMA 2 contains at
least three dense cores separated by $\sim$1\arcsec.

Finally, the region presents an extended ($\sim$3 arcmin) CO bipolar
outflow \citep{Sugitani1989} which has been mapped by \citet{Codella2001}. The
study by \citet{Nisini2001}, revealed several
H$_{2}$ jets inside the region. A more recent analysis \citep{Beltr'an2002,Beltr'an2004}
 shows that BIMA 1 and BIMA 2 are associated with
north-south and east-west bipolar molecular outflows, respectively.

\subsubsection{Analysis}\label{subsubsec:IC1396N_Data_used}

\begin{figure*} \centering
\includegraphics[width=6cm,angle=270]{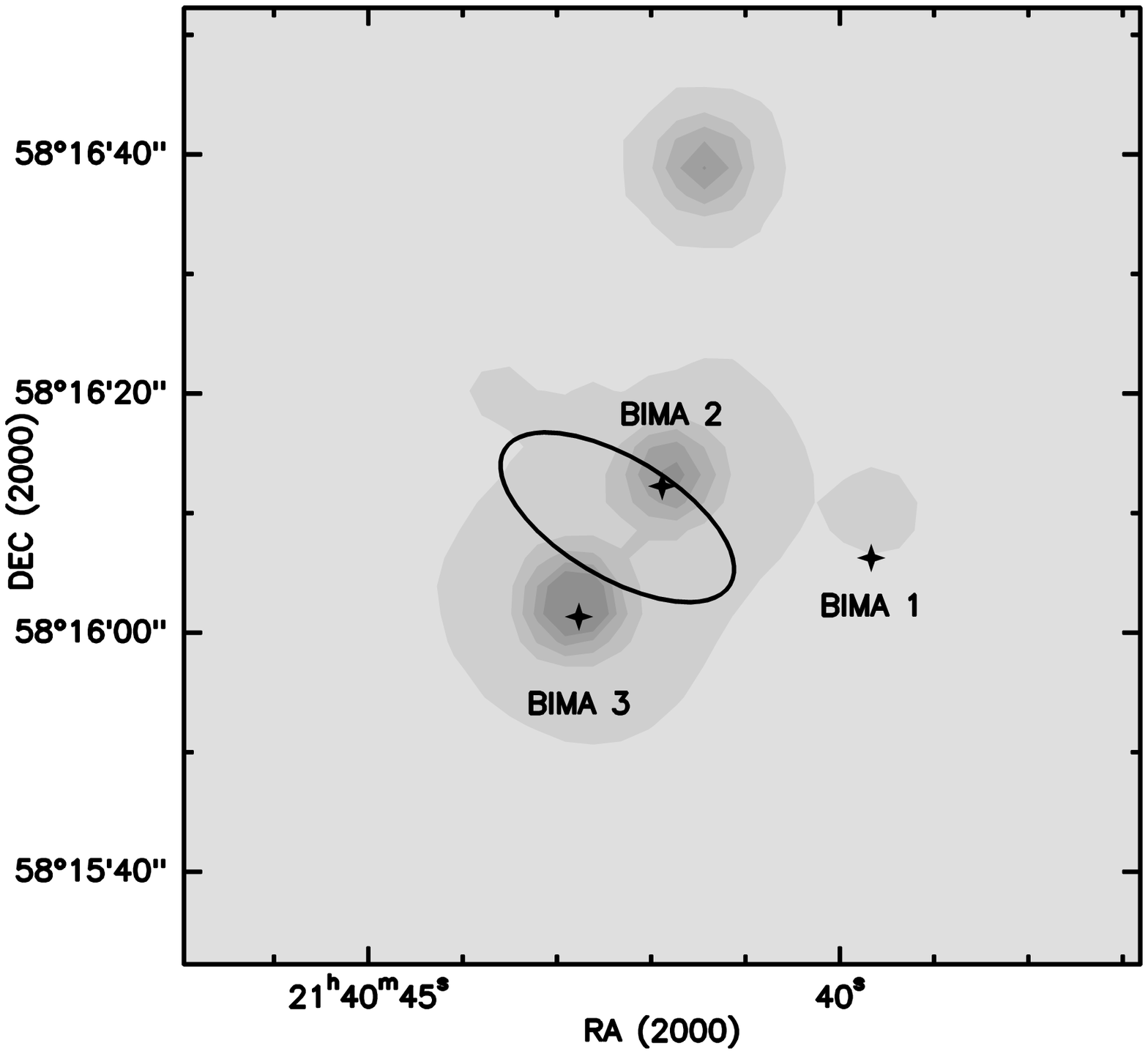}
\includegraphics[width=6cm,angle=270]{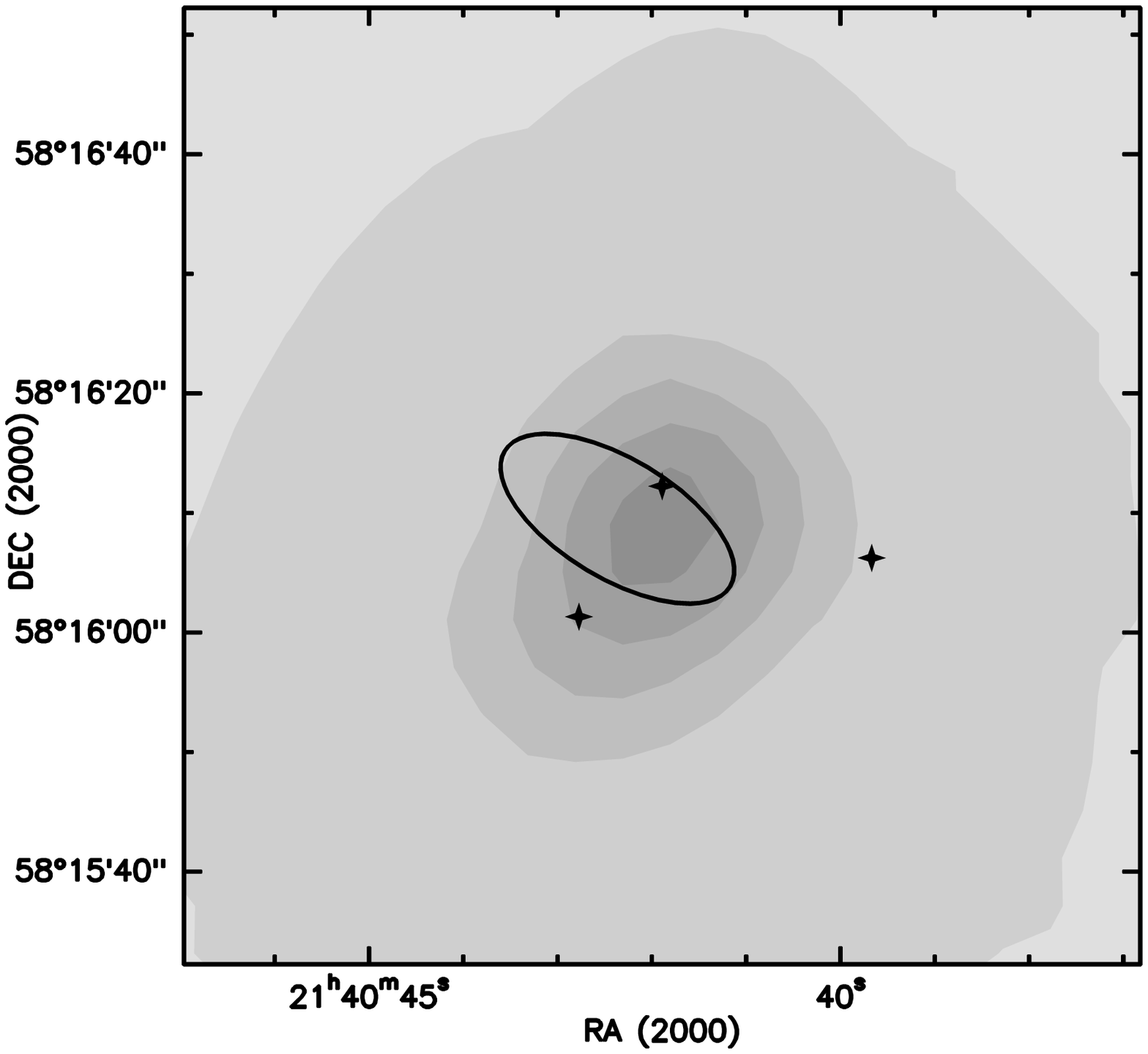}
\includegraphics[width=6cm,angle=270]{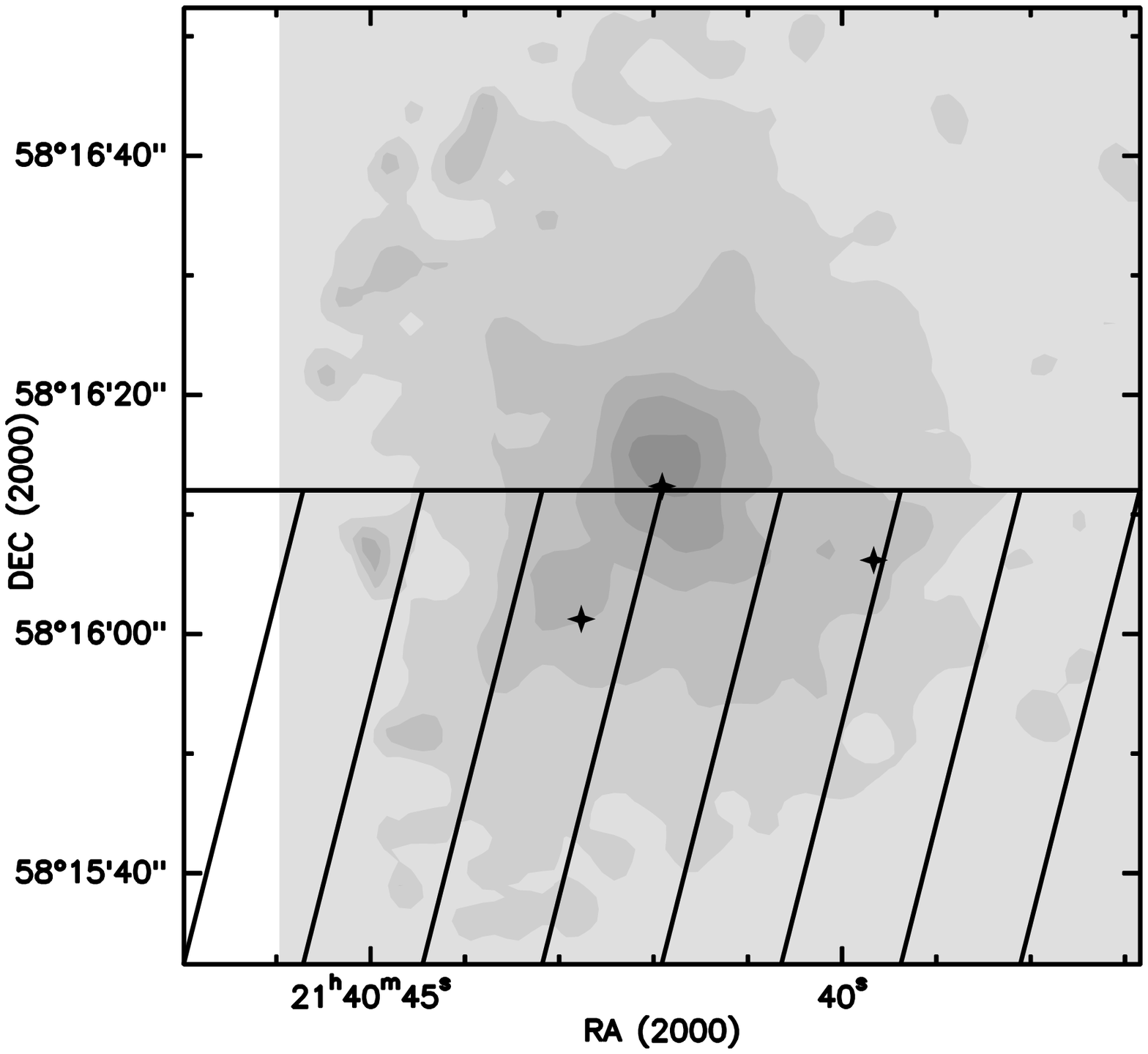}
\includegraphics[width=6cm,angle=270]{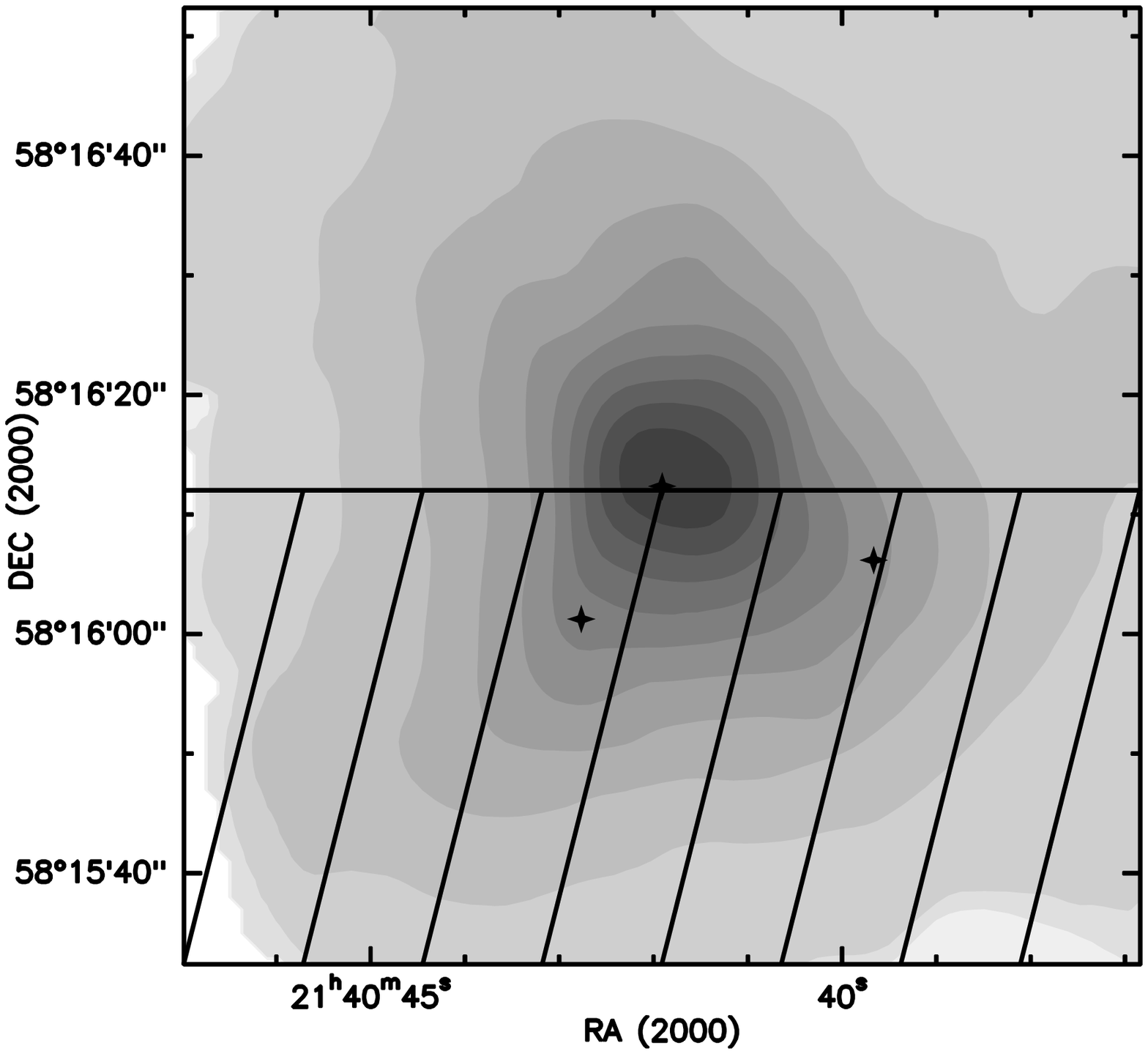}
\caption{Continuum emission maps around IC1396 N BIMA 2 at 24 $\mu$m (upper
  left panel), 70 $\mu$m (upper right panel), 450 $\mu$m (lower left
  panel), and 850 $\mu$m (lower right panel). The grey scale contours mark the
  continuum flux from 10 \% to 90 \% of the peak emission by steps of
  20 \%. The positions of BIMA 1, BIMA 2 and BIMA 3, as well as the
  uncertainty ellipse position of the IRAS observations are reported
  on the 24 $\mu$m map. The dashed regions on the 450 $\mu$m and 850 $\mu$m
  maps mark the regions excluded to derive the flux
  profiles. \label{IC1396N_maps}}
\end{figure*}

The continuum maps used for the IC1396 N BIMA 2 analysis are presented in
Fig. \ref{IC1396N_maps}.  The SCUBA maps were obtained in March 2002
as a part of the project m02au07. The brightness profiles (see
Fig. \ref{IC1396N_fits}) are derived excluding the regions
contaminated by the presence of BIMA 1 and BIMA 3 (dashed regions in
Fig. \ref{IC1396N_maps}). The Spitzer observations were obtained in
October 2004 as part of the programme ``Star Formation in Bright Rimmed
Clouds'' (AOR: 202, PI: Giovanni Fazio). The integrated fluxes used
for the analysis are reported in Table \ref{Integrated_fluxes} and in
Fig. \ref{IC1396N_fits}.  The integrated fluxes at 450 $\mu$m and 850 $\mu$m
were obtained by integration over a $\sim$ 35\arcsec. We varied the
integration radius from 20\arcsec to 90\arcsec, and found variations on the
fluxes $\lesssim$ 30\%. The large uncertainty in the integrated flux
at 24 $\mu$m is due to the proximity of BIMA 3 (see 24 $\mu$m map in
Fig. \ref{IC1396N_maps}). Note that due to the large IRAS beams at 60 $\mu$m
and 100 $\mu$m, BIMA 3 introduces significant spurious signal.
We thus used the IRAS and 70 $\mu$m Spitzer fluxes as upper
limits on the SED (Fig \ref{IC1396N_fits}).

\subsubsection{Best-fit}\label{subsubsec:IC1396N_best_fit}
\begin{figure*} \centering
\includegraphics[width=5cm,angle=90]{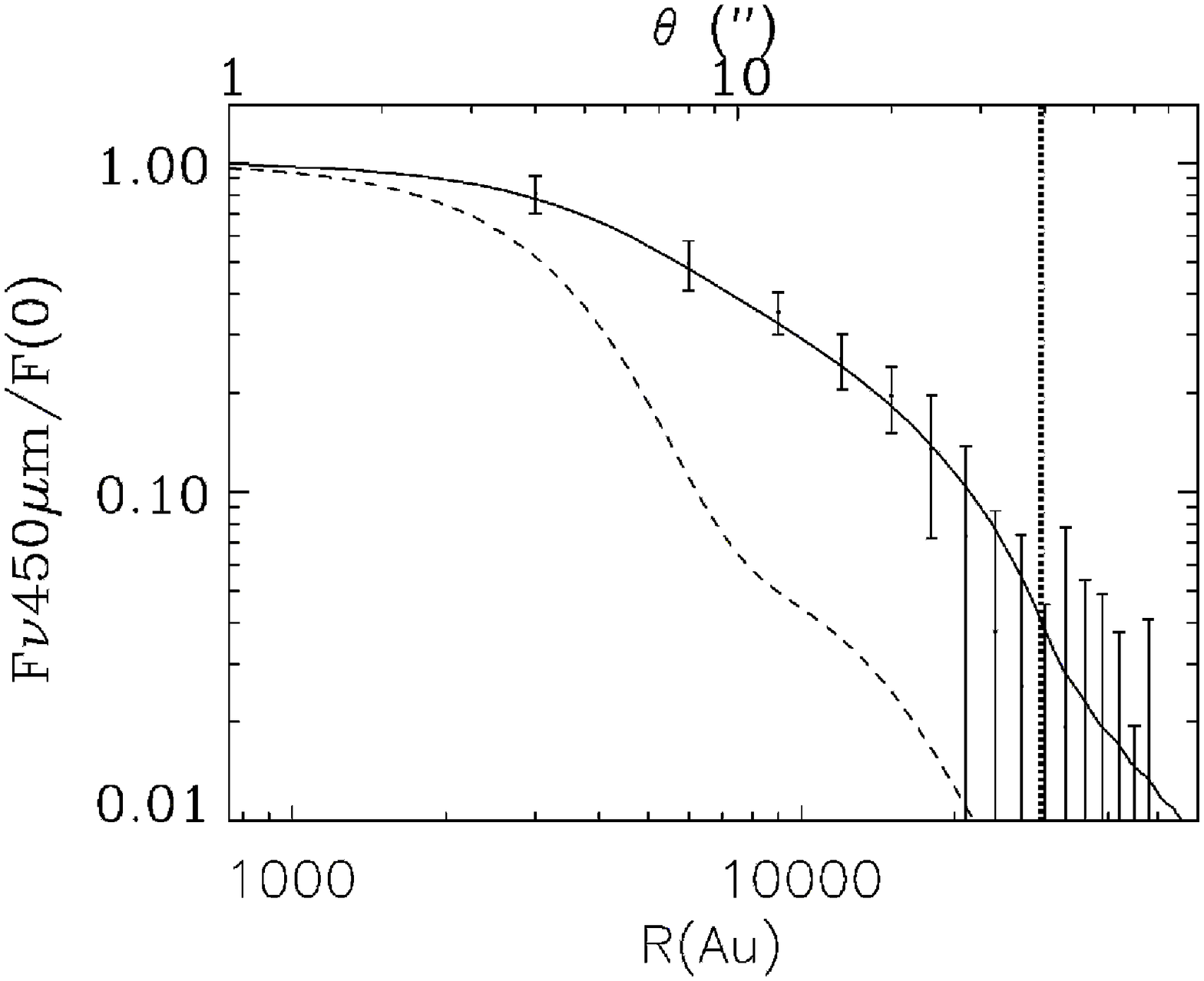}
\includegraphics[width=5cm,angle=90]{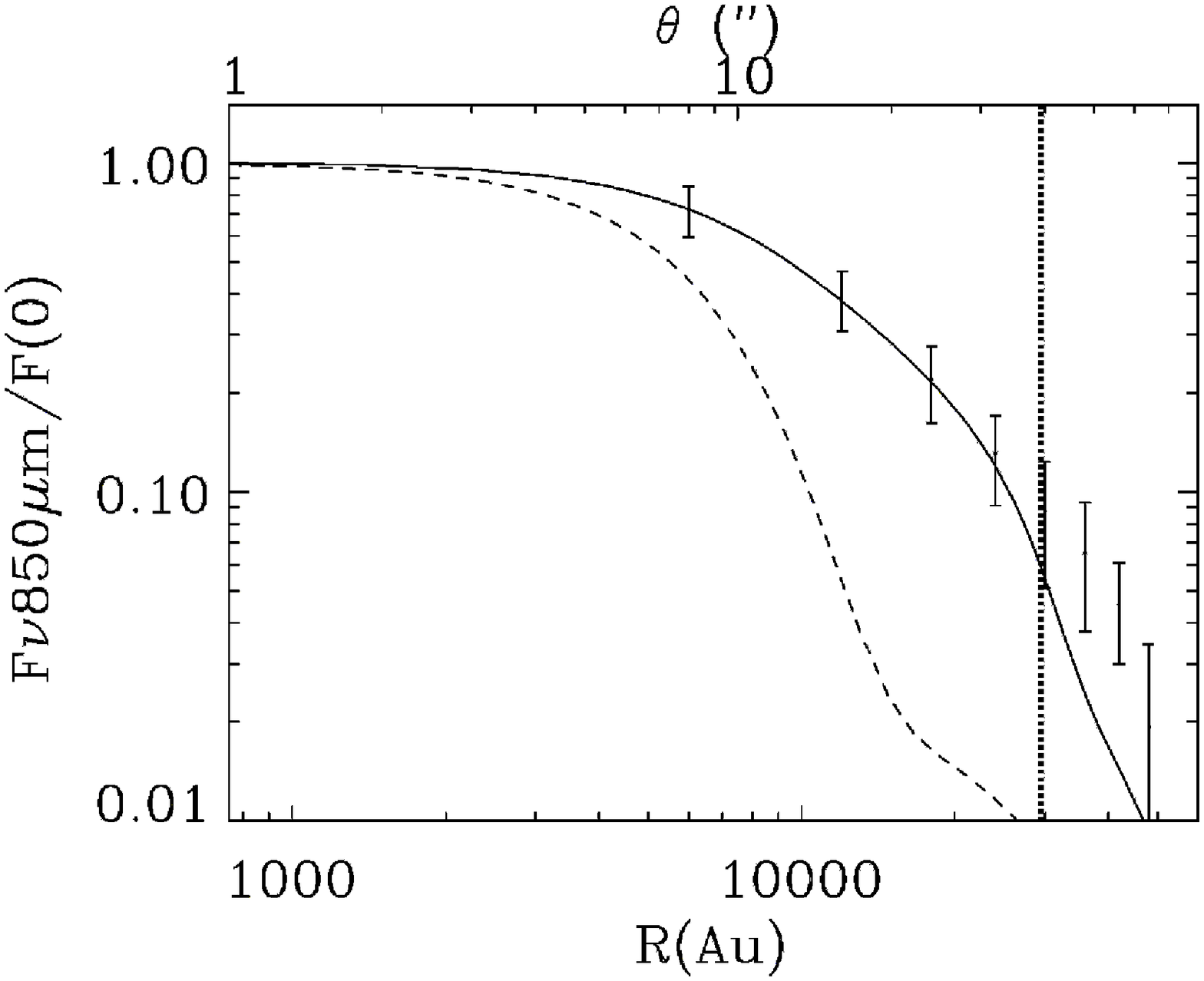}
\includegraphics[width=5cm,angle=90]{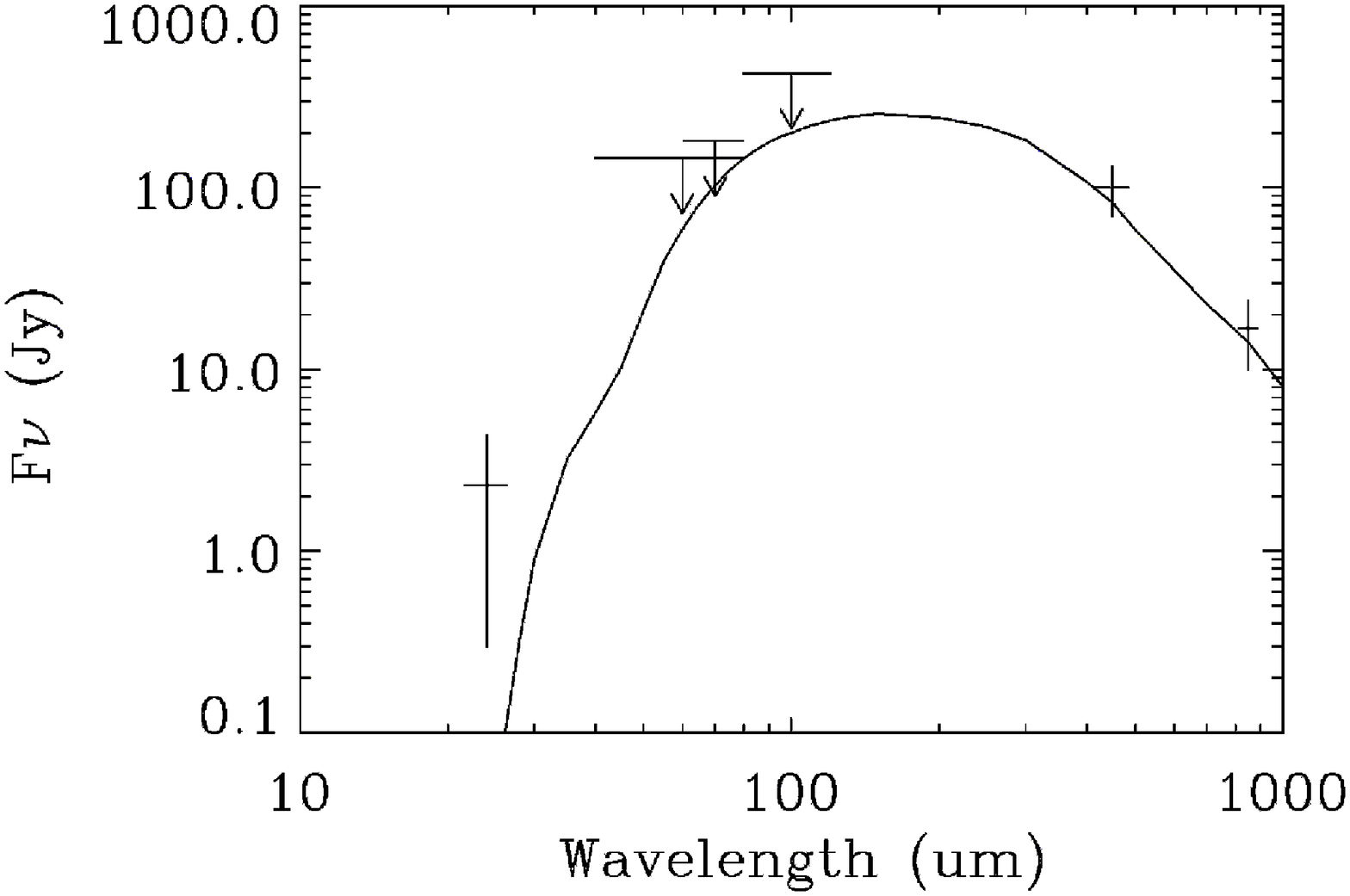}
\caption{Observed brightness profiles at 450 $\mu$m (upper left panel), 850
  $\mu$m (upper right panel) and the SED (lower centred panel) toward IC1396 N BIMA 2.
 The solid lines report the best-fit model (Table \ref{best_fit_phy_param}). The dashed lines
  represent the beam pattern of the telescope at 450 $\mu$m and 850
  $\mu$m. The vertical dotted lines show the outer radius of the best-fit model. The IRAS and 70 $\mu$m Spitzer fluxes are used as upper
  limits on the SED. \label{IC1396N_fits}}
\end{figure*}
\begin{figure*} \centering
\includegraphics[width=5cm,angle=90]{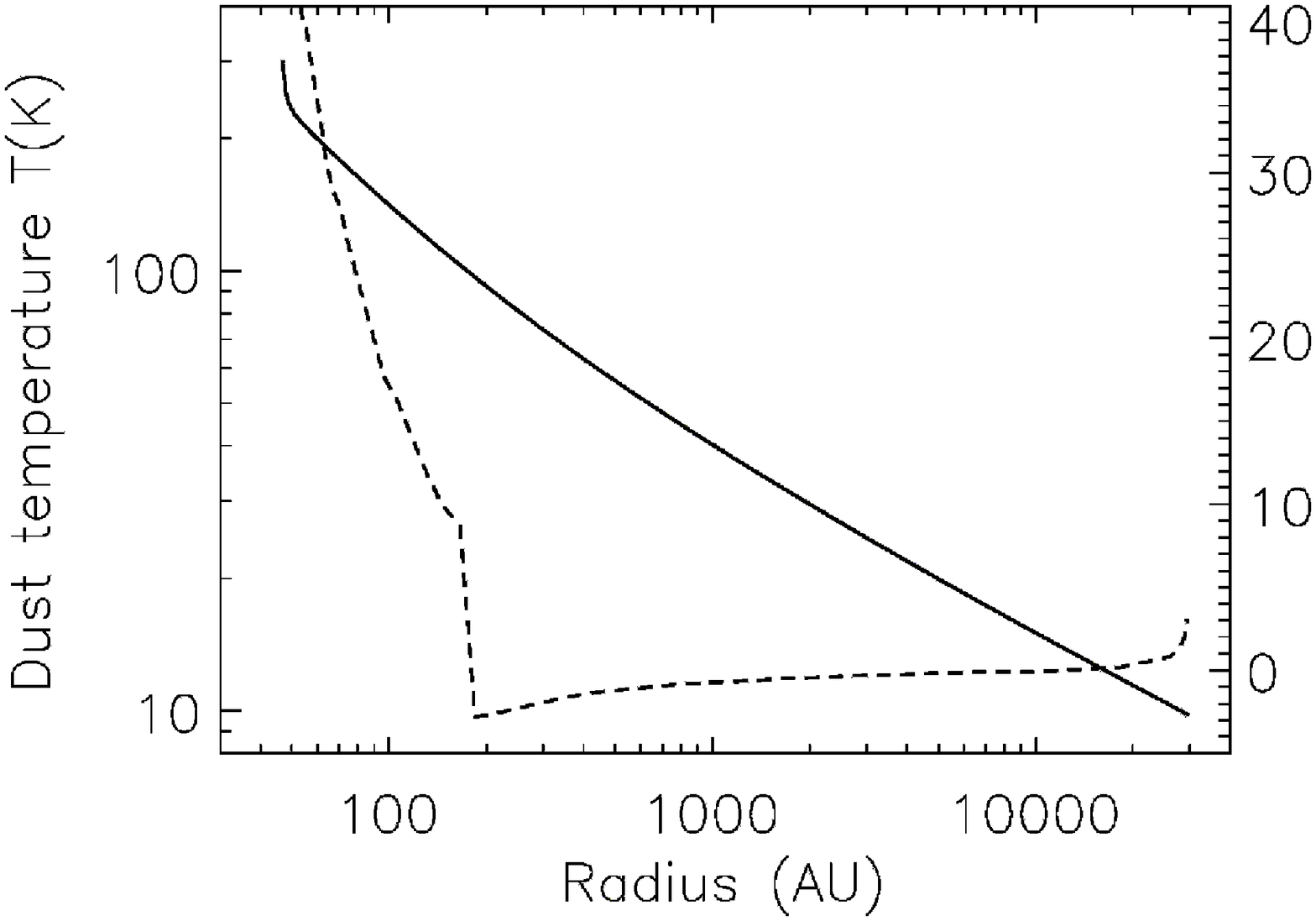}
\includegraphics[width=5cm,angle=90]{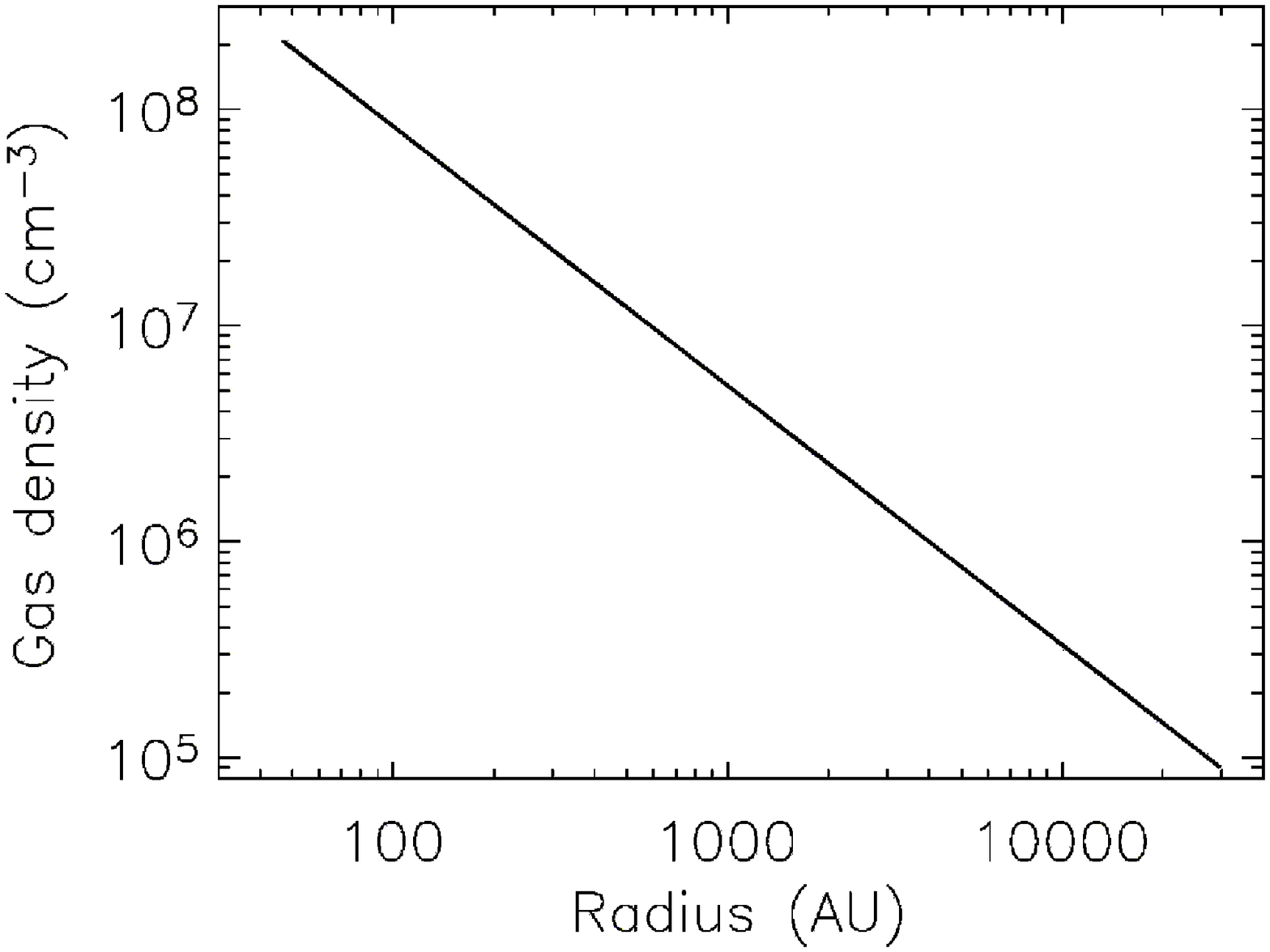}
\caption{IC1396 N BIMA 2 dust temperature (left panel) and H$_{2}$ density
  (right panel) profiles from the best-fit model. The dashed line
  represents the difference between the dust temperature and the gas
  temperature derived using X(H$_2$O)$_{in} =$ 1$\times$10$^{-5}$ and
  X(H$_2$O)$_{out} =$ 1$\times$10$^{-7}$ (scale on right axis). \label{IC1396N_T_N}}
\end{figure*}
Table \ref{best_fit_phy_param} presents the set of parameters $\alpha$,
Y, and $\tau_{100}$, which best reproduce the observations, and
summarises some relevant physical quantities of the model. Figure
\ref{IC1396N_fits} shows the relevant derived brightness profiles and
SED against the observations. The dust density and temperature
profiles of the best-fit model are reported in
Fig. \ref{IC1396N_T_N}. The observed flux profiles and SED fluxes at
450 $\mu$m and 850 $\mu$m are well reproduced by the model. Considering the
IRAS and 70 $\mu$m Spitzer fluxes as upper limits (see Sect.
\ref{subsubsec:IC1396N_Data_used}), the minimum $\chi_{SED}^{2}$ is
obtained for a source luminosity equal to 150 L$_\odot$, similar to
the value suggested by \citet{Beltr'an2002}. The observed flux at 24
$\mu$m is underestimated by about one order of magnitude. 
Considering the $\chi^{2}_{red}$ contour at 10\%
of the minimum $\chi ^{2}_{red}$ to estimate the uncertainty, we
obtained $r_{out}$ = (3.0$\pm$0.6) $\times10^{4}$ AU. 
Note that we obtain similar results without subtracting the dashed region of 
Fig. \ref{IC1396N_maps}, confirming the weak contribution
of BIMA 1 and 3 to the submillimetre emission.

Finally, contrary to the previous sources, the gas temperature profile
obtained for IC1396 N BIMA 2 is slightly decoupled from the dust
temperature in the inner region (see Fig. \ref{IC1396N_T_N}). The origin
of this thermal decoupling is discussed in detail in \citet{Crimier2009a}. 
Briefly, the decoupling is due to a competition between the water
abundance (the main coolant of the gas in the inner region) and the dust
density, which mainly regulates the balance energy in the inner part
of the envelope. This thermal decoupling is stronger in IC1396 N BIMA
2 (and NGC7129 FIRS 2) than in the other sources because of the lower
density in the inner region. The thermal decoupling reaches
a maximum value of about 40 K ($\sim$15\% of T$_{dust}$) for an
abundance of X(H$_2$O)$_{in} =$ 1$\times$10$^{-5}$ and about 10 K for
an abundance of X(H$_2$O)$_{in} =$ 1$\times$10$^{-6}$.

\subsection{NGC7129 FIRS 2}\label{subsec:NGC 7129 FIRS 2}

\subsubsection{Source background}

NGC7129 is a reflection nebula located in a complex and active
molecular cloud \citep{Hartigan1985,Miranda1993} and
estimated to be at a distance of 1250$\pm$50 pc \citep{Shevchenko1989} from the Sun.
 The region contains several Herbig AeBe stars, which
illuminate the nebula. NGC7129 FIRS 2 has been detected in the
far-infrared by \citet{Bechis1978} and \citet{Harvey1984}. FIRS 2
is not detected at optical or near-infrared wavelengths. Its position
coincides with a $^{13}$CO column density peak \citep{Bechis1978} and
a high-density NH$_{3}$ cloudlet \citep{Guesten1986}. It
is also close to an H$_2$O maser \citep{Rodriguez1980}. NGC7129 FIRS 2
has been classified as an IM Class 0 source by \citet{Eiroa1998}, who
carried out a multi-wavelength study of the continuum emission from 25 $\mu$m
to 2000 $\mu$m. These authors estimate a total mass and bolometric luminosity
 of $\sim$ 6 M$_{\sun}$ and $\sim$ 430 L$_{\sun}$,
respectively. In addition, Edwards $\&$ Snell (1983) detected a
bipolar CO outflow associated with FIRS 2. The interferometric study
of \citet{Fuente2001} has confirmed this bipolar
outflow and pointed out a quadrapolar morphology to the flow. This
quadrapolar morphology seems to be due to the superposition of two
flows, FIRS 2-out 1 and FIRS 2-out 2, likely associated with FIRS 2
and a more evolved star (FIRS 2-IR), respectively. 
Finally, \citet{Fuente2005a,Fuente2005}
 carried out an extensive chemical study of FIRS 2
providing the first detection of a hot core in an IM Class 0. Based on all
these studies, FIRS 2 is considered the youngest IM object known at
present.

\subsubsection{Analysis}

\begin{figure*} \centering
\includegraphics[width=6cm,angle=270]{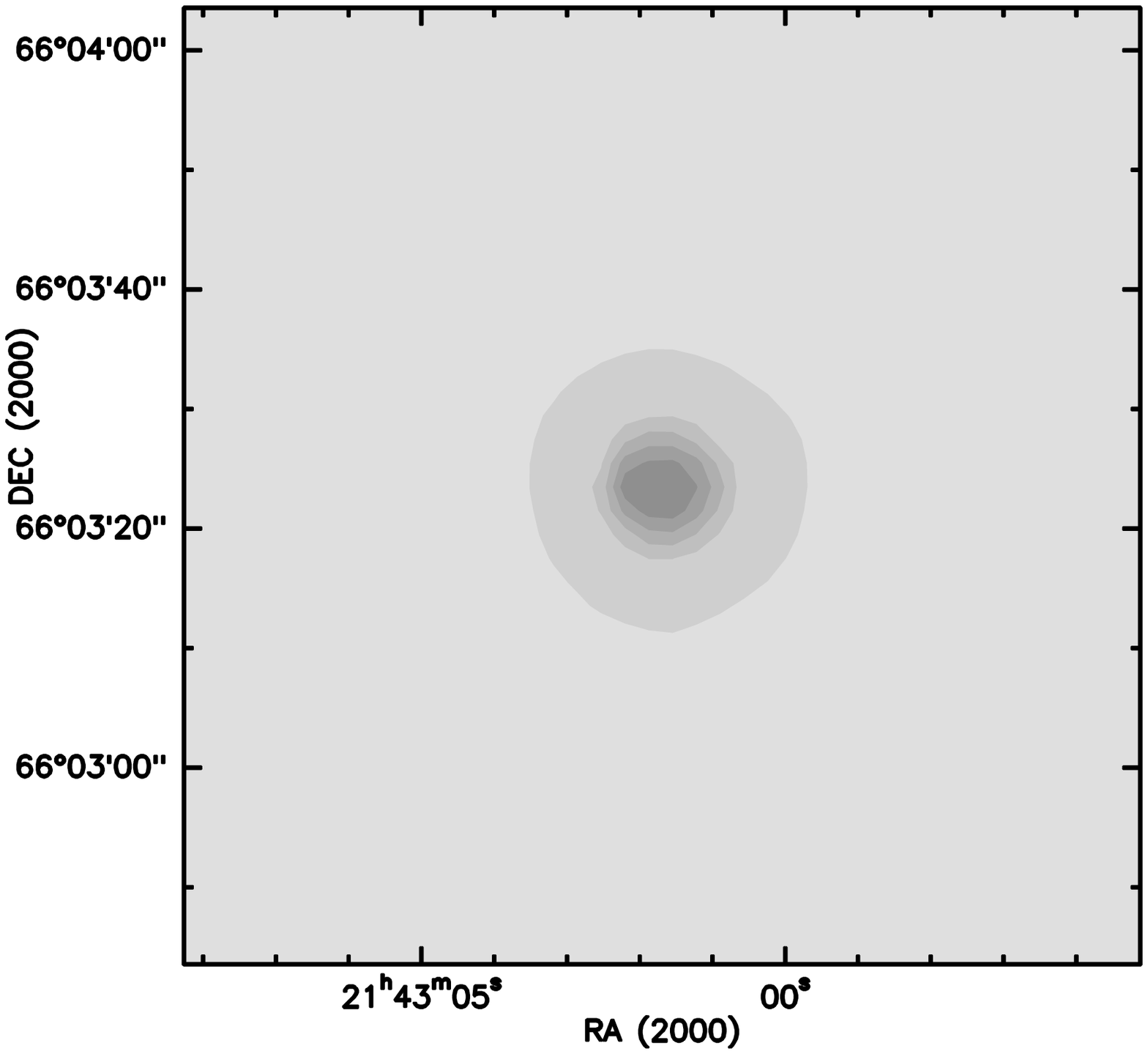}
\includegraphics[width=6cm,angle=270]{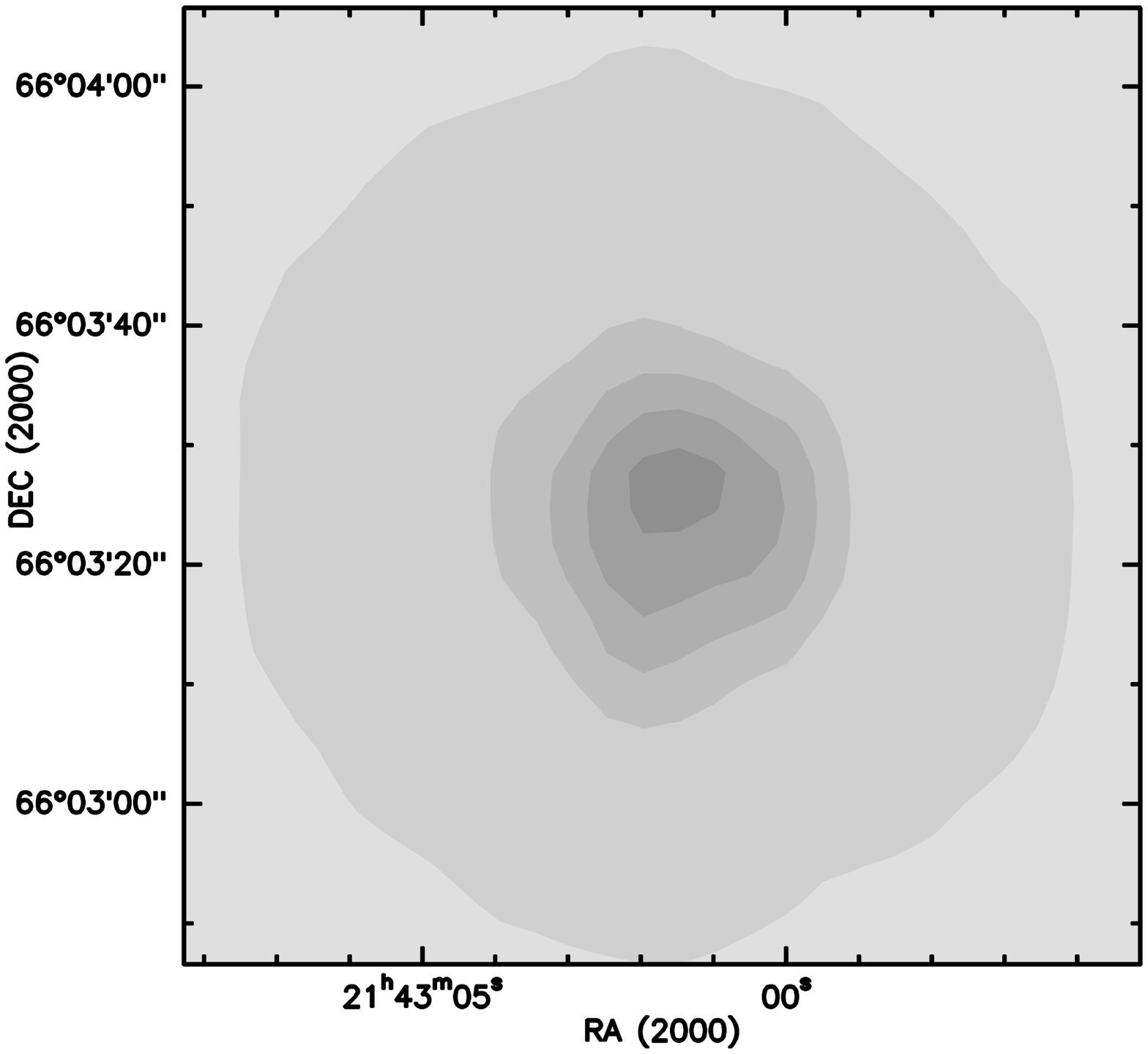}
\includegraphics[width=6cm,angle=270]{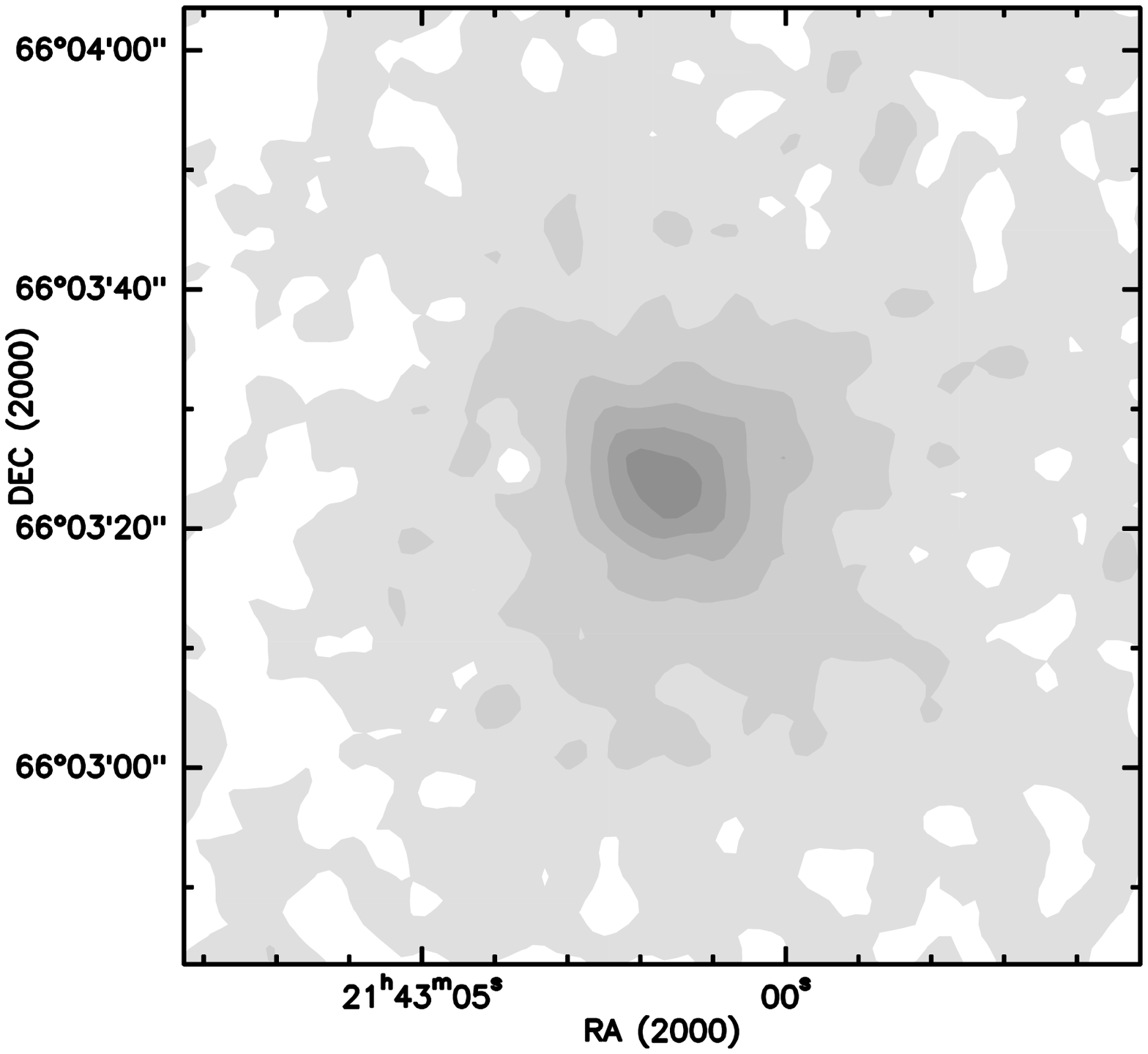}
\includegraphics[width=6cm,angle=270]{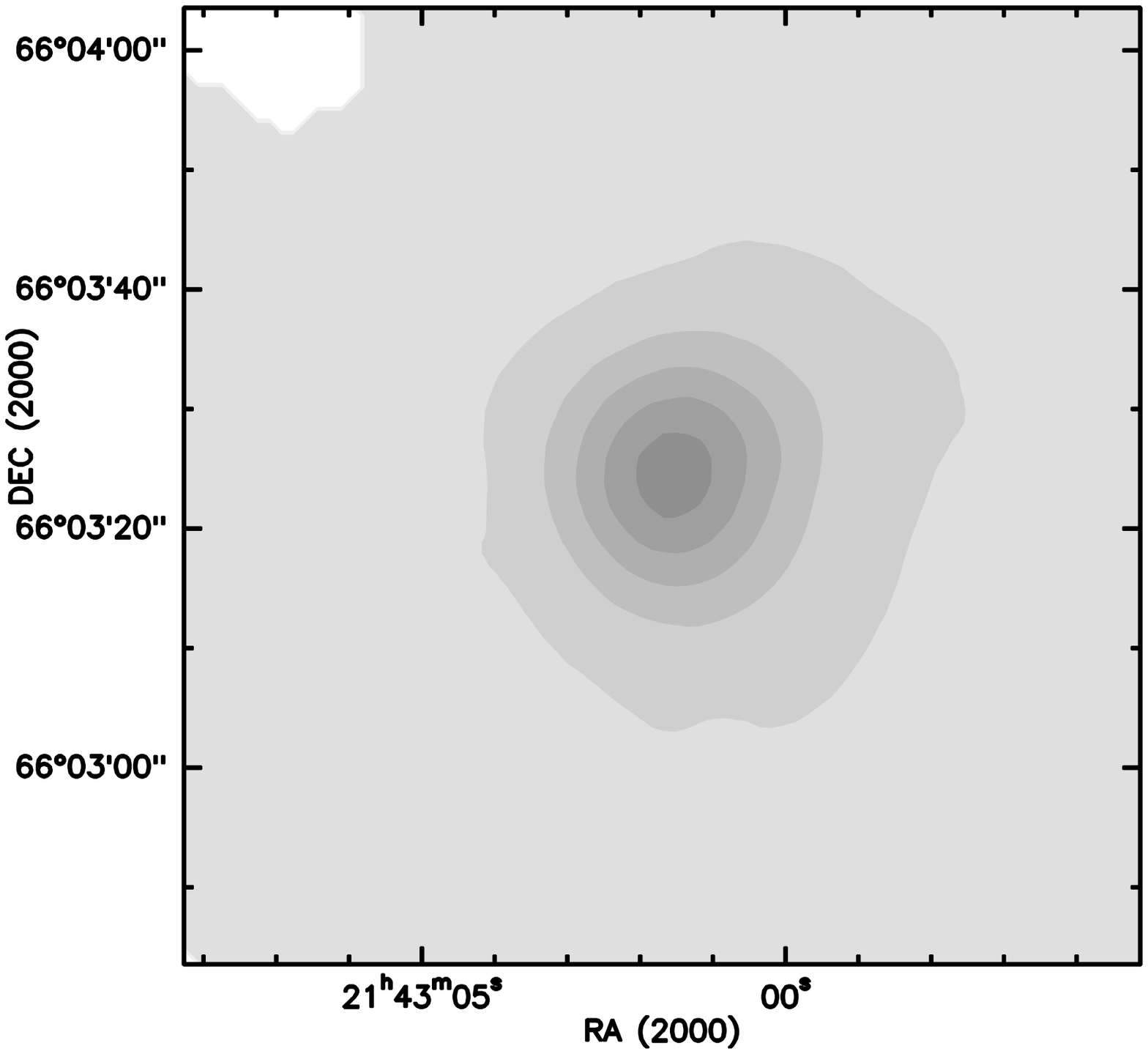}
\includegraphics[width=6cm,angle=270]{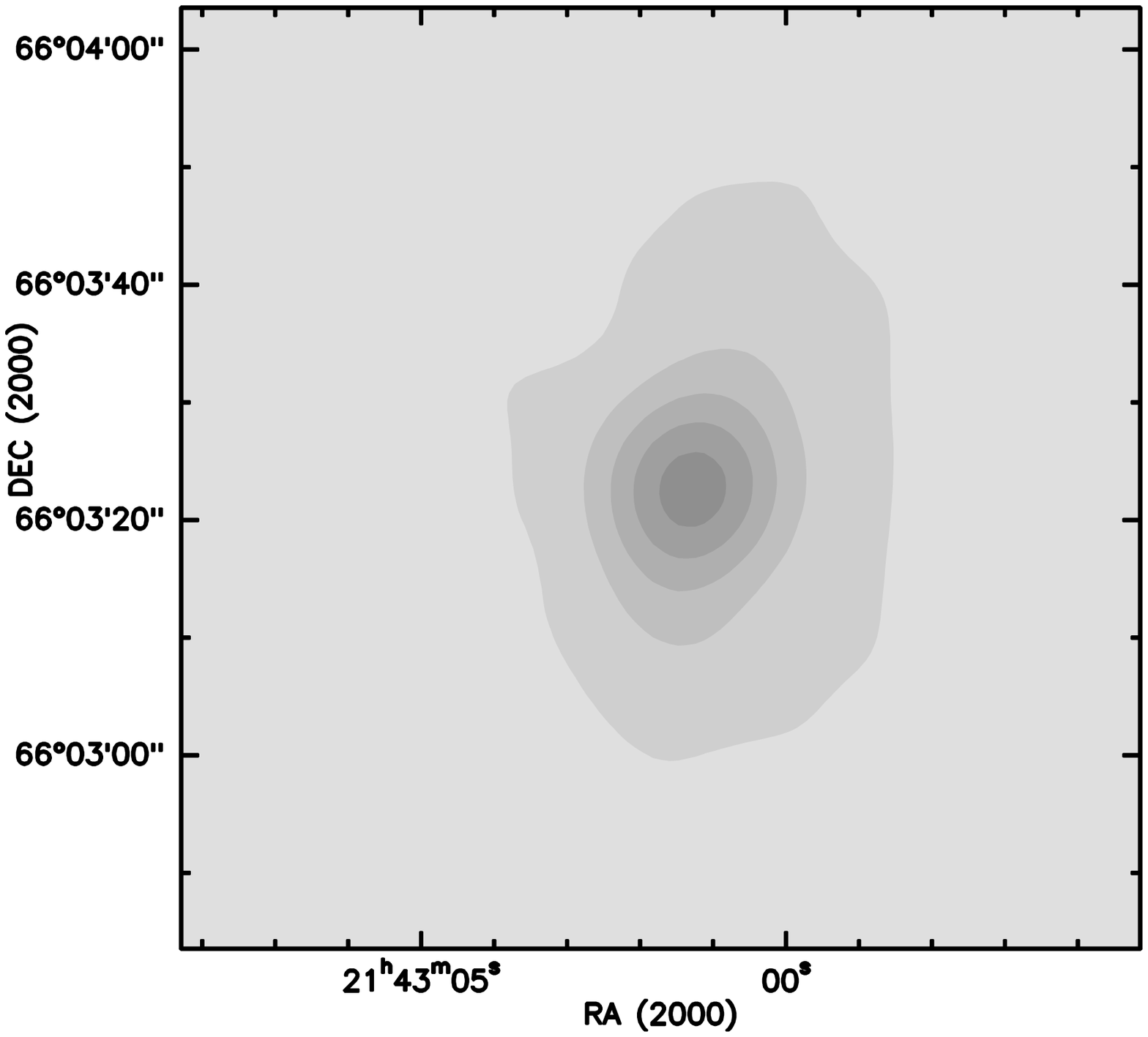}
\caption{Continuum emission maps around NGC7129 FIRS 2 at 24 $\mu$m (upper
  left panel), 70 $\mu$m (upper right panel), 450 $\mu$m (middle left
  panel), 850 $\mu$m (middle right panel), and 1300 $\mu$m (lower
  centred panel). The grey scale contours mark the continuum flux from 10 \% to
  90 \% of the peak emission by steps of 20 \%.  \label{NGC7129_maps}}
\end{figure*}

All the data and continuum maps used for the NGC7129 FIRS 2 analysis 
are reported in Table \ref{Integrated_fluxes} and Fig. \ref{NGC7129_maps}. T
he maps at 450 $\mu$m and 850 $\mu$m were obtained on the
1998 August 26 as a part of the project M98BU24. We also used the
flux profile and the integrated flux at 1300 $\mu$m extracted from the
single dish map at 1300 $\mu$m, presented in \citet{Fuente2001}.
Following \citet{Greve1998}, the beam at 1300 $\mu$m 
is assumed to be a combination
of three Gaussian curves, with HPBWs of 11\arcsec, 125\arcsec, and 180\arcsec, 
with amplitude ratios of 0.975, 0.005, and 0.001, respectively.  The
Spitzer observations were obtained on the 2003 September 19 as part of
the programme ``Protostars and Proto-Brown Dwarfs in a Nearby Dark
Cloud'' (AOR: 34, PI: Tom Megeath).
The integrated fluxes reported in Table \ref{Integrated_fluxes} and in
Fig. \ref{NGC7129_fits} were obtained by integration over a $\sim$ 40\arcsec
radius. The IRAS integrated fluxes at 60 $\mu$m and 100 $\mu$m were taken
from \citet{Eiroa1998}.

\subsubsection{Best-fit}

\begin{figure*} \centering
\includegraphics[width=5cm,angle=90]{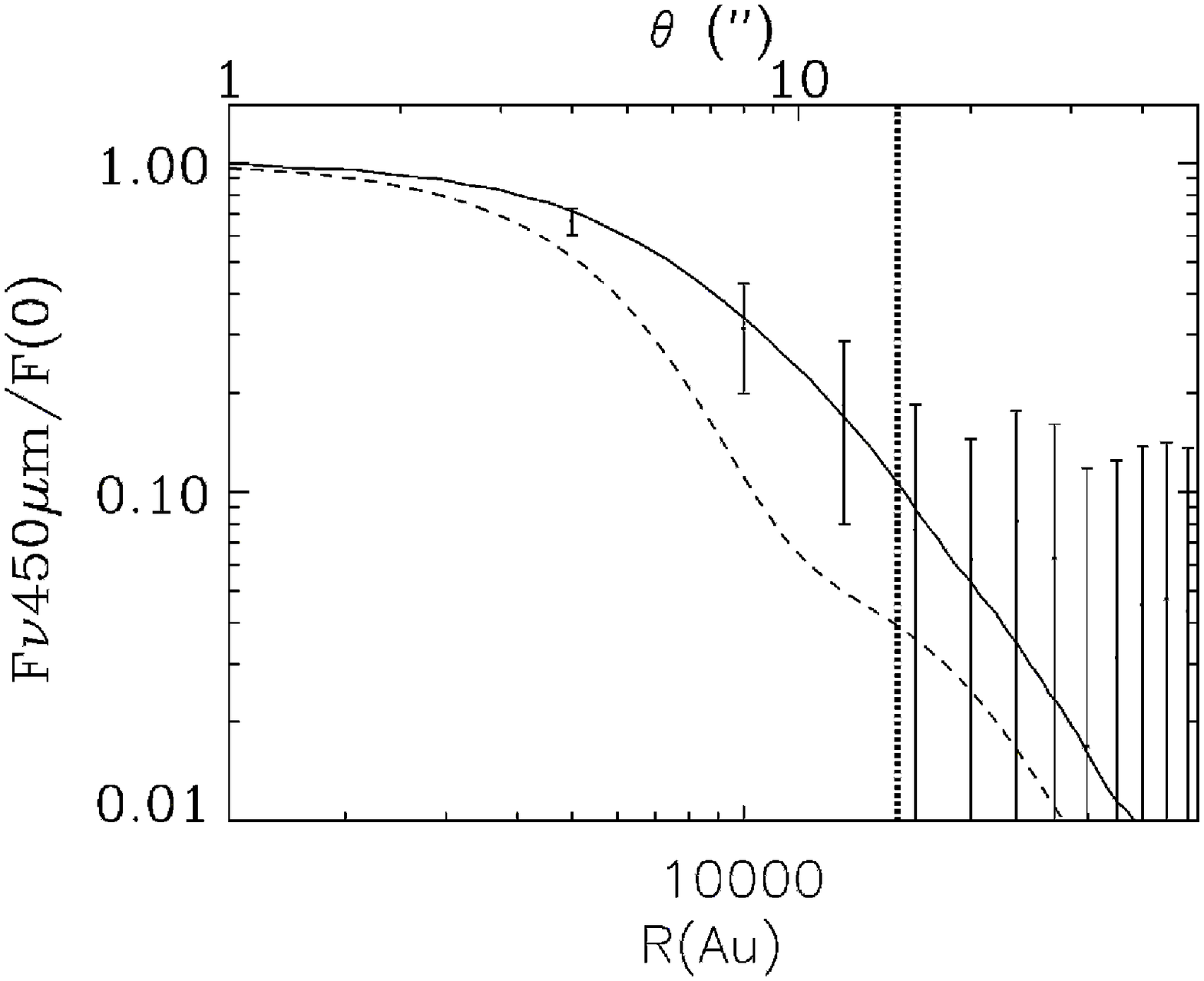}
\includegraphics[width=5cm,angle=90]{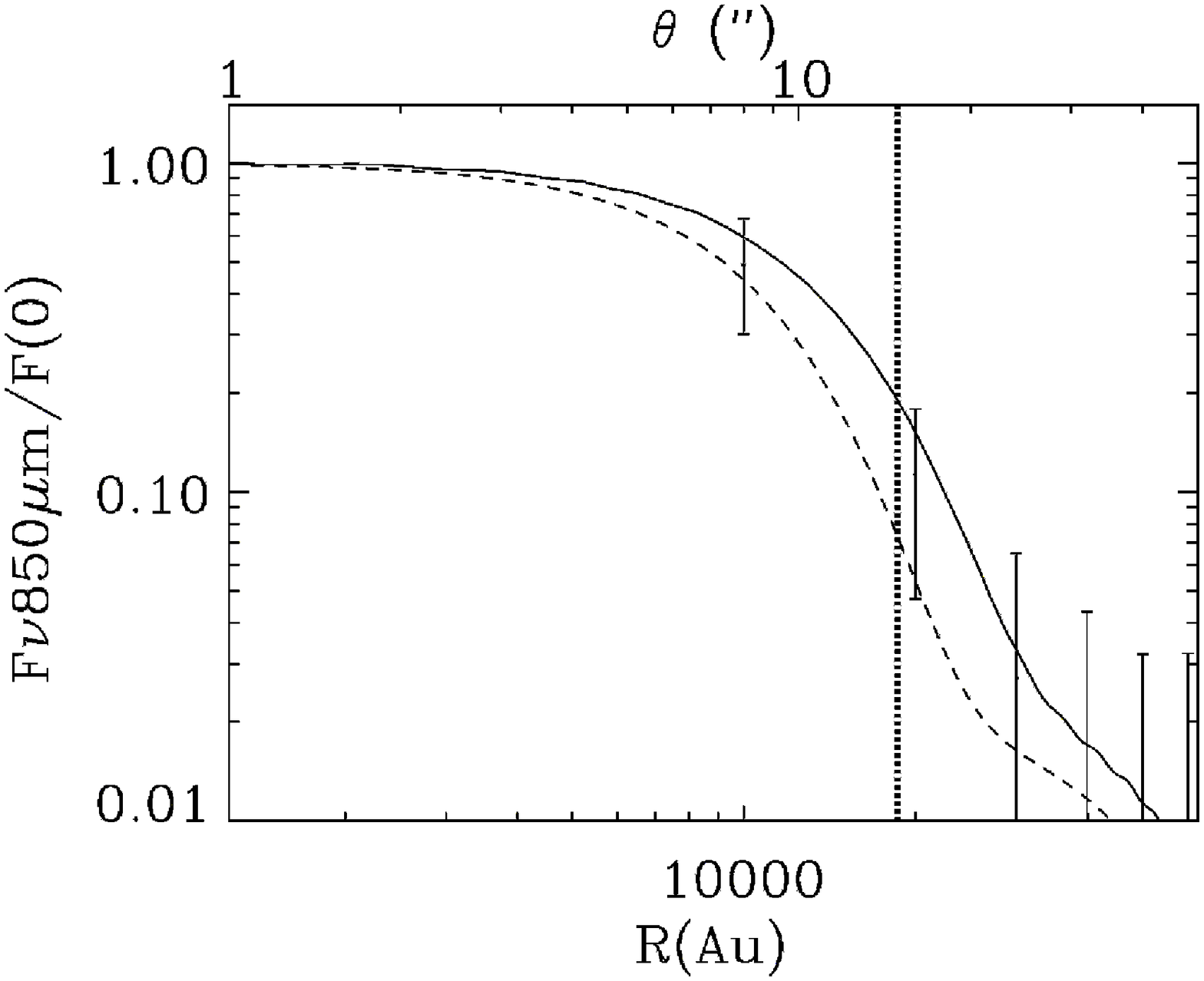}
\includegraphics[width=5cm,angle=90]{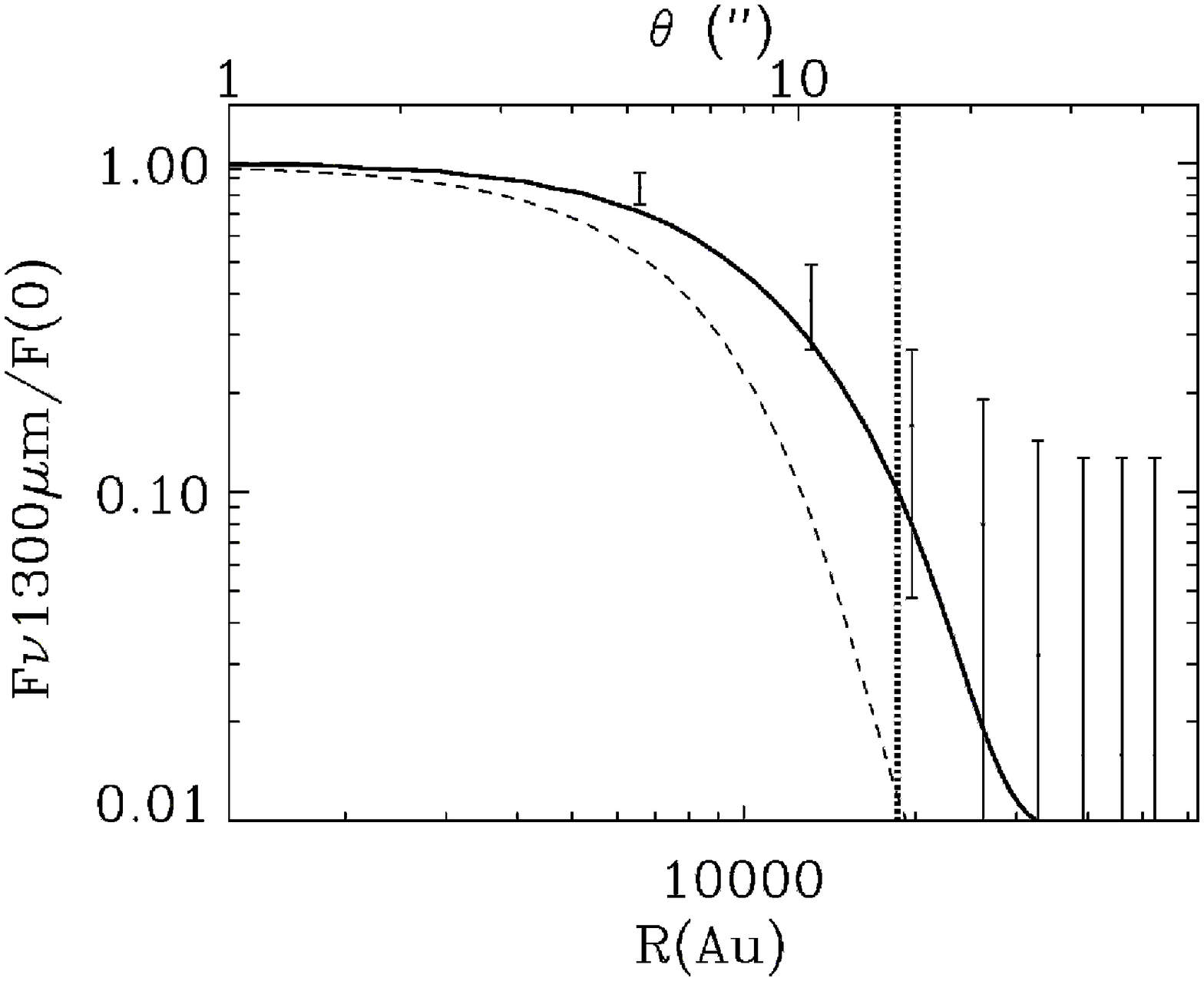}
\includegraphics[width=5cm,angle=90]{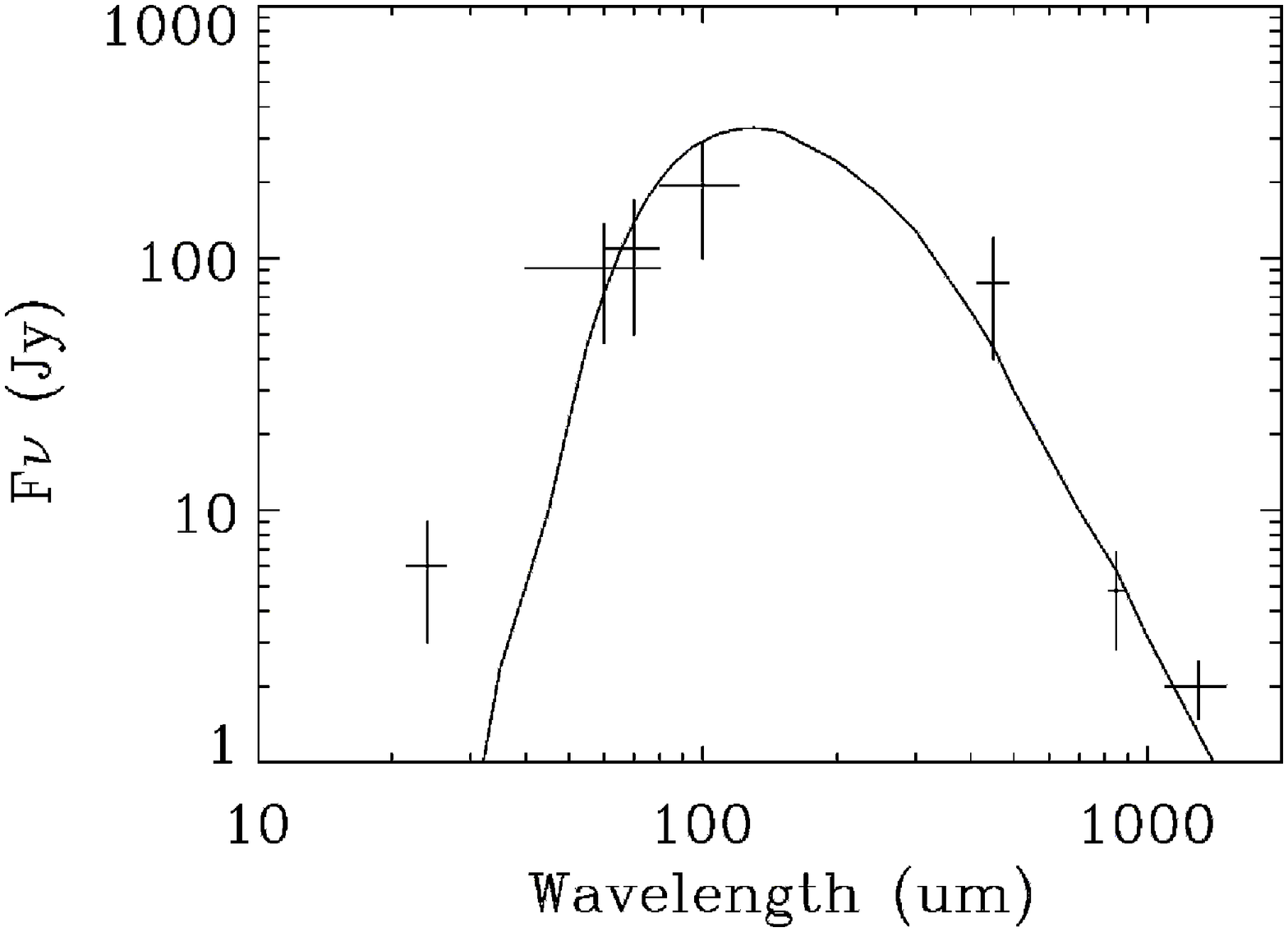}
\caption{Observed brightness profiles at 450 $\mu$m (upper left panel), 850
  $\mu$m (upper right panel), 1300 $\mu$m (lower left panel), and the SED
  (lower right panel) toward NGC7129 FIRS 2. The solid lines report the best-fit model (Table \ref{best_fit_phy_param}). The dashed lines represent the beam pattern of the
  telescope at 450 $\mu$m, 850 $\mu$m, 1300 $\mu$m. The vertical dotted lines show
  the outer radius of the best-fit model. \label{NGC7129_fits}}
\end{figure*}

\begin{figure*} \centering
\includegraphics[width=5cm,angle=90]{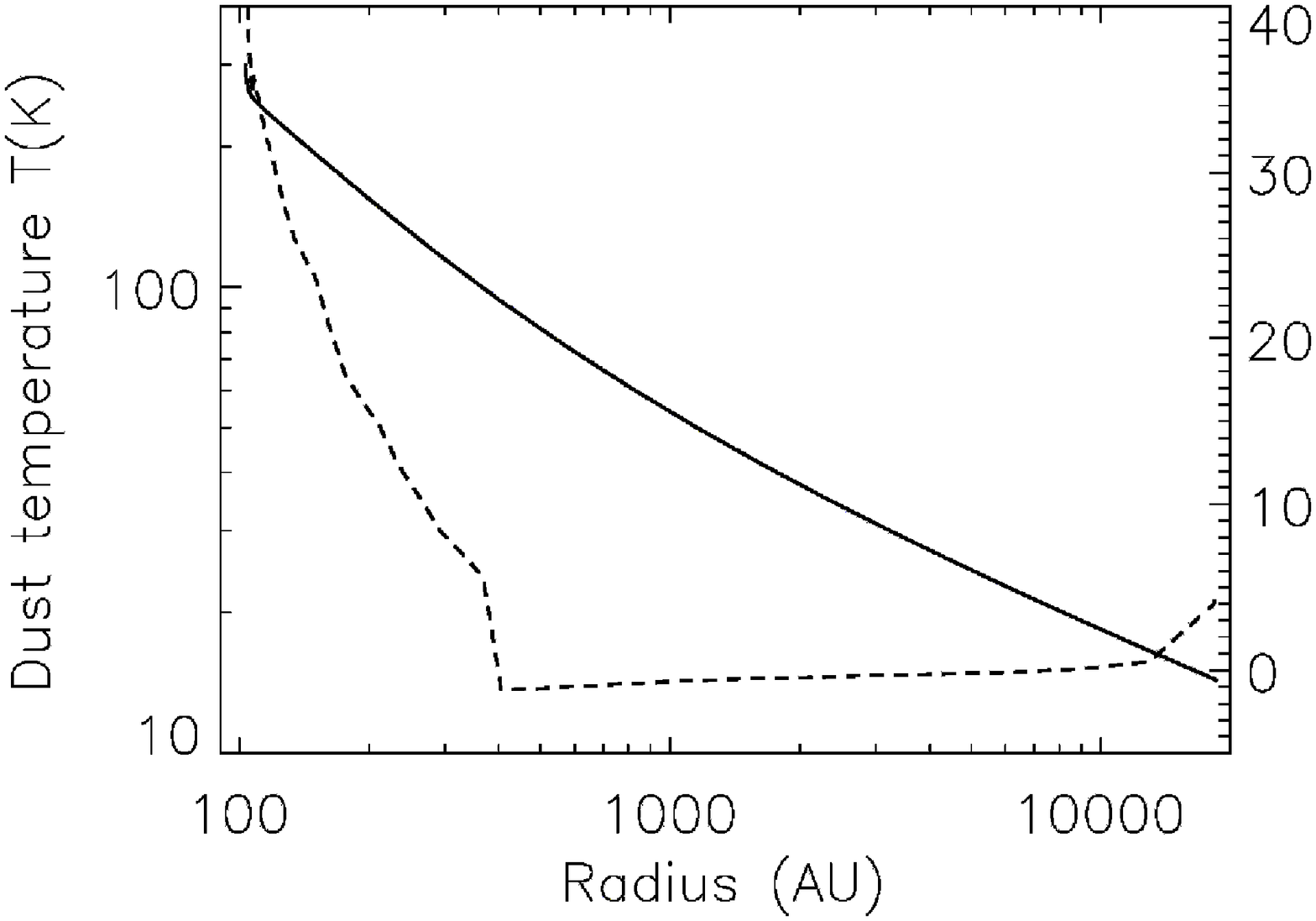}
\includegraphics[width=5cm,angle=90]{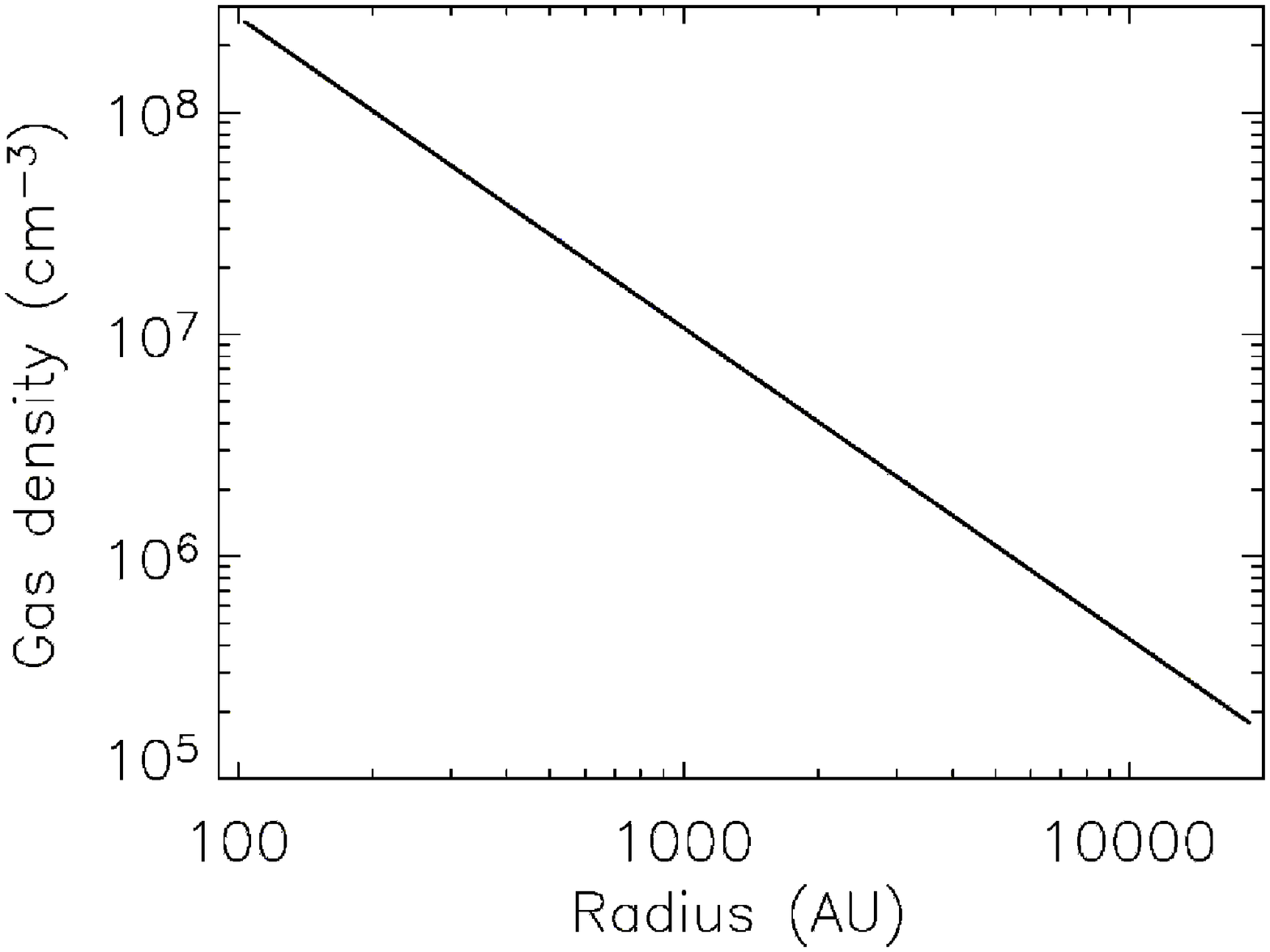}
\caption{NGC7129 FIRS 2 dust temperature (left panel) and H$_{2}$ density
  (right panel) profiles from the best-fit model. The dashed line
  represents the difference between the dust and gas temperature
  derived using X(H$_2$O)$_{in} =$ 1$\times$10$^{-5}$ and
  X(H$_2$O)$_{out} =$ 1$\times$10$^{-7}$ (scale on right axis).\label{NGC7129_T_N}}
\end{figure*}
Table \ref{best_fit_phy_param} presents the set of parameters $\alpha$,
Y, and $\tau_{100}$, which best reproduce the observations, and
summarises some relevant physical quantities of the model. Figure
\ref{NGC7129_fits} shows the relevant derived brightness profiles and
SED against the observations. The dust density and temperature
profiles of the best-fit model are reported in Fig. \ref{NGC7129_T_N}.
The observed flux profiles and SED fluxes from 60 $\mu$m to 850 $\mu$m are
well reproduced by the model. In minimising the $\chi _{SED}^{2}$, we
also varied the source luminosity from 400 L$_\odot$ to 600 L$_\odot$ and found
the best-fit for a source luminosity equal to 500 L$_\odot$.  The
observed flux at 24 $\mu$m is underestimated by about two orders of
magnitude. Considering the $\chi
^{2}_{red}$ contour at 10\% of the minimum $\chi ^{2}_{red}$ to
estimate the uncertainty, we obtained $r_{out}$ = (1.9$\pm$0.2)
$\times10^{4}$ AU.

Finally, similarly to IC1396 N BIMA 2, the gas temperature profile
obtained in NGC7129 FIRS 2 is slightly decoupled from the dust
temperature in the inner part (see Fig. \ref{NGC7129_T_N}).

\subsection{Serpens FIRS 1}\label{subsec:Serpens-FIRS_1}
\subsubsection{Source background}

Since its recognition as a very active star forming region \citep{Strom1974},
 the Serpens molecular cloud has been the target of many
observational studies. Several studies have been dedicated to evaluate
the distance of this cloud from the Sun, yielding a range of 
distances from $\sim$ 210
to 440 pc. Following the review by \citet{Eiroa2008},
which reports and compares the distance results from all the different
studies, we adopt a distance of $\sim$ 230$\pm$20 pc. Serpens is a
young protocluster whose members are in many different evolutionary
stages. About 100 embedded YSO, protostars, and prestellar clumps
\citep{Strom1976,Eiroa1992,Casali1993,Hurt1996,Giovannetti1998,Testi1998,Kaas1999},
 as well as molecular outflows \citep{White1995,Wolf-Chase1998,Williams2000}
 and H$_2$ emission \citep[e.g.][]{Eiroa1997,Herbst1997,Hodapp1999} 
 have been detected in this region.

Serpens FIRS 1 is located near the centre of the
Serpens main core and is the most luminous object embedded in the
cloud. Several continuum studies classify it as Class 0 source
with a bolometric luminosity estimated to range from 46 L$_{\sun}$ to
84 L$_{\sun}$ \citep{Harvey1984,Casali1993,Hurt1996,Larsson2000}.
 The latest estimation of its luminosity suggests
that Serpens FIRS 1 is on the low/intermediate mass protostar border. The envelope physical structure of Serpens FIRS 1 has been modelled by \citet{Hogerheijde1999} using millimetre interferometric observations of the continuum and molecular line observations. 
The Serpens FIRS 1 SED was first modelled by means of a
spherical envelope 1D radiative transfer code by \citet{Larsson2000}.
 Subsequently, \citet{Larsson2002} used a 2D radiative
transfer code and a torus model to reproduce the SED and line
observations from ISO. FIRS1 drives a molecular outflow, which is
orientated at a position angle of 50 degrees, coincident with a triple radio
source with symmetric lobes \citep{Rodriguez1989}.

\subsubsection{Analysis}

\begin{figure*} \centering
\includegraphics[width=6cm,angle=270]{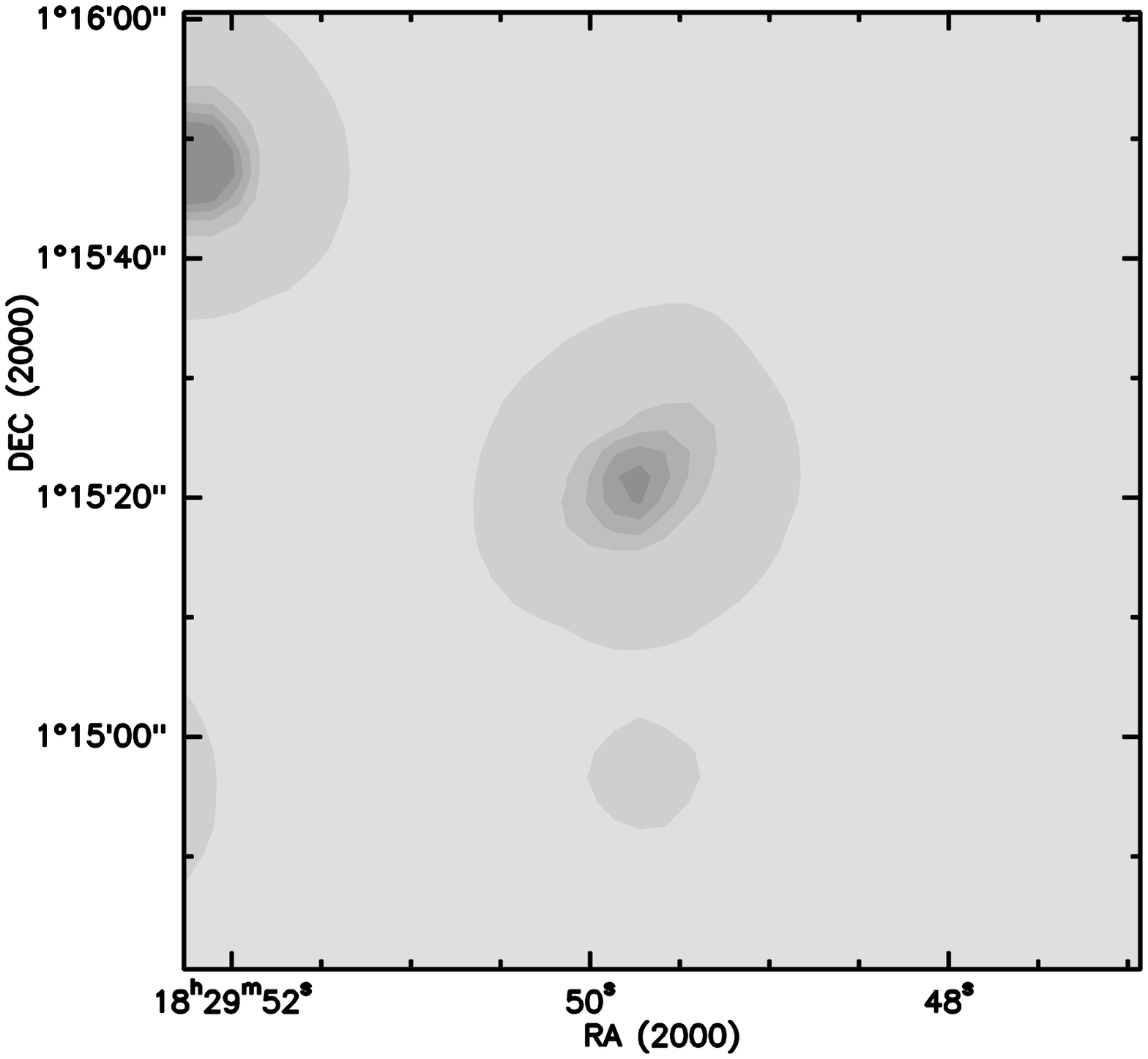}
\includegraphics[width=6cm,angle=270]{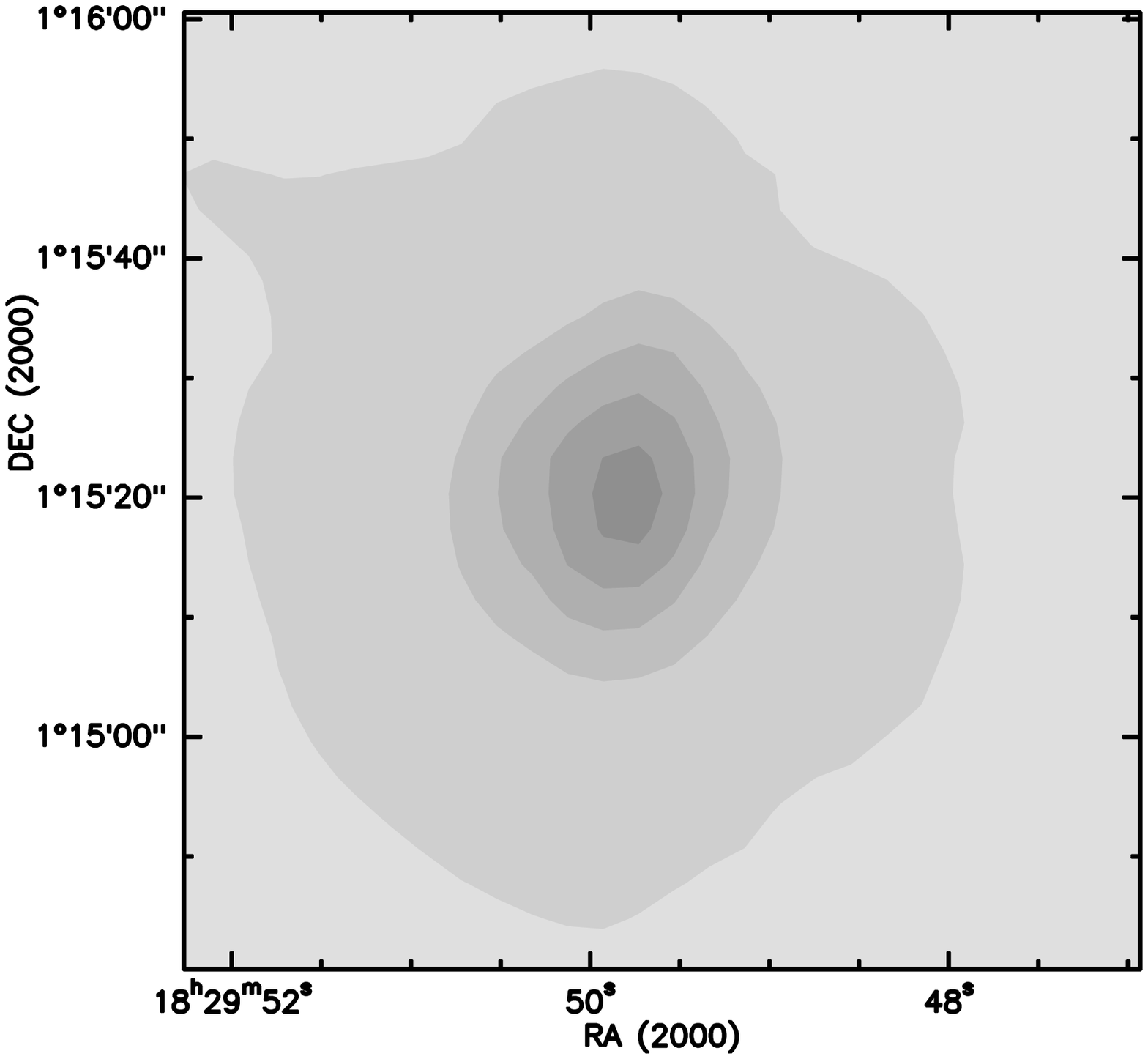}
\includegraphics[width=6cm,angle=270]{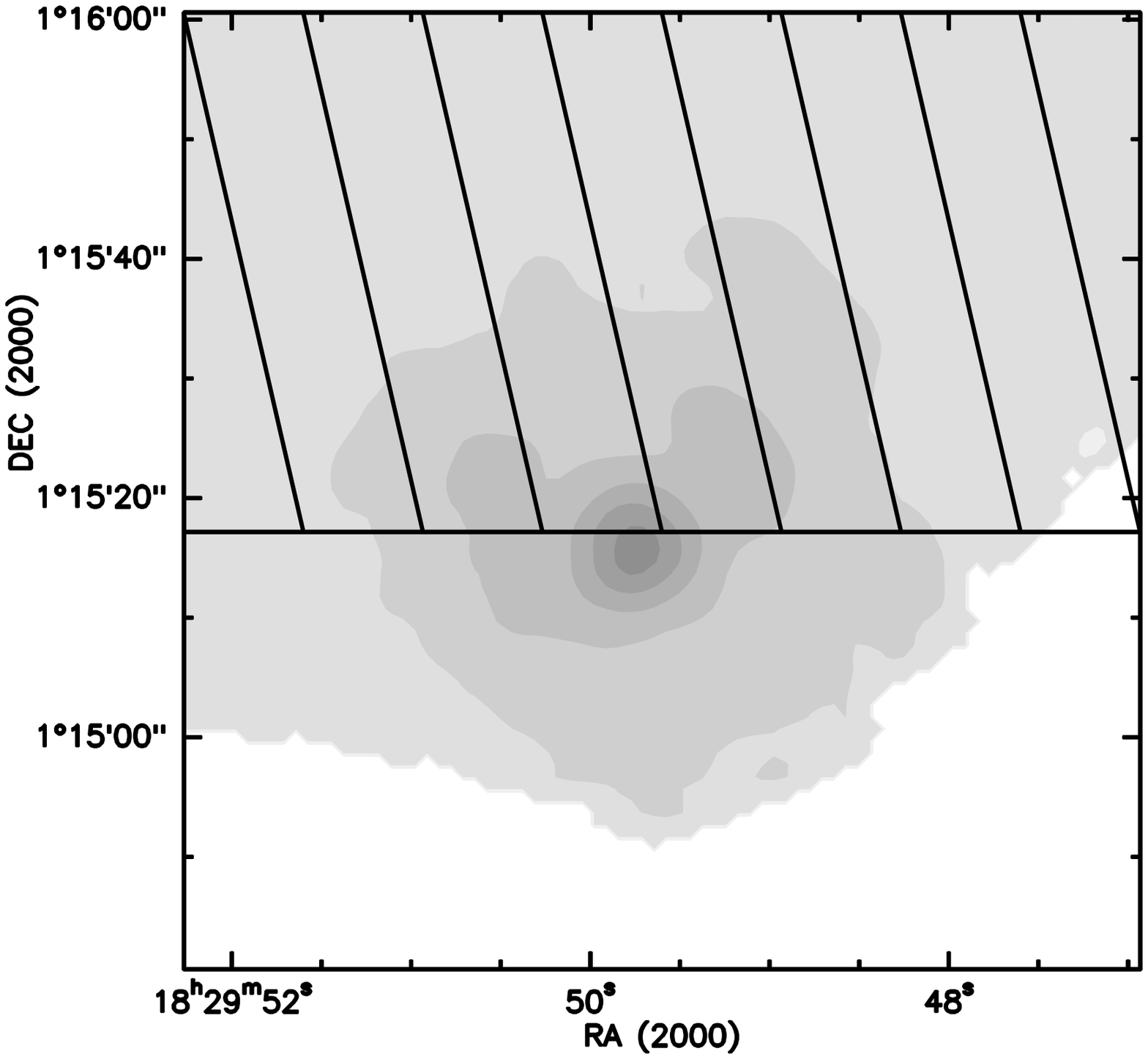}
\includegraphics[width=6cm,angle=270]{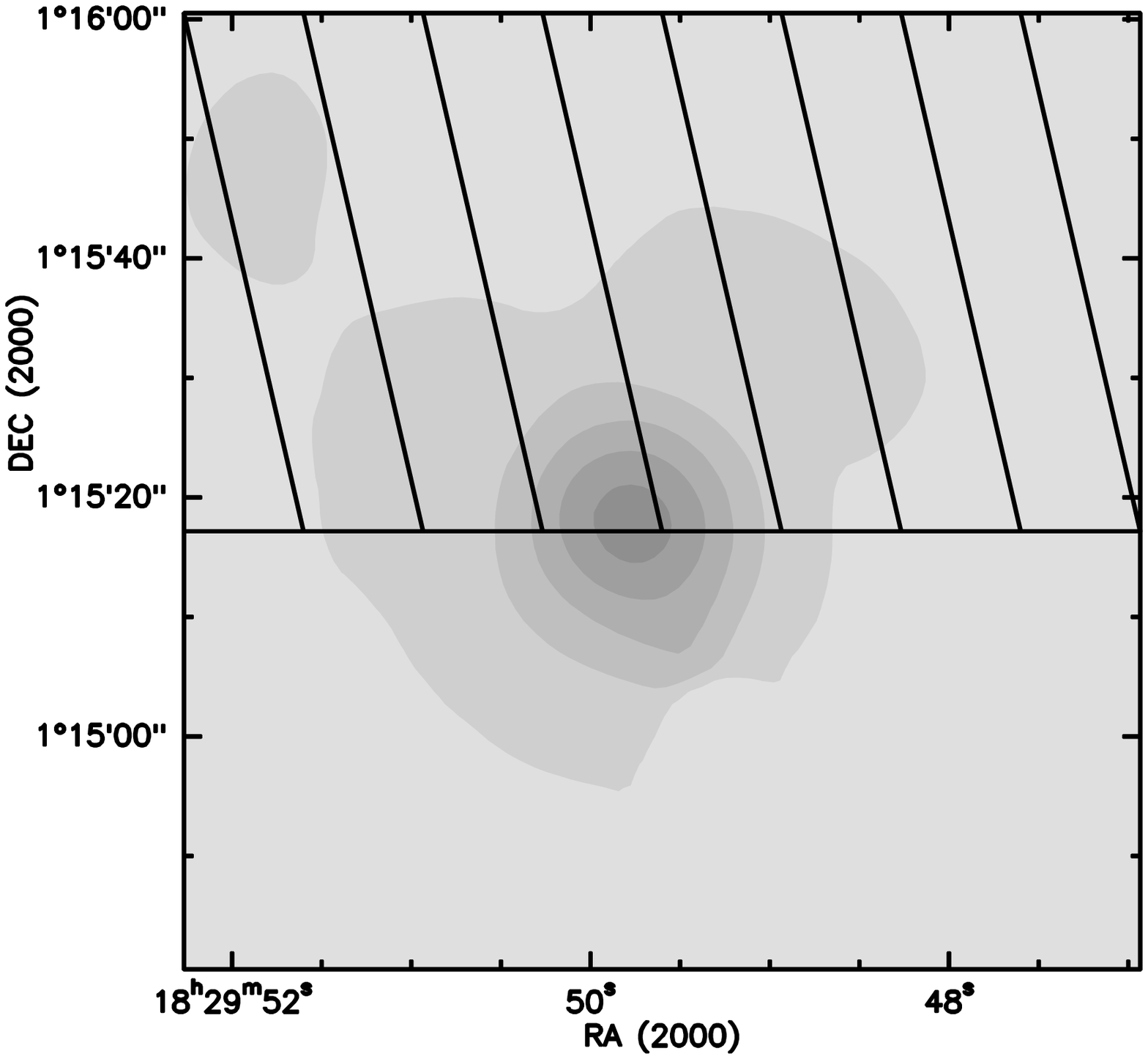}
\caption{Continuum emission maps around Serpens FIRS 1 at 24 $\mu$m
  (upper left panel), 70 $\mu$m (upper right panel), 450 $\mu$m (lower
  left panel), and 850 $\mu$m (lower right panel). The grey scale contours mark
  the continuum flux from 10 \% to 90 \% of the peak emission by steps
  of 20 \%. The dashed regions on the 450 $\mu$m and 850 $\mu$m show the
  regions excluded to derive the brightness profiles. \label{Serp_maps}}
\end{figure*}

The continuum maps used for the Serpens FIRS 1 analysis are presented
in Fig. \ref{Serp_maps}. The SCUBA maps were obtained in January 1998
as a part of the project m97bc30. The flux profiles (see
Fig. \ref{Serp_fits}) are derived excluding the dashed regions in
Fig. \ref{Serp_maps} to avoid the non-spherical extended emission. The
Spitzer observations were obtained in October 2004 as part of the
programme ``From Molecular Cores to Planets, continued'' (AOR: 174, PI:
Neal Evans). The integrated fluxes used for the analysis are reported
in Table \ref{Integrated_fluxes} and in Fig. \ref{Serp_fits}. The
integrated fluxes from 70 $\mu$m to 850 $\mu$m were obtained by integration
over a $\sim$ 40\arcsec. We also report on the SED the lower limits from
Plateau de Bure (PdB) interferometric fluxes at 1.3 mm and 3 mm
\citep{Fuente2007}.

\subsubsection{Best-fit}

\begin{figure*} \centering
\includegraphics[width=5cm,angle=90]{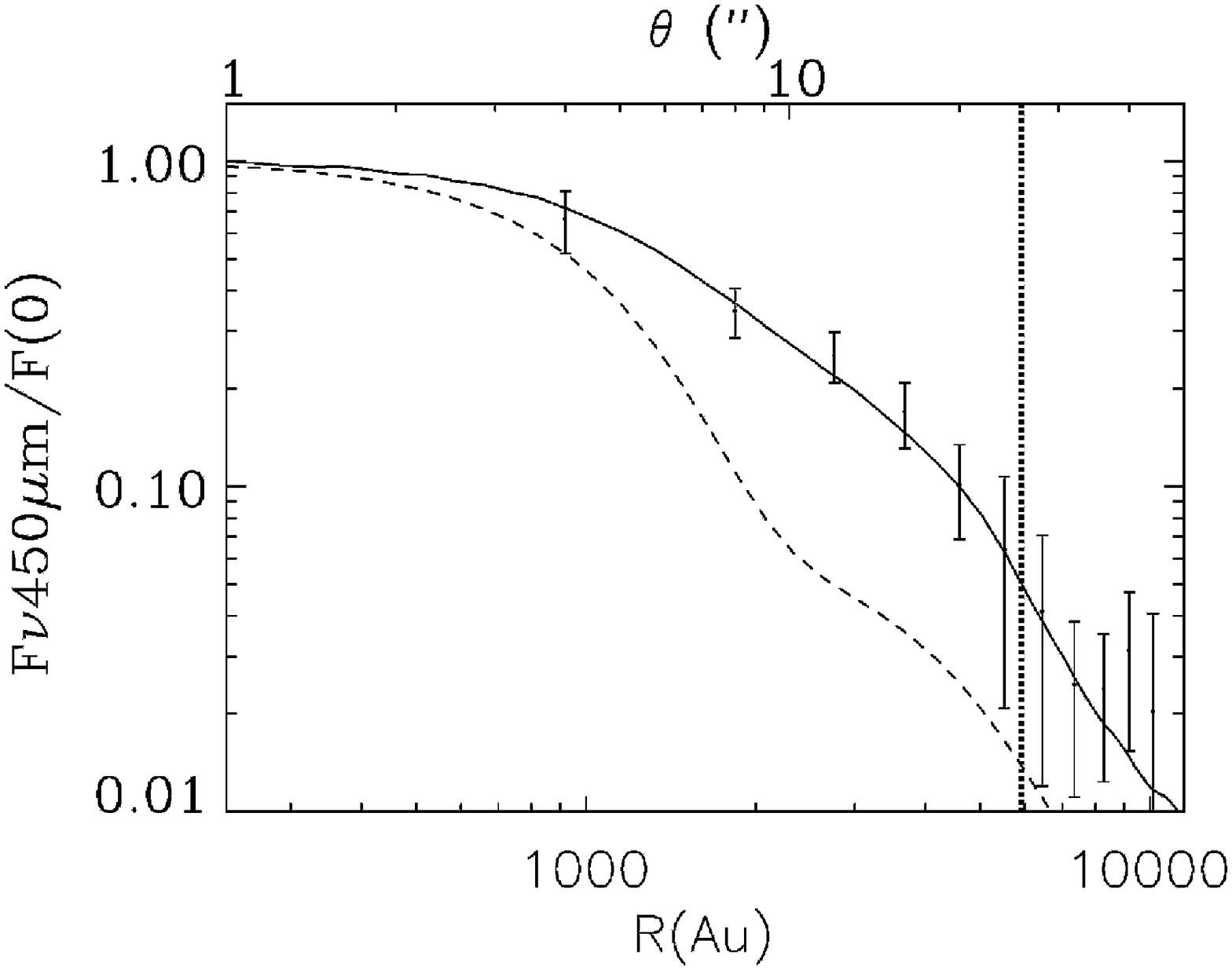}
\includegraphics[width=5cm,angle=90]{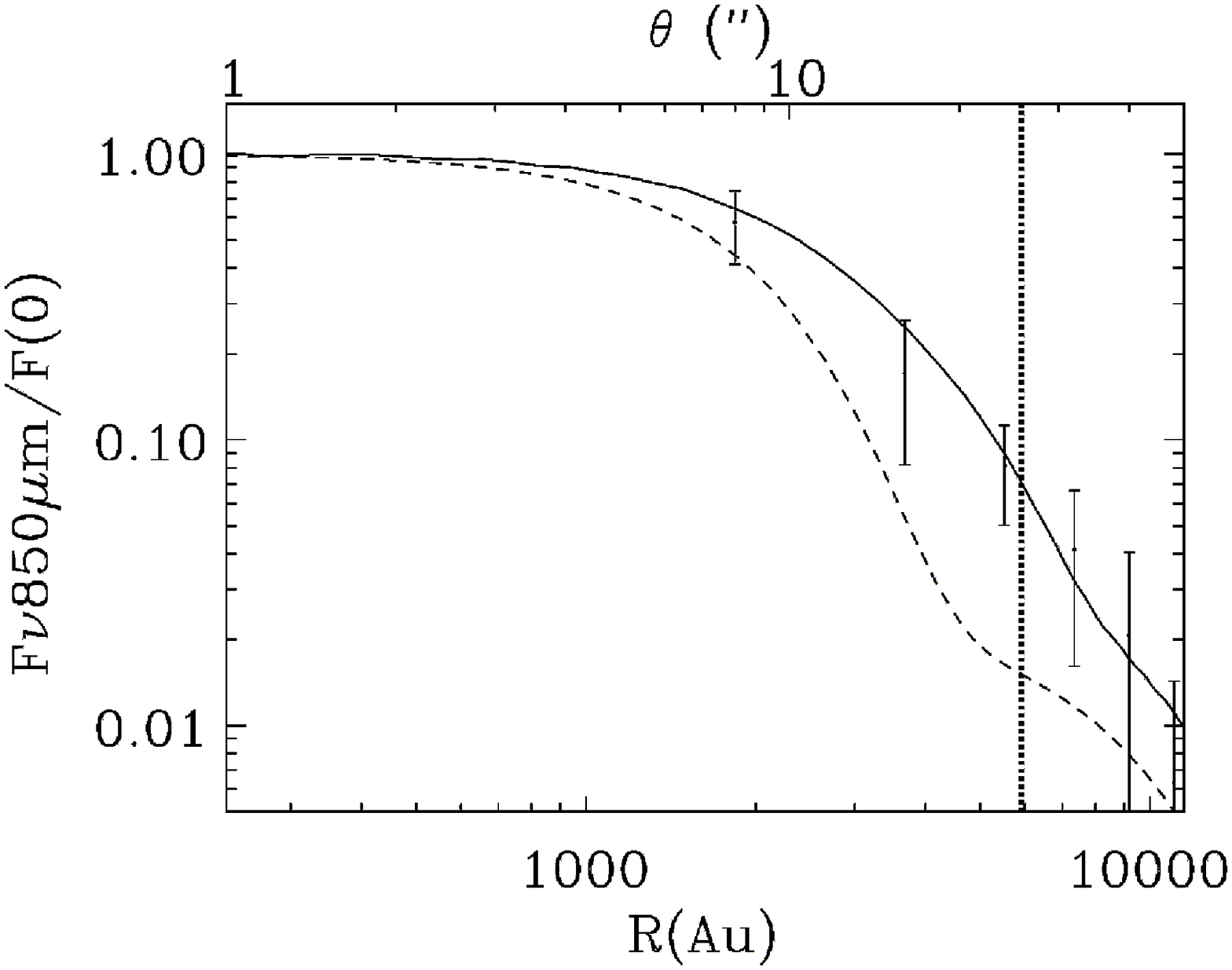}
\includegraphics[width=5cm,angle=90]{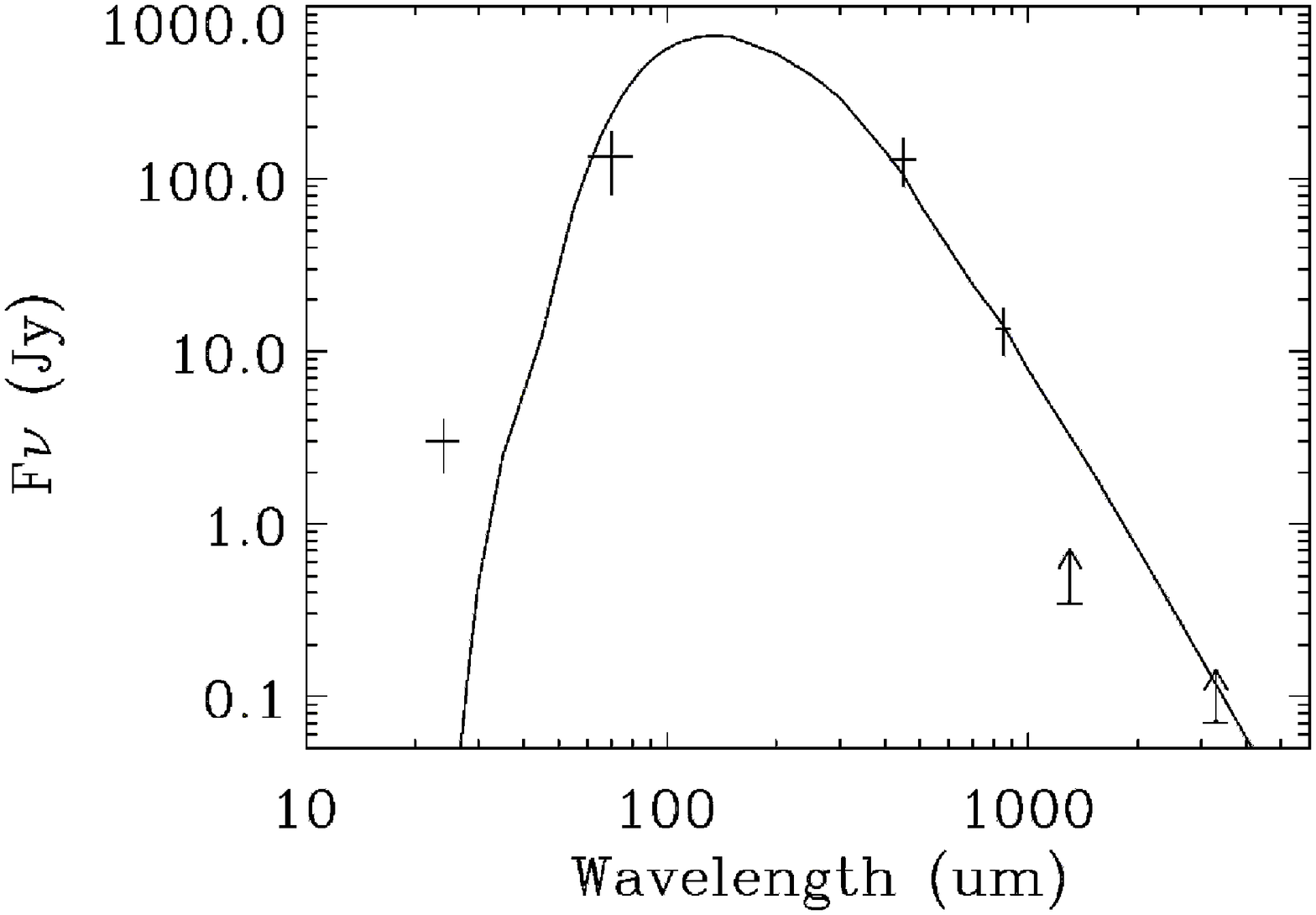}
\caption{Observed brightness profiles at 450 $\mu$m (upper left panel), 850
  $\mu$m (upper right panel) and the SED (lower centred panel) toward
  Serpens FIRS 1. The solid lines report the best-fit model (Table \ref{best_fit_phy_param}). The
  dashed lines represent the beam pattern of the telescope at 450 $\mu$m and
  850 $\mu$m. The vertical dotted lines show the outer radius of the
  best-fit model. The lower limits from Plateau de Bure
  interferometric fluxes at 1.3 mm and 3 mm \citep{Fuente2007} are
  also reported on the SED as lower limit. \label{Serp_fits}}
\end{figure*}

\begin{figure*} \centering
\includegraphics[width=5cm,angle=90]{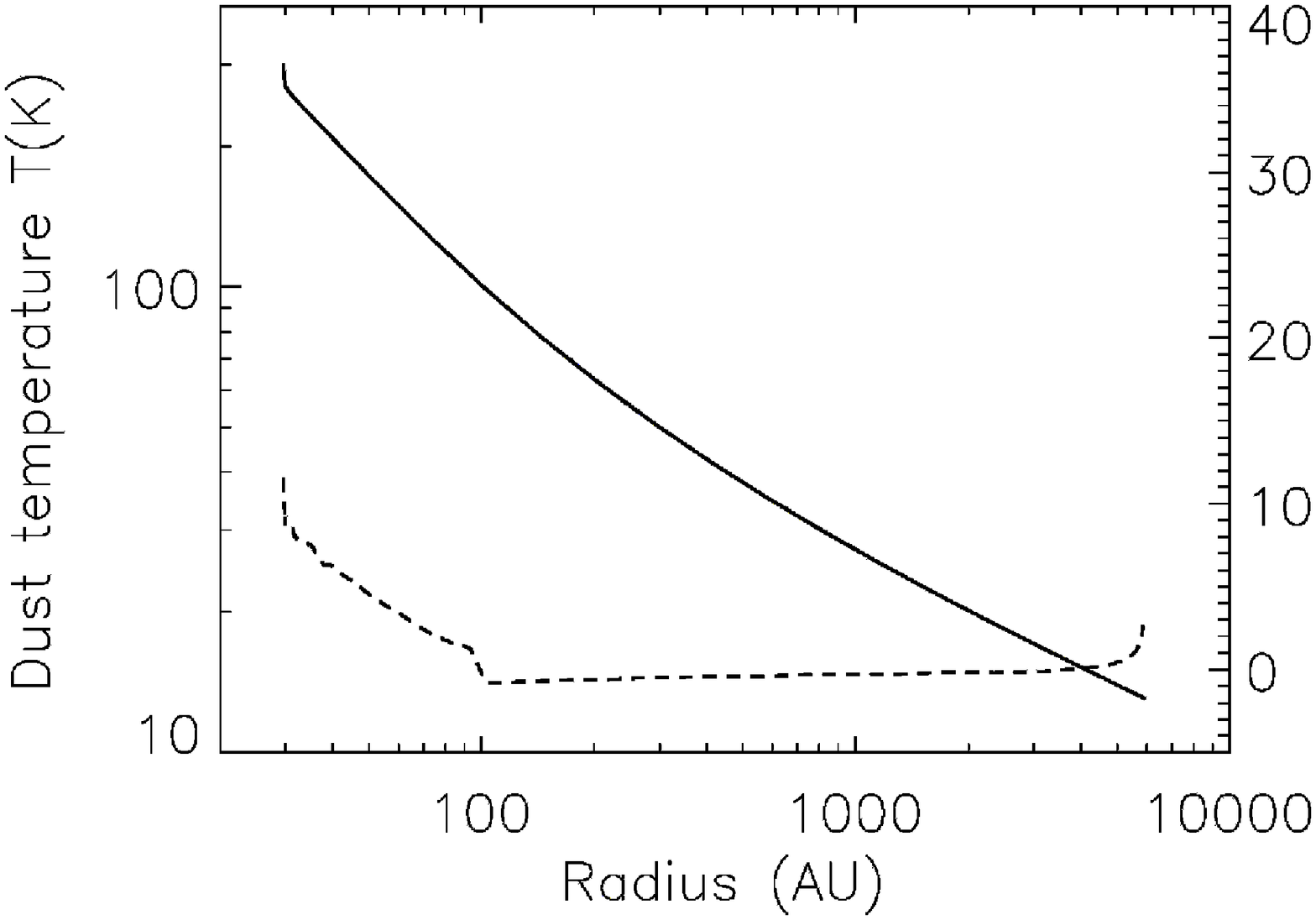}
\includegraphics[width=5cm,angle=90]{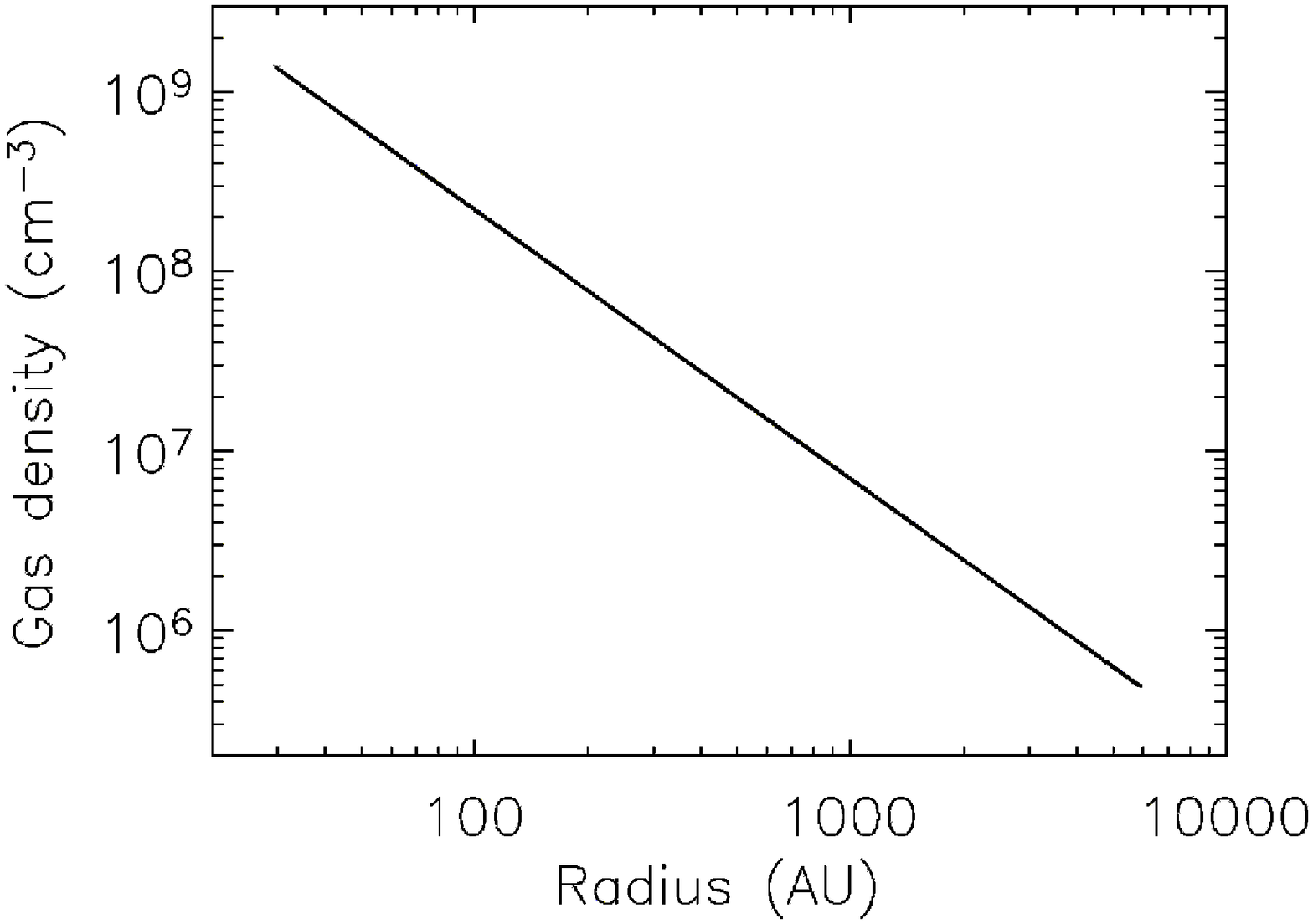}
\caption{Serpens FIRS 1 dust temperature (left panel) and H$_{2}$
  density (right panel) profiles from the best-fit model. The
  dashed line represents the difference between the dust temperature
  and the gas temperature derived using X(H$_2$O)$_{in} =$
  1$\times$10$^{-5}$ and X(H$_2$O)$_{out} =$
  1$\times$10$^{-7}$ (scale on right axis). \label{Serp_T_N}}
\end{figure*}

Table \ref{best_fit_phy_param} presents the set of parameters $\alpha$,
Y, and $\tau_{100}$, which best reproduce the observations, and
summarises some relevant physical quantities of the model. Figure
\ref{Serp_fits} shows the relevant derived brightness profiles and SED
against the observations. The dust density and temperature profiles
of the best-fit model are reported in Fig. \ref{Serp_T_N}. The model
well reproduces the observed flux profiles and SED fluxes from 70 $\mu$m to
850 $\mu$m and agrees well with the lower limits from Plateau
de Bure interferometric fluxes at 1.3 mm and 3 mm \citep{Fuente2007}.
In minimising the $\chi_{SED}^{2}$, we also varied the source
luminosity from 20 L$_\odot$ to 60 L$_\odot$ and found the best-fit for a source
luminosity equal to 33 L$_\odot$.  Note that the estimated value of
the bolometric luminosity is much lower than the values derived in
previous studies toward this source \citep{Harvey1984,Casali1993,Hurt1996,Larsson2000}.
This is due to the adopted smaller
distance to the source.  Considering the $\chi ^{2}_{red}$ contour at
10\% of the minimum $\chi ^{2}_{red}$ to estimate the uncertainty, we
obtained $r_{out}$ = (5.9$\pm$0.3) $\times10^{3}$ AU.

\citet{Hogerheijde1999} studied the envelope physical structure of Serpens FIRS 1 using interferometric continuum observations at OVRO\footnote{Owens Valley Radio Observatory} at $\lambda =$ 3.4-1.4 mm and single-dish line observations from JCMT and CSO\footnote{Caltech Submillimetre Observatory}. They adopted a bolometric luminosity of 77 L$_\odot$ and fixed the outer radius of the envelope R$_{out} =$ 8000 AU for a distance of 400 pc. \citet{Hogerheijde1999} derived an envelope with density profile index $\alpha =$ 2$\pm$0.5 in good agreement with our results. By scaling their results to the same distance adopted here (230 pc) their adopted luminosity, $\sim$ 25 L$_\odot$, and the envelope radius R$_{out}, \sim$ 4600 AU, are relatively consistant with our results (Table \ref{best_fit_phy_param}) : L $\sim$ 33 L$_\odot$, R$_{out} \sim$ 5900 AU. Also their derived gas temperature profile slightly differs from ours: for example they obtained T$_{gas} \sim$ 40 K at 200 AU while we obtained T$_{gas} \sim$ 65 K. \citet{Larsson2000} studied the SED of Serpens FIRS 1, derived from the ISO LWS maps and photometric data between $\sim$ 1 and 10$^5$ $\mu$m retrieved from the literature. These authors used i) a
single temperature/optically thin approximation and ii) a radiative
transfer model to analyse the SED. In their models they adopted a
distance of 310 pc, a source diameter of 7\arcsec and a power law density
profile index, $\alpha =$2. They succeeded in reproducing the SED between 
40 $\mu$m and 3 mm. By scaling their results to the same distance adopted here 
(230 pc), they obtain a luminosity of $\sim$
40 L$_{\odot}$, against our estimate of 33 L$_\odot$, and an envelope
mass of 1.1-3.3 M$_{\odot}$, against our estimate of 5
M$_{\odot}$. However, as noticed by the authors themselves, their mass
estimate is rather a lower limit, given the optically thin
approximation they used.

Finally, the observed flux at 24 $\mu$m is underestimated by about two
orders of magnitude. This phenomenon is also observed 
 in the \citet{Larsson2002,Larsson2000} studies. We discuss
 the situation at 24 $\mu$m in
more detail in the the following section, Sect. \ref{sec:Discussion}.

\end{appendix}

\end{document}